\documentclass[preprint]{aastex}

\usepackage{multirow}
\usepackage{ccaption}

\newcommand\spitzer{{\it Spitzer}}
\newcommand\lath{\hbox{$^{h}$}}
\newcommand\latm{\hbox{$^{m}$}}
\newcommand\lats{\hbox{$^{s}$}}

\begin{document}

	\title{The \spitzer\ survey of interstellar clouds in the Gould Belt. II. The Cepheus Flare observed with IRAC and MIPS} 
	\shorttitle{The Cepheus Flare observed with IRAC and MIPS}
	
	\author{ Jason M.~Kirk\altaffilmark{1},	Derek Ward-Thompson\altaffilmark{1}, James Di Francesco\altaffilmark{2}, Tyler L. Bourke\altaffilmark{3}, Neal J. Evans II\altaffilmark{4}, Bruno Mer{\'{\i}}n\altaffilmark{5}, Lori E. Allen\altaffilmark{3}, Lucas A. Cieza\altaffilmark{6}, Michael H. Dunham\altaffilmark{4}, Paul Harvey\altaffilmark{4}, Tracy Huard\altaffilmark{7}, Jes K. J{\o}rgensen\altaffilmark{8}, Jennifer F. Miller\altaffilmark{7}, Alberto Noriega-Crespo\altaffilmark{9}, Dawn Peterson\altaffilmark{3}, Tom P. Ray\altaffilmark{10}, Luisa M. Rebull\altaffilmark{9}}
	
	\altaffiltext{1}{School of Physics and Astronomy, Cardiff University, Queens Buildings, The Parade, Cardiff, CF24 3AA, United Kingdom; {jason.kirk@astro.cf.ac.uk}, {derek.ward-thompson@astro.cf.ac.uk} }
	\altaffiltext{2}{National Research Council of Canada, Herzberg Institute of Astrophysics, 5071 West Saanich Road, Victoria, BC V9E 2E7, Canada; {james.difrancesco@nrc-cnrc.gc.ca} }
	\altaffiltext{3}{Smithsonian Astrophysical Observatory, 60 Garden Street, MS42, Cambridge, MA 02138; {bourke@cfa.harvard.edu}, {leallen@cfa.harvard.edu}, {dpeterson@cfa.harvard.edu} }
	\altaffiltext{4}{Department of Astronomy, University of Texas at Austin, 1 University Station, C1400, Austin, TX 78712-0259; {nje@bubba.as.utexas.edu}, {mdunham@astro.as.utexas.edu}, {pmh@astro.as.utexas.edu} }
	\altaffiltext{5}{Herschel Science Centre, European Space Astronomy Center (ESA), P.O. Box 78, E-28691 Villanueva de la Ca\~{n}ada, Madrid, Spain; {Bruno.Merin@sciops.esa.int} }
	\altaffiltext{6}{Institute for Astronomy, University of Hawaii, Manoa, HI 96822; {lcieza@ifa.hawaii.edu} }
	\altaffiltext{7}{Astronomy Department, University of Maryland, College Park, MD 20742; {thuard@astro.umd.edu}, {jfm@astro.umd.edu} }
	\altaffiltext{8}{Argelander-Institut f\"{u}r Astronomie, University of Bonn, Auf dem H\"{u}gel 71, 53121 Bonn, Germany; {jes@astro.uni-bonn.de} }
	\altaffiltext{9}{\spitzer\ Science Center, California Institute of Technology, Pasadena, CA, 91125; {alberto@ipac.caltech.edu}, {rebull@ipac.caltech.edu} }
	\altaffiltext{10}{School of Cosmic Physics, Dublin Institute for Advanced Studies, 5 Merrion Square, Dublin 2, Ireland; {tr@cp.dias.ie} }
	
	\shortauthors{Kirk et al.}
	
	\begin{abstract}
	We present \spitzer\ IRAC ($\sim$2 deg$^{2}$) and MIPS ($\sim$8 deg$^{2}$) observations of the Cepheus Flare which is associated with the Gould Belt, at an approximate distance of $\sim$300~pc. Around 6500 sources are detected in all four IRAC bands, of which $\sim900$ have MIPS 24~\micron\ detections. We identify 133 YSO candidates using color-magnitude diagram techniques, a large number of the YSO candidates are associated with the NGC 7023 reflection nebula. Cross identifications were made with the Guide Star Catalog II and the IRAS Faint Source Catalog, and spectral energy distributions (SED) were constructed. SED modeling was conducted to estimate the degree of infrared excess. It was found that a large majority of disks were optically thick accreting disks, suggesting that there has been little disk evolution in these sources. Nearest-neighbor clustering analysis identified four small protostellar groups (L1228, L1228N, L1251A, and L1251B) with 5-15 members each and the larger NGC 7023 association with 32 YSO members. The star formation efficiency for cores with clusters of protostars and for those without clusters was found to be $\sim8\%$ and $\sim1\%$ respectively. The cores L1155, L1241, and L1247 are confirmed to be starless down to our luminosity limit of $L_{\rm bol}=0.06$~L$_{\odot}$.
	\end{abstract}
	
	
\section{Introduction}

	The Gould Belt is a band of stars and molecular clouds that encircle the sky at an inclination of $\sim$20\degr\ to the Galactic Plane \citep{1847herschel, 1879gould}. It is the locus of star formation within 140--500 pc of the Sun and includes many, well-known star formation regions (Serpens, Ophiuchus, Orion, etc). The \spitzer\ Gould Belt Survey \citep[SGBS;][]{2008allen} is a \spitzer\ GO-4 legacy project designed to extend the earlier successful \spitzer\ Cores to Disks \citep[c2d;][]{2003evans,2008evans} legacy project and complete a census of star formation within 500~pc. In addition, SGBS and c2d data complement data from the upcoming JCMT \citep{2007scubagb} and {\it Herschel} \citep{2005andre} Gould Belt surveys. The \spitzer\ Space Telescope is a 85-cm-diameter cryogenically cooled satellite telescope designed to operate throughout the infrared regime \citep{2004werner}. Its two instruments used in this study are the Infrared Array Camera \citep[IRAC;][]{2004fazio} which can observe at 3.6--8.0~\micron, and the Multiband Imaging Photometer for \spitzer\ \citep[MIPS;][]{2004rieke}, which can observe at 24--160~\micron. In this paper we focus on \spitzer\ data from one part of the Gould Belt known as the Cepheus Flare region.

\begin{figure*}
	\centering{
		\includegraphics[width=0.98\textwidth]{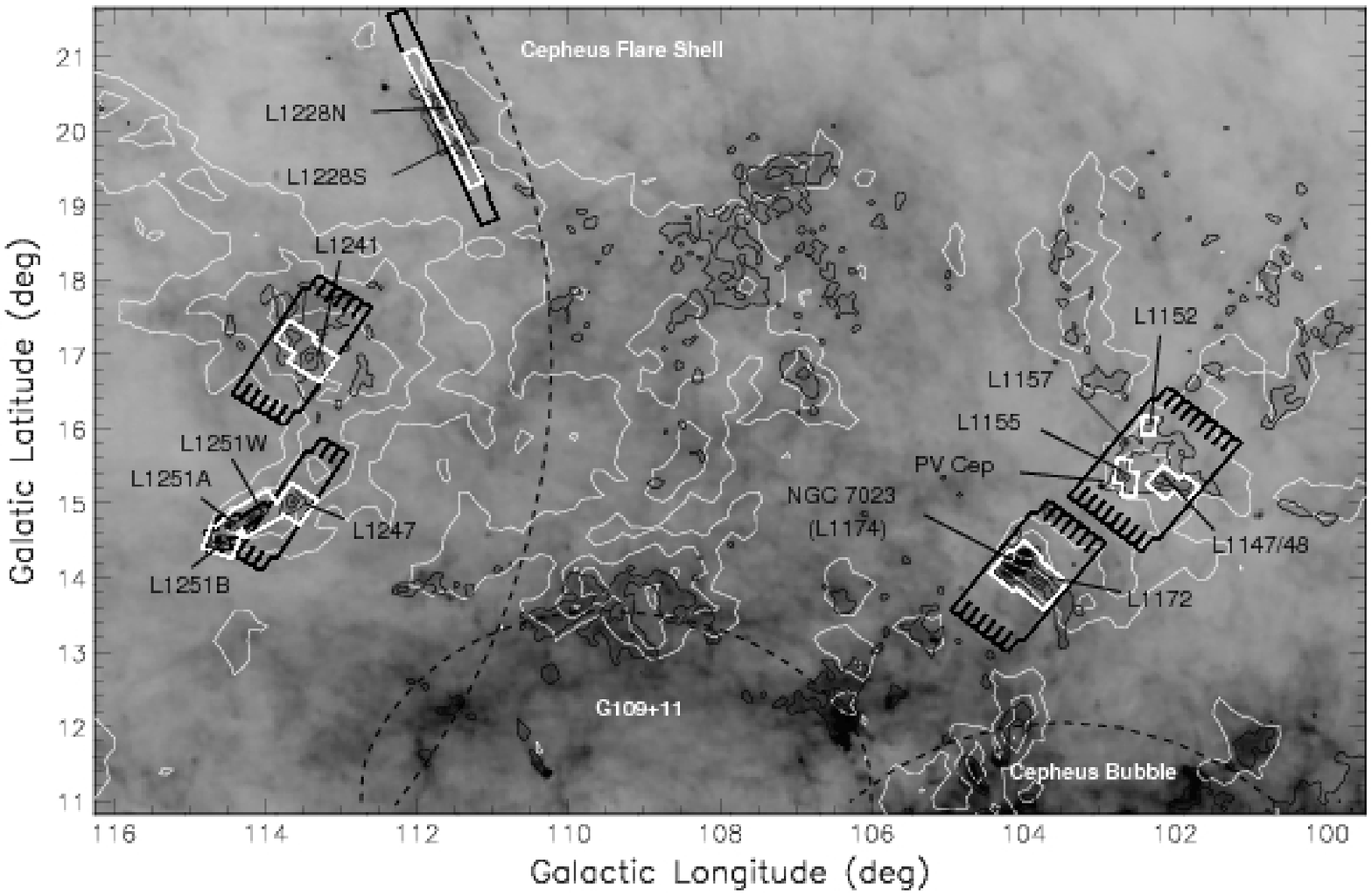}
		}
	\caption{A finding chart for the $\sim$300~pc Cepheus Flare region showing the positions of the regions discussed in this paper. The areas mapped with \spitzer\ are shown as black (24~\micron\ MIPS) and white (3.6~\micron\ IRAC) footprints. The labels are the names of selected Lynds dark nebulae \citep{1962lynds}. Also shown are the reflection nebula NGC 7023 and the high velocity YSO PV Cep that is reported to have been ejected from it \citep{2004goodman}. The underlying gray-scale map shows IRIS (Improved Reprocessed IRAS Survey) 100~\micron\ emission \citep{2005miville}. The white contours show integrated CO emission at levels of  5.5, 12, and 18.5 K kms$^{-1}$ \citep{2001dame}. The black contours show visual extinction towards the Cepheus region generated from the Digitized Sky Survey with levels at $A_V$=1, 2 and 3 mag \citep{2005dobashi}. The dashed-black lines denote large scale loops and shells (see text for details). The L1221 region is not shown, but see Table \ref{tab:assocprop} for its relative position. }
	\label{fig:coverage}
\end{figure*}

	The Cepheus Flare is a complex of nebulae that extends 10-20\degr\ out of the plane of the Galactic disk at a Galactic longitude of $\sim$110\degr\ \citep{1934hubble}. Star formation towards the Flare can be broken down into that associated with the wall of the Local Bubble at $\sim$160 pc, that associated with the sweep of the Gould Belt at $\sim$300~pc, and that associated with the Perseus arm of the galaxy at $\sim$800~pc \citep{1997yonekura,2006kiss}. Figure \ref{fig:coverage} shows a visual extinction \citep{2005dobashi}, CO \citep{2001dame}, and IRAS 100~\micron\ \citep{2005miville} finding chart towards the Cepheus Flare. It reveals five associations of dark clouds, each located at the middle distance, including L1148+L1152+L1155, L1172+L1174, L1228, L1241 and L1247 + L1251. In this paper, we present IRAC and MIPS data obtained by the SGBS program towards these dark cloud associations and the related dark cloud L1221 which is outside the region shown in Figure \ref{fig:coverage}.
	
	The white contours in Figure \ref{fig:coverage} show the integrated $^{13}$CO emission surrounding the dark cloud associations \citep{2001dame}. Table \ref{tab:assocprop} lists the cloud positions along with their inferred masses and $^{13}$CO line widths from \citet{1997yonekura}. \cite{1998kun} matched sources in the IRAS point source catalogs with H$\alpha$ data to identify a catalog of pre-main sequence stars and dense cores in Cepheus. Their distribution of sources showed that star formation was occurring along the cloud edges. Similarly, Figure \ref{fig:coverage} shows that regions of high extinction are also distributed towards the edges of the clouds (the lowest CO contours). 
		 
	The Cepheus Flare is an expansive and sparsely filled region. Studies have attempted to define structure within the Flare by grouping the scattered objects into associations and groups. \citet{1997yonekura} mapped Cepheus in CO and identified several large associations of dark clouds. The clouds presented in this paper correspond to Yonekura Group A. \cite{2006kiss} studied the cloud morphology in Cepheus using their own 256 square degree extinction map. They identified eight cloud complexes across the wider area.  
	
	The Cepheus Flare is bounded by a series of shells and loops. Three of these are shown in Figure \ref{fig:coverage} as dashed lines. The Cepheus Flare Shell is an expanding supernova bubble at a distance comparable to the clouds in this paper. Its center is located at $l\sim120^{\circ}$, $b\sim17^{\circ}$ (approximately 7\degr\ east of L1241) and it has a radius of $\sim9.5^{\circ}$ (50~pc at 300~pc) \citep{1989grenier, 2006olano}. It is possible that the star formation within L1251 has been triggered by the passage of this shell \citep{2006olano, 2008kun}. The dark cloud L1228 is coincident with the current radius of the shell \citep{2008kun}. Other star formation regions associated with the Cepheus Flare Shell include L1333 ($l=129^{\circ}$, $b=+13^{\circ}$) at a distance of 180~pc \citep{1998obayashi}. L1333 is in the neighboring constellation of Cassiopeia, on the opposite side of the Cepheus Flare Shell \citep{2008kun}. The Cepheus Flare Shell and the older, larger Loop III supernova shell appear to be concentric spheres \citep{2008kun}. 
	
	The Cepheus Shell should not be confused with the Cepheus Bubble which is an expanding dust ring that surrounds the Cep OB2 association \citep{1998patel, 2000abraham}. \spitzer\  observations of three clusters associated with Cep OB2 found that 10\% of detected disks were ``transition'' objects \citep[their spectral energy distributions, SEDs, are essentially photospheric except for an infrared excess at the longest wavelengths; ][]{2006aguilar}. Clouds coincident with this Bubble are excluded from our analysis as they are believed to be associated with more distant material \citep{2000abraham}. Other star formation groups in the direction of Cepheus include S140, Cep OB3 and Cep OB4, These are part of the Yonekura Distant Group, which is at a distance of 600-800~pc \citep{1997yonekura}. The infrared loop G109+11 was identified by \citet{2004kiss} who later associated it with ``Void \#2'' in their extinction survey and a bright rim of excess 12~\micron\ emission \citep{2006kiss}.  
	
	In Section, \ref{distance} we discuss distance estimates to the Cepheus dark cloud associations. In Section, \ref{obs} we outline the observation strategy, the data reduction procedure, and present the assembled catalog and false-color images of the region. In Section \ref{ysos}, we use several techniques to identify YSO candidates based on their infrared colors. We classify the candidates based on their spectral indices and present color-color diagrams of the resulting list. In Section \ref{photometric}, we analyze the photometry of the YSO candidates, adding additional data where available, to produce spectral energy distributions. From these we calculate the Cepheus Flare luminosity function and conduct basic SED modeling. In Section \ref{extended}, we look at the relation of the YSO candidates to their surroundings and present extinction and MIPS 160~\micron\ maps as well as the results of clustering analysis and comparisons with existing surveys. In Section 7, we discuss a star formation scenario for the Cepheus Flare region. In Appendix \ref{regions}, we discuss each of the individual dark cloud associations and compare results with archival 850~\micron\ submillimeter continuum data.

\section{Distance}
\label{distance}

	The Cepheus Flare contains a number of different components whose distributions on the sky overlap, but whose velocities and distances are different (see \citealt{2008kun} for a review). The dark cloud associations shown in Figure \ref{fig:coverage} were selected from the \citet{2005dobashi} extinction map because they each had peak $A_{\rm V} > 3$ and are within a distance of $\sim$500~pc. The selected clouds divide into a Galactic West Group comprising the L1172+L1174 and L1148+L1152+L1155 associations at a Galactic longitude of 101--105$^\circ$ and a Galactic East Group comprising the L1228, L1241, and L1247+L1251 associations at a Galactic longitude of 111--115$^\circ$. There is a noticeable lack of regions with $A_{\rm V} > 3$ in the span 106--110$^\circ$ although it does not mean that this span is completely devoid of young stars \citep{2005tachihara, 2008kun} or of molecular material as evidenced by the CO and extinction maps. However, the lack of $A_{\rm V} > 3$ extinction means that this region falls outside of our selection rules for \spitzer\ imaging. The arc of the Cepheus Flare Shell would appear to divide the active Galactic East star formation group and the barren central molecular mass of the Cepheus Flare. Thus the  Cepheus Flare Shell appears to be triggering star formation as its sweeps across the Flare.  
	
	The Galactic West Group of dark cloud associations is surrounded by common low-level integrated molecular emission suggesting that they are at a similar distance (e.g., see Figure \ref{fig:coverage}). Many of the distance estimates to these clouds ultimately rely on distance estimates to the Herbig AeBe star HD 200775 embedded within L1174. This is the driving source of the NGC 7023 reflection nebula, but estimates of its distance have been hampered by uncertainties in its spectral type. \citet{1981whitcomb} surveyed the state of the art at the time of their publication and found a range of distances from 350 to 600 pc before adopting a distance of 440~pc. This distance was used by many studies of molecular cores \citep[e.g.][]{1987myers,1999wardthompson}. This value is in agreement with the {\it Hipparcos} distance of $430^{+160}_{-90}$~pc to HD 200775 \citep{1997ancker,1999bertout}.
	
	Estimates of the distance to L1174 based on diagrams of color excess-distance modulus give estimates of $300\pm20$~pc \citep{1989shevchenko} and $288\pm25$~pc \citep{1992straizys}. The second estimate is based on a group of reddened stars in the proximity of the nebula, but does not include the actual distance to HD 200775. For that star, \citet{1992straizys} derived alternative distances of 275~pc and 417~pc and adopted the former based on the correlation with their first estimate. \citet{1992straizys} estimated a distance to the L1148/55 molecular ring of 325$\pm$13~pc. Given the uncertainty of the HD 200775 distance \citep{2008kun} we adopt the \citeauthor{1992straizys} distances to L1174 and L1148+L1152+L1155.  

	The Galactic West Group of dark cloud associations are separated from the Galactic East Group of associations by $\sim10\degr$. At the adopted distance of these clouds this is $\sim50$~pc. \citet{1998kun} used Wolf Diagrams to estimate distances of $200^{+100}_{-20}$~pc for L1228 and $300^{+50}_{-10}$~pc for L1241 and L1247/L1251. The distance to L1251 matches the distance adopted by \cite{2006lee}. Despite their proximity on the sky, L1228 and L1251 are actually on opposite surfaces of the projected Cepheus Flare Shell and thus have different distances \citep{2008kun}. For L1221, we use the distance adopted by \citet{2008young}. Table \ref{tab:assocprop} lists the adopted distances of the dark cloud associations. The mass of each cloud, $M_{\rm cloud}$, from \citet{1997yonekura} has been adjusted to our adopted distances  and is listed in column 7.

\section{Observations and Data Processing}
	\label{obs}

	The areas with $A_{\rm V} > 3$ identified from the Dobashi maps were cross referenced with the \spitzer\ data archive and the ``isolated cores'' lists from the c2d program to create two lists of targets. The first list comprised those regions for which public data existed that was compatible with the processing requirements of the c2d/SGBS analysis pipeline and the second list comprised those regions that had not yet been observed with \spitzer. It was this ``second list'' that became the basis for \spitzer\ observations undertaken specifically for the SGBS. Table \ref{tab:aors} lists the complete set of \spitzer\ observations (Astronomical Observation Requests -- or AORs) included in this paper. The first column gives the name of the central cloud and the second column lists the program identifier (PID). The majority of the archival observations come from the c2d (PID \#139) isolated cores program with the exception of L1228 which was taken as part of the Galactic First Look Survey (PID \#104).  The new regions are listed under the SGBS PID \#30574. The extents of the regions listed in Table \ref{tab:aors} are shown in Figure \ref{fig:coverage} by the rectangular footprints (white for IRAC, black for MIPS).

	\subsection{\spitzer\ Gould Belt data} 
	
		Two epochs separated by 5-6 hours were used to take two complete sets of observations of the new SGBS clouds. The redundancy in coverage allowed for the rejection of transient phenomena (including foreground asteroids), the recovery of coverage lost by the blanking of bad pixels, and full mapping at MIPS 70~\micron\ where half the array was not working (each epoch was offset by half the array width). The unique AOR numbers for each epoch's observations are listed in columns 3 and 4 for IRAC and columns 7 and 8 for MIPS. For the archival observations, only one epoch was obtained. Columns 5 and 9 list the respective dates the IRAC and MIPS observations were taken. Likewise, columns 6 and 10 list the version of the \spitzer\ Science Center (SSC) data pipeline that had been used to process the data prior to them being downloaded from the archive. 
		
		The total area mapped was 2.11 square degrees with IRAC and 8.33 square degrees with MIPS. MIPS mapped a larger area than IRAC because its minimum scan-lengths were longer than the diameter of the compact regions of high extinction.  The regions for which we have four-band IRAC fluxes are slightly smaller than the areas listed in Table \ref{tab:assocprop}. This is because the IRAC detectors are paired into two different pointing offsets resulting in slightly different coverage patterns between 3.6/5.8~\micron\ and 4.5/8.0~\micron. The IRAC observations have a total integration time of 48 seconds per point split, between the two epochs. Each epoch consisted of two dithers. The SGBS MIPS observations were taken with the fast-scanning mode using a 240 arcsec step size and a cross scan overlap. This gave a total per point integration time of 31.4 seconds at 24 and 70~\micron\ and 6.1 seconds at 160~\micron. 
			
		The basic calibrated data (BCD) were downloaded from the \spitzer\ archive and then ingested into the c2d/SGBS analysis framework (for a full description see \citealt{2007evans} or separately \citealt{2006harvey} for IRAC and \citealt{2007rebull} for MIPS). In brief, the data were inspected and custom data masks created to identify bad pixels. The data were then corrected for instrumental effects including bleed over, saturation, and banding effects for IRAC and jailbar and stim flash latents for MIPS. The improved data were mosaiced with the MOPEX package \citep{2006makovoz} and source extraction with the c2dphot tool was performed independently at each wavelength and epoch \citep{2007evans}. The separate epoch/wavelength source lists were band-merged together with the 2MASS catalog \citep{2006skrutskie} and cross identifications were made at better than 2 arcsec accuracy. Each source was then characterized spectrally and for the quality of detection. The catalog was ``band-filled'' to produce upper limit flux estimates for sources that were not detected at all wavelengths \citep{2007evans}. 

		Table \ref{tab:bands} lists the original survey/instrument, filter, nominal wavelength, flux zero-point $S_{\rm 0pt}$, and limiting magnitude of the ten bands from the band-merged catalog (2MASS $J$--$K_{s}$; IRAC 3.6--8.0~\micron\ and MIPS 24--160~\micron) plus additional bands from the second {\it HST} Guide Star Catalog (GSC-II, $B$--$I$), the {\it IRAS} point source catalogs (12--100~\micron), and SCUBA data archive (450-850~\micron). Data for these additional bands were only added for YSO candidates, as described in Section \ref{photometry}. The photometric system is based on the Vega magnitude system using the flux zero points taken from the c2d delivery documentation \citep{2007evans}. These are within 1-2 per cent of the SSC's IRAC\footnote{\url{http://ssc.spitzer.caltech.edu/irac/calib/}} \citep{2005reach} and MIPS\footnote{\url{http://ssc.spitzer.caltech.edu/mips/calib/}} zero points. We use the c2d zero points to maintain compatibility with the c2d delivery documentation \citep{2007evans}. The uncertainties for individual flux measurements are $>4$\% and $>8$\% for IRAC and MIPS (24~\micron\ and 70~\micron) respectively. These uncertainties do not include the zero point / absolute calibration uncertainities of 1.5\% for IRAC, 4\% for MIPS 24~\micron, 20\% for MIPS 70~\micron\ \citep{2007evans}. The absolute \spitzer\ calibration uncertainties are lower than the individual measurement uncertainties by a factor of 2 or more, except for the MIPS 70~\micron\ band which has a significantly higher calibration uncertainty. The limiting magnitudes are identified in Section \ref{catalogue} as the turnover in the source count distributions.
		
		\subsection{Archival \spitzer\ data}

		The c2d data used in this paper comes from their final data release\footnote{\url{http://ssc.spitzer.caltech.edu/legacy/c2dhistory.html}} (DR4) and uses the same target selection and observation strategy as the SGBS program. A significant difference was that the c2d cores data were taken using the MIPS small map photometry mode rather than the fast-scanning technique. As a result, the 70~\micron\ coverage on the c2d cores can actually be smaller than the IRAC area \citep{2007evans}. No 160~\micron\ data were taken for the c2d isolated cores (requesting 160~\micron\ data imposes a scheduling limitation as it can only be taken in particular ``cold'' campaigns). The L1148+L1152+L1155 region contains 3 c2d cores in close proximity so the entire area was covered by a single SGBS MIPS fast-scan to obtain complete 160~\micron\ coverage. The only regions for which 160~\micron\ data were not taken were L1221 and L1251.
	
		At the time of writing, the c2d studies on the individual first list cores have already been published or are in the process of being published. L1148 was studied by \citet{2005kauffmann} who discovered a new very low luminosity (VeLLO) Class 0 YSO. L1155C, L1152, and L1228N were studied by \citet{2008chapman} to examine the extinction law from 3.6 to 24~\micron. Although L1228N was observed by c2d, we use the First Look Galactic data-set as it covers both L1228N and L1228S. An analysis of the small YSO group in L1228S was published by \citet{2004padgett}. L1221 was extensively studied by \citet{2008young} including follow-up observations with the VLA. L1251B was studied by \citet{2006lee} who compared their YSO detections with continuum submillimeter data. This was then followed-up by a molecular line study \citep{2007lee}. The dataset was built with the aim of maximizing coverage with a (relatively) uniform sensitivity. We chose not to add deeper data \citep[e.g., the PID \#3656 data used by ][]{2008chapman} as we wish to have all regions covered to approximately the same depth. 
				
		The L1152, L1155, and L1221 studies used data from the DR4 c2d data release \citep{2008chapman,2008young}, the same one used by this paper. The L1251 study, however, used data from an earlier data release (DR3) that were processed through version S11 of the SSC pipeline \citep{2006lee}. Likewise, the First Look study of L1228 used data from version S9.1 of the SSC pipeline \citep{2004padgett}. The SSC pipeline is constantly being revised to improve calibration and reduce instrumental artifacts\footnote{See \url{http://ssc.spitzer.caltech.edu/archanaly/plhistory/} for a revision history.}. Improvements in the calibration will change the photometry for regions where a newer pipeline has been used than in the original study. This is most significant for the L1228 region where it was found that fluxes measured from S14/16 data had increased by an average of 16\% against the S9 version of the data. The greatest increase was at 70~\micron\ which increased by almost 40\%.
		
	\subsection{Images of extended emission}
		\label{images}

\begin{figure*}
	\centering{
	\includegraphics[height=0.8\textheight]{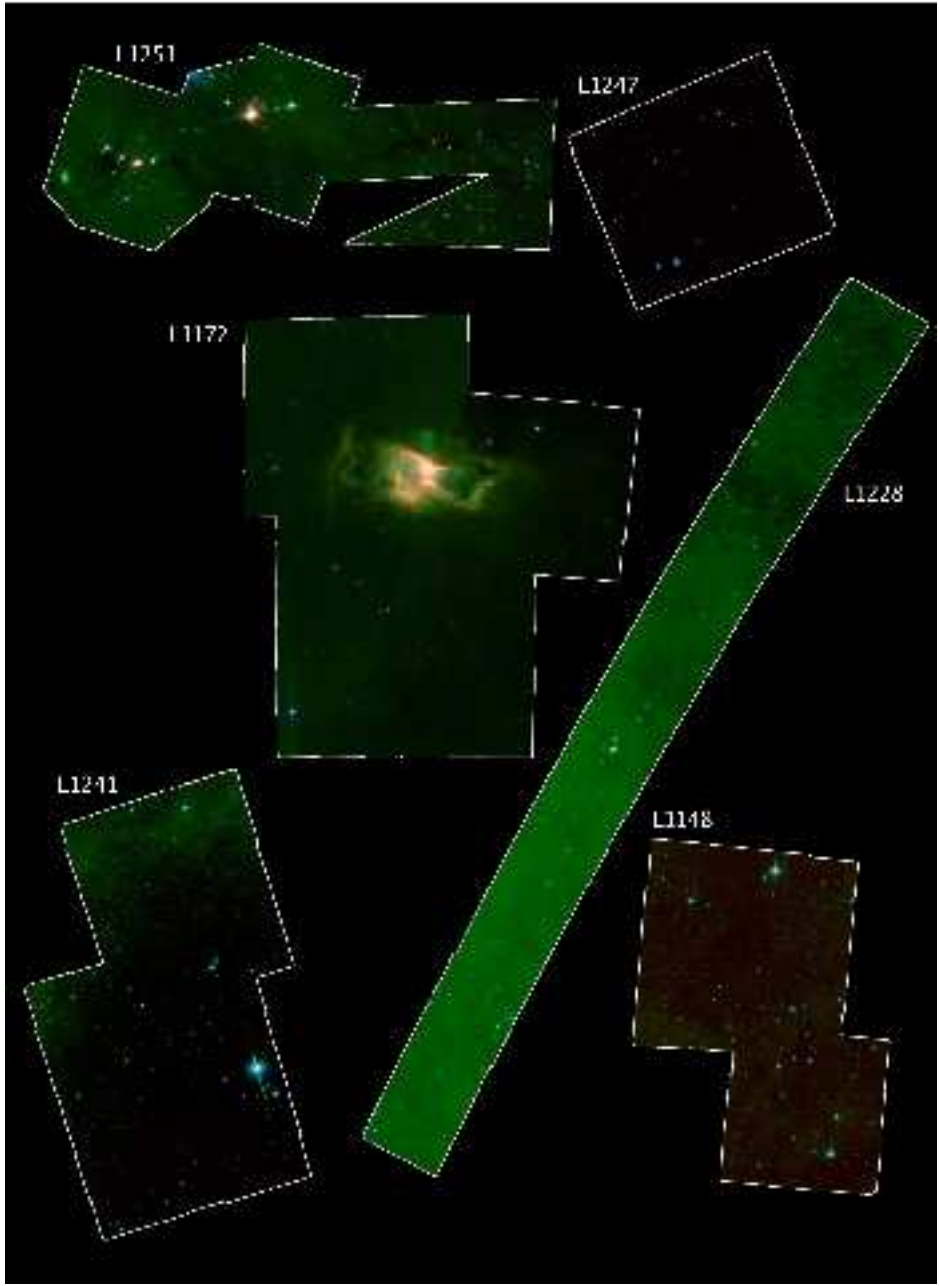}
	}
	\caption{\label{fig:fc245} IRAC and MIPS composite RGB images 4.5~\micron\ (blue), 8.0~\micron\ (green), and 24~\micron\ (red) emission towards the regions in this survey. The nebulosity in L1172 is the NGC 7023 reflection nebula. The red source at the center of L1148 is the source discovered by \citet{2005kauffmann}. }
\end{figure*}

		Figure \ref{fig:fc245} shows RGB color composites of 4.5~\micron\ (blue), 8.0~\micron\ (green), and 24~\micron\ (red) towards the regions observed for the SGBS. The green haze that covers a large area of the images comes from 8.0~\micron\ background which is believed to arise predominantly from 7.7~\micron\ PAH emission \citep{2006flagey}. In L1241, the haze appears along the edge of the image and is anti-correlated with regions of higher 160~\micron\ emission (see Section \ref{extended}). While the mosaic processing removed a number of instrumental effects, it was unable to correct for the bleeding at 8~\micron\ that appears in the color composite images as green smearing upwards from the bright center of NGC 7023. The red point source in the L1148 map is the heavily reddened L1148-IRS source studied by \citet{2005kauffmann}.
		
\begin{figure*}
	\centering{
	\includegraphics[width=\textwidth]{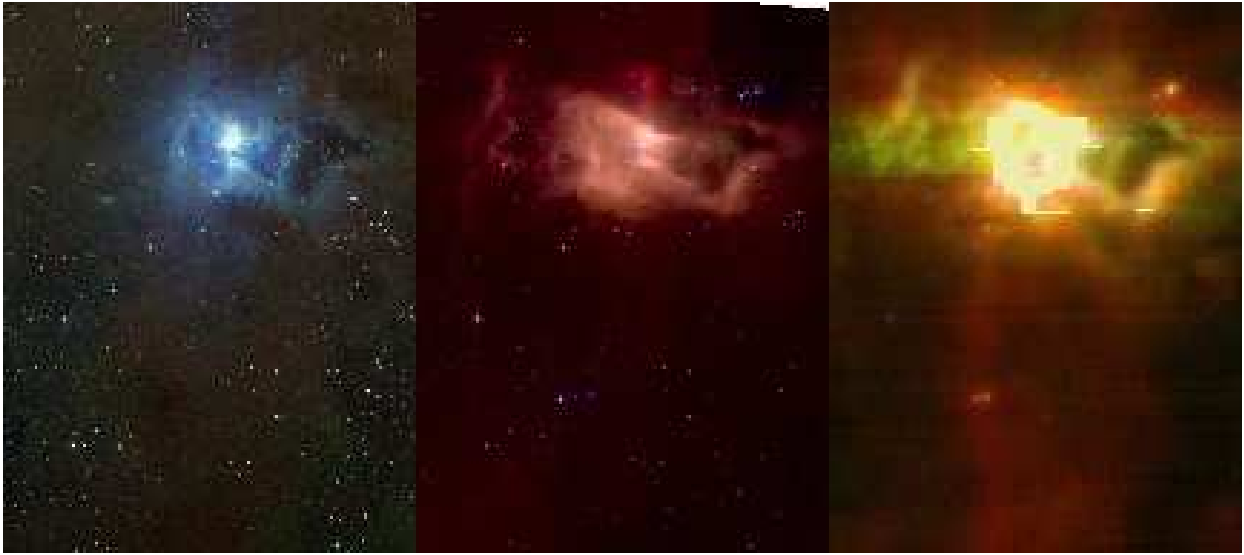}
	}
	\caption{\label{fig:ngc7023fc} Three composite RGB maps showing how the pattern of emission towards the NGC 7023 region changes between the optical, infrared, and far-infrared regimes. Left: POSS-II $B_{\rm J}$ (blue), $R_{\rm F}$ (green), and $I_{\rm N}$ (red); Middle: IRAC 3.6~\micron\ (blue), 5.8~\micron\ (green), and 8.0~\micron\ (red); Right: MIPS 24~\micron\ (blue), 70~\micron\ (green), and 160~\micron\ (red). The dense gas appears in extinction in the optical image and emission in the far-infrared MIPS image. Embedded YSOs are hidden by the extinction in the optical image, but they appear as bright point-sources in the IRAC infrared image. }
\end{figure*}

		The brightest extended structure at all wavelengths in this survey is the NGC 7023 reflection nebula. Figure \ref{fig:ngc7023fc} shows three-color composite images covering the visible (Digitized Sky Survey-II; left image), infrared (IRAC; middle image) and far-IR/submillimeter wavelengths (MIPS; right image). The extended cold dust emission, seen in silhouette in the first panel and emission in the last panel, forms a faint cross with HD 200775 at its intersection. The vertical axis of the cross is formed by the north-south filament that includes L1172 and L1174. The east-west axis is formed by the reflection nebula forming in this filament. The outflow from HD 200775 is currently inactive, but it has generated an asymmetric east-west bi-conical cavity that is filled with hot atomic gas \citep{1998fuente}. The larger western lobe of this cavity appears as a hole in the left hand panel through which background stars are clearly visible. The eastern lobe is truncated by dense material which is being heated and photo-dissociated by the YSO \citep{1998fuente}.
		
		The reflection nebula is confined to the north by a dense filament. This same structure is shown in strong 3.6~\micron\ emission on the leading edge of colder dense material revealed by 850~\micron\ dust emission (see Appendix \ref{regions}). To the south of NGC 7023 is the L1172D dark nebula that appears in silhouette in the visible image, but reveals three embedded sources in the mid-infrared image. The dust itself emits at the longest wavelengths and shows the bright point source associated with the L1172~SMM~1 embedded protostar. The extended 8~\micron\ emission (the red haze in the central image) is anti-correlated, however, with the high visual extinction in the first image and the dense gas in the third image.

	\subsection{Source Catalog}
		\label{catalogue}

		Tables \ref{tab:stat} and \ref{tab:filterstats} list the number of objects detected in various combinations of the IRAC or MIPS bands within the SGBS Cepheus Delivery Catalog, which contains a subset of the data included in this paper. Specifically it includes the five regions L1148, L1172, L1228, L1241 and L1247+L1251. The other regions (L1155, L1152, and L1221) are available from the SSC as c2d data products. Source detection for SGBS data is defined as an object with a peak intensity $\geq$3 $\times$ the local rms in its respective band. A 2MASS detection constitutes an object seen at $\geq$10 $\times$ the local rms in both the $H$ and $K_s$ bands \citep[a design constraint of the 2MASS catalog itself; see ][]{2006skrutskie}. 
		
		Table \ref{tab:stat} shows that a total of 71076 sources were detected in at least one IRAC band and of those 6518 were detected in all four bands. Table \ref{tab:filterstats} lists the number of source detections for each of the four IRAC bands and the first two MIPS bands for several different signal-to-noise thresholds. It shows that the majority of these sources were predominantly detected at shorter wavelengths. An important distinction between Tables \ref{tab:stat} and \ref{tab:filterstats} is that the former only refers to the area common to all \spitzer\ wavelengths whereas the latter lists statistics for the entire region of the Delivery Catalog. 

\begin{figure}
	\centering{
		\includegraphics[width=\columnwidth]{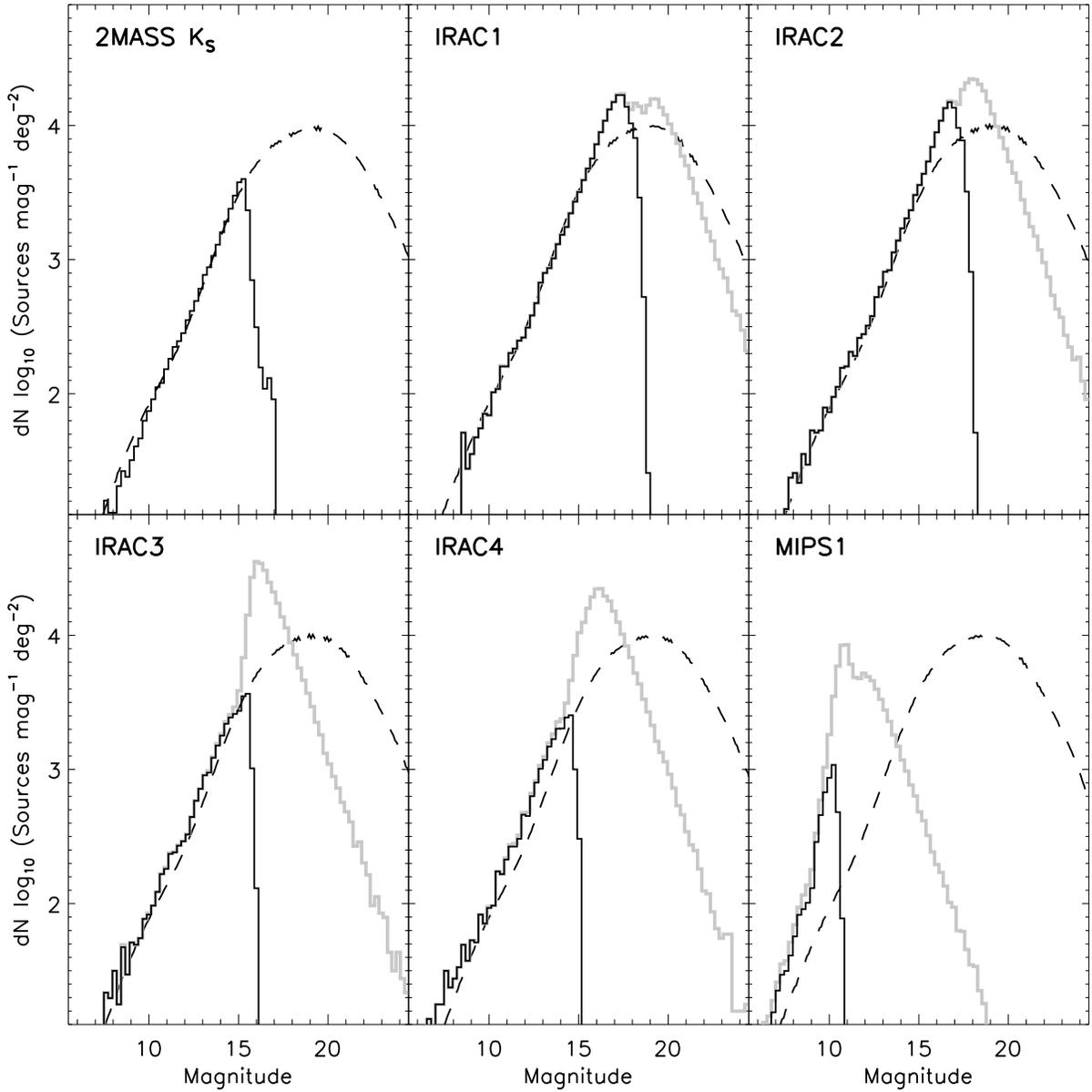}
	}
	\caption{\label{fig:sourcecount} Source counts per square degree per magnitude for six bands from the Cepheus Delivery Catalog. The gray line shows all sources whilst the black line shows sources with a S/N$>$3. The turn over in each black line is taken as the limiting magnitude for that band. The 2MASS catalog only includes sources with S/N$>$10 so no gray line plot is shown for the $K_{s}$ band. The dashed black line is the expected background source count towards Cepheus calculated from the \citet{1992wainscoat} model of Galactic infrared source counts with a visual extinction of $A_V=1$. }
\end{figure}

		Figure \ref{fig:sourcecount} shows source count histograms of the number density of sources per magnitude interval per square degree for six bands from the delivery catalog. The gray curve shows all sources. The limiting magnitudes listed in column 5 of Table \ref{tab:bands} are taken as the magnitudes where the black curves in Figure \ref{fig:sourcecount} turn over. The total source count histogram at 3.6~\micron\ is double-peaked. The 4.5~\micron\ source count histogram shows a second peak, but the effect is far less pronounced. Double-peaks were also found in the source count histograms for Perseus \citep{2006jorgensen} and Chamaeleon II \citep{2007porras}. \citet{2006jorgensen} showed that the double peak was not due to the presence of background galaxies by comparing counts of on-cloud and off-cloud sources and that it must derive from something, like extinction, that is common to the entire region. The solid black curves on Figure \ref{fig:sourcecount} plot only those sources that meet our detection criteria. The peaks of the black curves for 3.6~\micron\ (IRAC1) and 4.5~\micron\ (IRAC2) are coincident with the first of the double-peaks.  
		
		Over-plotted on Figure \ref{fig:sourcecount} are the source count histograms of Galactic infrared sources towards Cepheus estimated from the model of \citet{1992wainscoat} updated by Carpenter (private communication) to apply to the \spitzer\ bands. The source counts agree with the model at low magnitudes before diverging at high magnitudes. The divergence from the model increases with increasing wavelength. An excess of sources above the model can be attributed to the detection of extragalactic sources that are not present in the Wainscoat model. The majority of the 24~\micron\ (MIPS1) sources are in excess of the Wainscoat model and are therefore probably extragalactic in origin.

\section{Candidate YSOs in Cepheus}
\label{ysos}	

	\subsection{Identification of YSO Candidates}
		\label{candidates}
			
		Young stellar objects (YSOs) and background galaxies can have similar infrared colors, but they can be differentiated by the fainter apparent magnitudes of the galaxies. There are several published schemes for identifying YSO candidates from their \spitzer\ colors, each relying on specific subsets of the \spitzer\ wavelength coverage and the equivalent 2MASS fluxes \citep[e.g., ][]{2006harvey,2007harvey,2007rebull,2007allen,2008gutermuth}. These schemes are commonly calibrated against a known galaxy catalog. SWIRE (the \spitzer\ Wide-area Infrared Extragalactic Survey, another of the \spitzer\ legacy programs) was specifically designed to observe the infrared extragalactic background along sight-lines that avoided as much foreground (Galactic) contamination as possible \citep{2003lonsdale}. Thus, it provides an ideal calibration dataset to use with any galaxy-rejecting protocol. 
		
		IRAC observations were targeted towards regions of high visual extinction with the expectation that these regions will harbor the highest concentration of YSOs. A side-effect of the different operational modes of IRAC and MIPS is that the MIPS data cover an area four times as large as the IRAC data. A galaxy-rejection scheme that used IRAC-only data would be unusable across 3/4 of our surveyed area. Therefore we adopt different schemes for the common (IRAC+MIPS) and the MIPS only areas. A significant factor in both schemes is a 24~\micron\ MIPS detection. Since that band is not as sensitive as the IRAC bands, a third scheme is adopted for sources detected only with IRAC.
	
		Table \ref{tab:ysofluxes} lists the YSO candidates identified towards the regions in our survey. The first three sections of Table \ref{tab:ysofluxes} list the YSO candidates identified by the 5-Band, IRAC-only, and 2MASS/MIPS schemes respectively (described below). The first column lists the source number. The second column lists the \spitzer\ catalog position identifier of each source (a contraction of the source's seven-digit right ascension and declination without delimiters). All positions are given in J2000. Column 3 lists cross-identification names from the literature. Columns 4--9 list the flux and statistical errors or where appropriate the band-filled upper limit. Column 10 lists a series of flags denoting which candidate identification methods identify the source. The letters F, I and M are shown for sources identified by the 5-Band, IRAC and 2MASS/MIPS methods respectively. A source may satisfy the rule sets of more than one method, but it is assigned to a group based on the order F$>$I$>$M. This catalog includes data previously published in a series of c2d papers examining individual dark clouds. Column 11 lists references to these and other studies. 
		
		\subsubsection{5-Band Identifications}
		
		\citet{2007harvey} used a five-band (4.5, 8.0, and 24~\micron, plus 2MASS $H$ and $K_{s}$) scheme to estimate the probability that a given source is a background galaxy. The unnormalized probability $P_{\rm gal}$ is the product of a series of probabilities that a given source is a background galaxy. Each of these probabilities is based on the selection criteria from a different color-magnitude diagram. The final value of $P_{\rm gal}$ is moderated by a series of additional factors which include whether the source is extended at 3.6\micron\ or 4.5\micron\ (see Table 1 of \citealt{2007harvey} for a full list). Additionally, the scheme requires that a source be detected at each IRAC band and at MIPS 24~\micron, irrespective of whether that band was actually used to construct $P_{\rm gal}$.
		
		Under the 5-band scheme, a catalog is filtered to reject those sources that can be adequately modeled by a stellar photosphere. $P_{\rm gal}$ is calculated for the remaining sources and these are then filtered to retain those sources with a suitably low value of $P_{\rm gal}$, i.e., rejecting sources that are statistically likely to be background galaxies. From their study of Serpens, \citet{2007harvey} found that an upper limit of $\log(P_{\rm gal})<-1.47$ rejected the galaxies from their SWIRE control catalog. This method creates a catalog of YSO candidates that is largely free from background galaxies. The catalog will be luminosity limited partially because it requires [24]$>$10 (fainter objects at [24] are preferentially background galaxies, but this cut off will also exclude faint YSOs).
				
\begin{figure}
	\centering{
		\includegraphics[width=0.7\columnwidth]{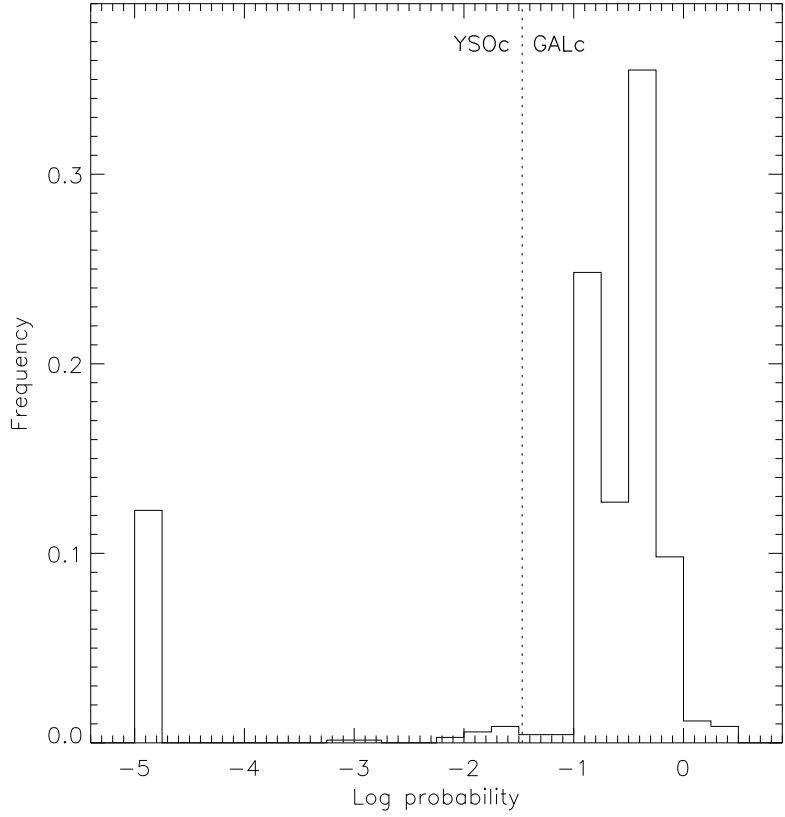}
	}
	\caption{\label{fig:loggal} A histogram of $\log(P_{\rm gal})$ for the sources in the Cepheus catalog that were detected in each IRAC band and MIPS 24~\micron. The dashed line shows the $\log(P_{\rm gal})=-1.47$ criterion established by \citet{2007harvey} as the divide between YSO candidates (YSOc) and galaxy candidates (GALc). The majority of the galaxy candidates appear as a distribution at $-1<\log(P_{\rm gal})<0$, whereas the majority of the YSO candidates appear as a separate peak at $\log(P_{\rm gal})=-5$. }
\end{figure}

		Figure \ref{fig:loggal} shows a histogram of $\log(P_{\rm gal})$ for sources in the Cepheus catalog with detections in all IRAC bands and MIPS 24~\micron. The peak at -5 contains the majority of the YSO candidates identified by this method. There is also a small tail of sources to the left of the -1.47 divide (the vertical dashed line). There are 12 sources in the tail with $-2.5<\log(P_{\rm gal})<-1.47$ and 3 (25\%) of them are previously known YSO candidates. We choose to retain the canonical dividing line for YSO candidates for consistency with the c2d studies \citep{2007harvey, 2008evans} and because moving the division by any significant amount would exclude the three known sources. If half of the tail sources were background galaxies, it would represent a contamination of $\sim5\%$ in the total number of YSO candidates identified by the 5-Band method. 
		
\begin{figure*}
	\centering{
		\includegraphics[width=\textwidth]{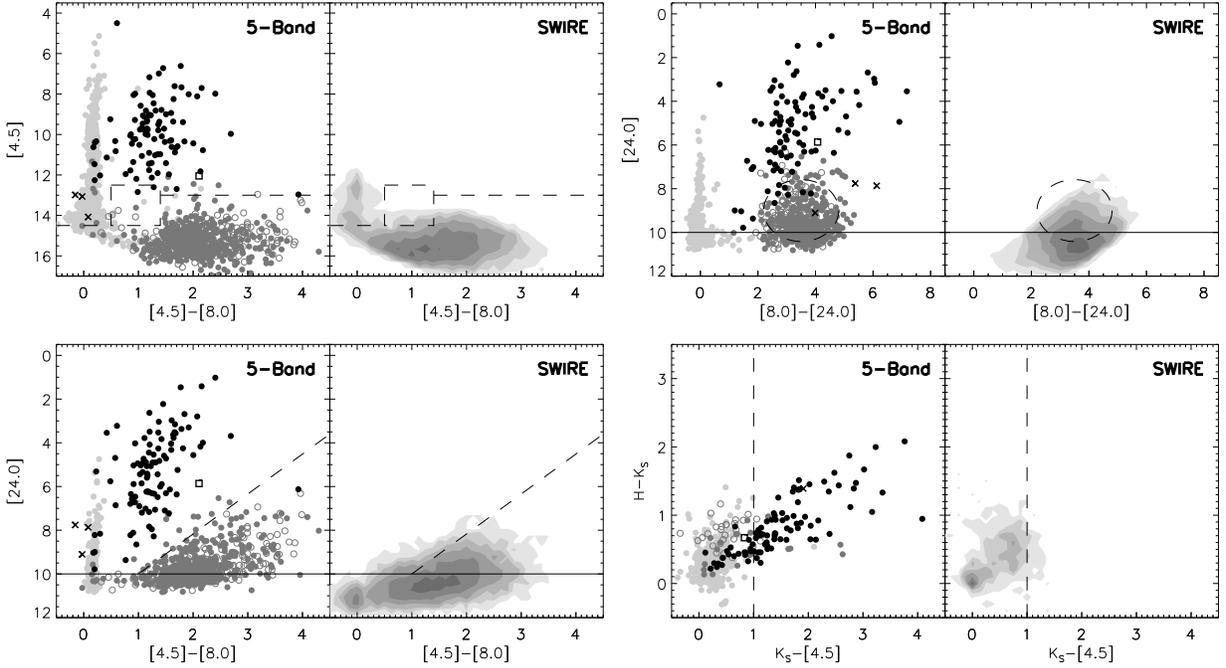}
	}
	\caption{\label{fig:5band-cc} Color-magnitude diagrams after \citet{2007harvey} showing the separation of YSO and galaxy candidates and the dividing lines that were used to construct $P_{\rm gal}$. Left panels: SGBS Cepheus data. The markers show point-line/extended YSOs (black filled-circles/crosses) and point-like/extended background galaxy candidates (dark gray filled/open-circles) that have been separated via their $P_{\rm gal}$ value. Stellar sources (light gray filled circles) with the same S/N requirements as the YSOs and galaxies are shown for comparison. The open square is a source with only an 8~\micron\ flux upper-limit, but otherwise satisfies the 5-band YSO selection criterion. Right panels: Contour plots of the number density of sources identified as neither YSOs or stars from the SWIRE catalog. Dashed black lines show the ``fuzzy'' borders used to calculate $P_{\rm gal}$ while solid lines show the hard $[24.0]<10$ mag cut-off. }
\end{figure*}

		Figure \ref{fig:5band-cc} shows a sequence of color-color plots for sources detected in all four IRAC bands and shows the cut-offs \citet{2007harvey} used to construct $P_{\rm gal}$. Dashed-lines show ``fuzzy'' limits while solid lines show hard limits. The left panel in each pair shows the SGBS data. Black filled-circles and crosses are respectively point-like and extended sources identified as YSO candidates by $P_{\rm gal}$, whilst dark-gray circles are sources identified as point like (filled circles) and extended (open circles) galaxies. Also shown as pale-gray points are sources that were identified as stellar photospheres via their SEDs. The right-hand panel of each pair of plots in Figure \ref{fig:5band-cc} shows a contour plot of the SWIRE Catalog after it has been processed in a manner similar to the SGBS Catalog. It can be seen how the limits used in the calculation of $P_{\rm gal}$ have been chosen to reject regions of parameter space where there are significant numbers of SWIRE galaxies. These plots show that the galaxies are tightly clustered and that there is no significant overlap between the different candidate types. The 5-Band scheme identified 98 YSO candidates and 1 possible YSO candidate with a band-filled (upper limit) flux at 8~\micron\ (shown as an open square in Figure \ref{fig:5band-cc}). Sources that satisfy the 5-Band Scheme are listed with an F flag in column 10 of Table \ref{tab:ysofluxes}.
		
		\subsubsection{IRAC-only Identifications}
		
		For those sources which have no detection at 24~\micron, we must use selection rules based on IRAC colors alone. For this, we use the selection rules from \citet{2008harvey}. They define ([4.5]-[8.0] $<$ 0.5 and $[8.0]>13-([4.5]-[8.0])$) to reject galaxies and stars. The upper row of Figure \ref{fig:non5band-cc} shows a plot of [4.5]-[8.0] versus [8.0] for sources without a MIPS 24~\micron\ detection and comparative plots of the 5-Band and SWIRE sources. Of the YSOs identified by the 5-Band scheme, 12 would be misidentified by the IRAC-only scheme as stars due to their small [4.5]-[8.0] color and six bright galaxies would have crept into the sample.
		
		Sources that satisfy the IRAC-only scheme are listed with an I flag in column 10 of Table \ref{tab:ysofluxes}. A source flagged with a lowercase f has a 24~\micron\ upper limit flux, but would otherwise satisfy the 5-Band scheme. A comparison of sources with I and f flags shows that the IRAC-only scheme and 5-Band scheme are consistent for sources without an 24~\micron\ detection. A total of 10 additional YSO candidates was identified by the IRAC-only scheme. 
			
		An additional constraint can be imposed in color-space as external galaxies are often rich in PAH emission compared to YSO sources. \citet{2008gutermuth} proposed using [4.5]-[8.0] vs. [3.6]-[5.8] and [5.8]-[8.0] vs. [4.5]-[5.8] color-color diagrams to reject background galaxies with strong PAH emission. Sources that match these rules are shown with triangle rather than circle markers in Figure \ref{fig:non5band-cc}. No IRAC YSO candidates have evidence of PAH emission from this technique, but two 5-Band YSO candidates are flagged. One of these is L1148 IRS.	
		
		\subsubsection{2MASS/MIPS Identifications}
		
\begin{figure*}
	\centering{
		\includegraphics[width=\textwidth]{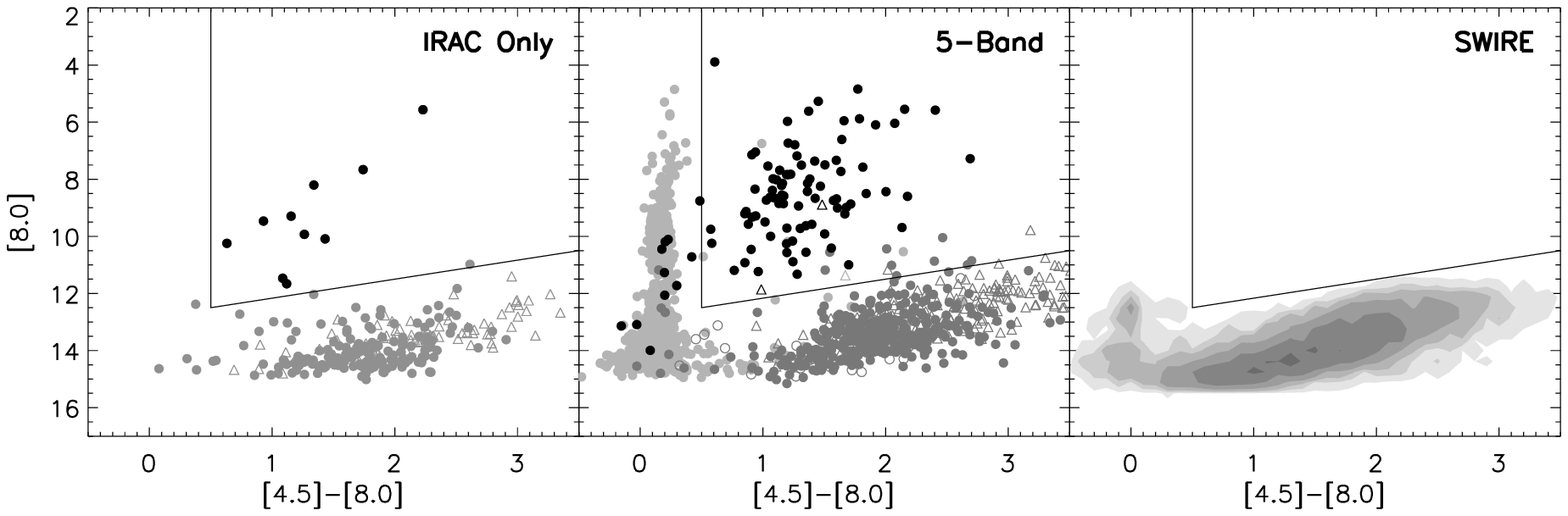}
		\includegraphics[width=\textwidth]{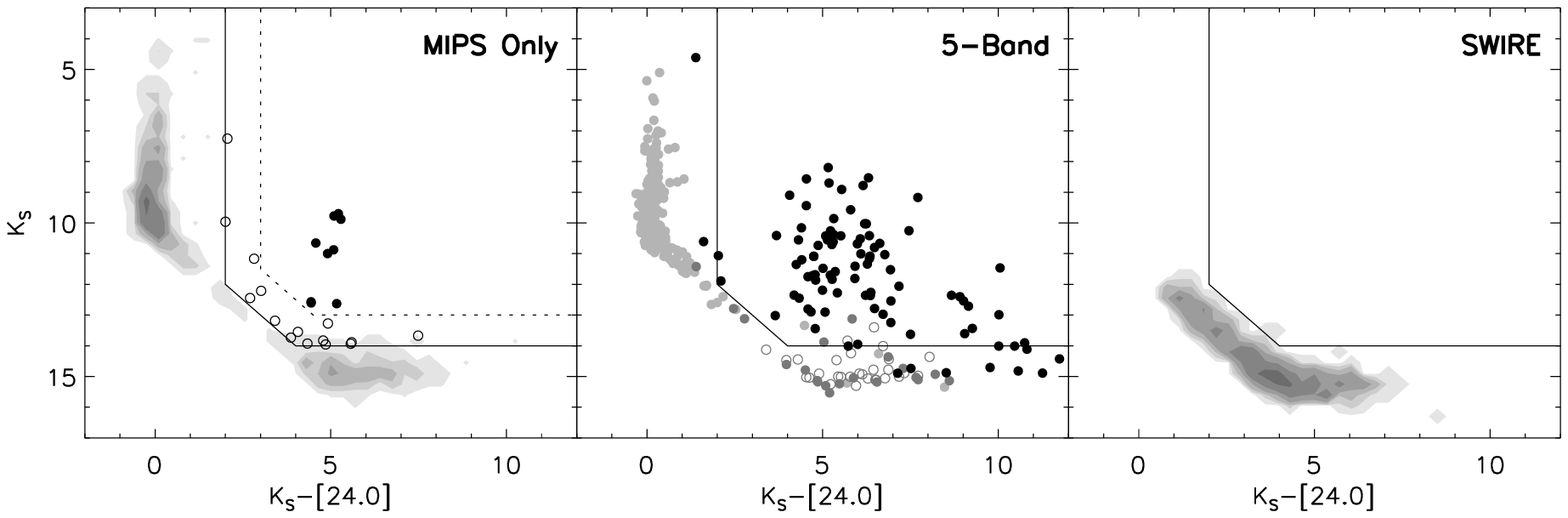}
		}
	\caption{\label{fig:non5band-cc} Color-magnitude diagrams for IRAC and 2MASS/MIPS detections showing the dividing lines used to separate YSO candidates from other sources. Top row: IRAC-only detections. The left panel shows the IRAC-only data, the middle and right panels the 5-Band and SWIRE galaxies as per Figure \ref{fig:5band-cc}. The  triangle markers show PAH-strong sources as identified by the selection rules from \citet{2008gutermuth}. Lower row: 2MASS/MIPS only detections. Left panel shows 2MASS/MIPS only detections, middle and right panels are 5-Band detections and SWIRE galaxies as per Figure \ref{fig:5band-cc}. The solid lines show the selection cut-offs. Dashed lines show a 1 mag offset from the selection cut-offs and open markers show YSO candidates within that zone.  Non-YSOs in some panels have been shown as a density plot for clarity.  }
\end{figure*}

		An inherent strength of the 5-Band scheme is that it relies on a broad range of color combinations, but it can only be used in regions where there is IRAC coverage. \cite{2007rebull} proposed a scheme to identify YSOs using MIPS 24~\micron\ and 2MASS $K_{s}$ and the selection rules $K_{s}-[24]>2$ and $K_{s}<14$. To this we add the limit $[24]<10$ for consistency with the 5-Band scheme. The lower row of Figure \ref{fig:non5band-cc} shows color-magnitude plots of $K_{s}-[24]$ versus $K_{s}$ for sources without an IRAC detection and comparative plots of 5-Band and SWIRE sources. Column 10 of Table \ref{tab:ysofluxes} includes an M flag for sources that would have been identified as a YSO candidate by this method. A significant fraction of 2MASS/MIPS YSO candidates were found within 1 magnitude of the selection limits. These are marked as open circles on Figure \ref{fig:non5band-cc} and listed with a lowercase m in Column 10. 

		The 2MASS/MIPS scheme would have misidentified 19 of the 5-Band YSO candidates as background galaxies and stars. From Figure \ref{fig:non5band-cc}, we can see that the majority of these appear in the region of background galaxies with a $K_{s}-[24]$ color of $\sim10$. This is approximately equivalent to the spectral index of an embedded protostellar source (a Class I YSO, see following section and Figure \ref{fig:alpha}). This shows a further advantage of the 5-Band scheme in that it is more sensitive to embedded sources as it does not rely so heavily on $K_{s}$. Based on the MIPS scheme, 4--5 galaxies could have crept into the YSO sample. A total of 24 additional YSO candidates were identified by the 2MASS/MIPS scheme that were not identified by the 5-Band and IRAC-Only schemes.		
				
		\subsubsection{The Final Catalog}
		
		The morphologies of all YSO candidates were checked at each wavelength listed in Table \ref{tab:bands} to screen for obvious galaxy candidates or possible artifacts. During the visual checking, an unusual extended source was found at 21\lath 02\latm 21.2\lats +68\degr 04\arcmin 36\arcsec. It is a flat spectrum source across the 2-24~\micron\ waveband, but rises at longer wavelengths. It is detected in all three MIPS wavebands. It was coincident with a spur in the extended 160~\micron\ emission from the NGC 7023 nebula and a small increase in visual extinction. It corresponds with a 1.4~GHz (20cm) source NVSS 210221+680436 \citep{1998condon}. Its colors of [5.8]-[8.0] = 2.8 and [3.6]-[4.5] = 0.0 place it well away from the color-color clustering of the other YSO candidates (see Section \ref{ccdiagrams}). This object has been excluded from further consideration as a YSO as it is probably a background galaxy.
		
		A total of 133 YSO candidates was identified by the three schemes. The vast majority of these (99) were identified by the 5-Band scheme and a further 10 were identified by the IRAC scheme. The 2MASS/MIPS data identified a further 24 candidates. Although the 2MASS/MIPS data covered a larger area than the 5-Band scheme, most of it was at a lower visual extinction. Thus, it could be argued that we would not expect to detect the same density of YSOs. The position of each YSO candidate was checked against the SIMBAD database. The SIMBAD name and other selected literature names for with each source are listed in column 3 of Table \ref{tab:ysofluxes}. A total of 59 YSO candidates were found to have antecedent catalog names.
		
		It should be noted that not all known YSOs and YSO candidates coincident with our mapped area were identified by the three YSO candidate identification schemes. Section \ref{knownYSOs} below discusses the completeness of our YSO sample and Table \ref{tab:simbadYSOs} lists 15 known sources missed by our selection criteria. These include luminous YSOs like HD 200775 (the driving source of the NGC 7023 nebula) and PV Cep \citep[a ``run-away'' YSO, ][]{2004goodman} which were excluded because they saturate the \spitzer\ detectors. In addition sources like XMMU J223727.7+751725 were excluded since they have no detectable infrared (2MASS/\spitzer) flux. From this list 10 are coincident with entries in the Cepheus catalog, but which were not identified by the three \spitzer\ schemes. Photometry for these 10 sources is listed in the fourth section of Table \ref{tab:ysofluxes}. We refer to these 10 sources as non-\spitzer\ identified YSO candidates. 
		
		Hereafter, we assume that these 143 (133 \spitzer\ plus 10 previously identified) YSO candidates are actually YSOs and analyze their properties accordingly.  
		
	\subsection{YSO Classification}
		\label{classification}

		YSOs can be separated into a series of four evolutionary classes depending on either their infrared spectral index \citep{1989wilking} or the mean frequency of their SED \citep{1993myers}. The spectral index classification of a source as a Class II or Class III YSO is generally correlated with its respective classification as a Classical T Tauri (CTTS) or Weak-Line T Tauri (WTTS) star \citep{1989wilking, 1994andre}. Class 0 and Class I protostars are younger embedded YSOs and are differentiated from each other by the amount of material remaining in their envelopes. The envelope appears as a colder component to the SED and peaks towards the submillimeter. The original definition of a Class 0 protostar was an object whose submillimeter luminosity ($\lambda >$ 350~\micron) contributed greater than 5\% to the source's total bolometric luminosity, which is equivalent to a protostar-to-envelope mass ratio of less than one \citep{1993andre}. The second method of YSO classification uses a bolometric temperature calculated for a blackbody which has the same mean frequency as the source's SED. The value of the bolometric temperature decreases from Class III to 0 as the SED is increasingly dominated by long wavelength emission. The results for this method are described in Section \ref{seds}. 
			
		Table \ref{tab:ysoprop} lists the derived properties of the YSOs from Table \ref{tab:ysofluxes}. The first column lists the index number of the YSO. Column 2 lists the dark cloud from Figure \ref{fig:coverage} with which the YSO candidate is associated. Locations that have a visual extinction less than one magnitude in the \citet{2005dobashi} maps or are not part of an identifiable YSO group (see Section \ref{clustering}) are listed as ``off-cloud.'' Column 3 lists the spectral index of the source, $\alpha_{\rm IR}$, as given by 
		\begin{equation}                                                                      
		\alpha_{\rm IR} = \frac{d\log(\lambda S_{\lambda})}{ d\log\lambda} \label{eqn:alpha}
		\end{equation}
		where $S_{\lambda}$ is the monochromatic flux density at wavelength $\lambda$. The index was calculated by a least-squares fit to all available data in the range 2MASS $K_{s}$ (2.2~\micron) to MIPS 24~\micron.
		
		The mapping between the YSO evolutionary sequence and spectral index proposed by \citet{1989wilking} and amended by \citet{1994greene} and \citet{1994andre} are Class I ($\alpha_{\lambda}\leq0.3$), ``flat'' ($0.3>\alpha_{\lambda}\geq-0.3$), Class II ($-0.3>\alpha_{\lambda}\geq-1.6$) and Class III ($-1.6>\alpha_{\lambda}$). The decreasing spectral index is attributed to the decline in the amount of circumstellar dust with advancing YSO evolution. The associated class for each source is listed in column 4 of Table \ref{tab:ysoprop}. ``Flat'' spectrum sources have spectra intermediate between embedded and T Tauri stages. Class 0 YSOs cannot be distinguished from Class I YSOs via this method as it is not sensitive to the long wavelength part of the SED where the majority of the envelope emission radiates \citep{2008enoch}. 
				
\begin{figure}
	\centering{
		\includegraphics[width=0.7\columnwidth]{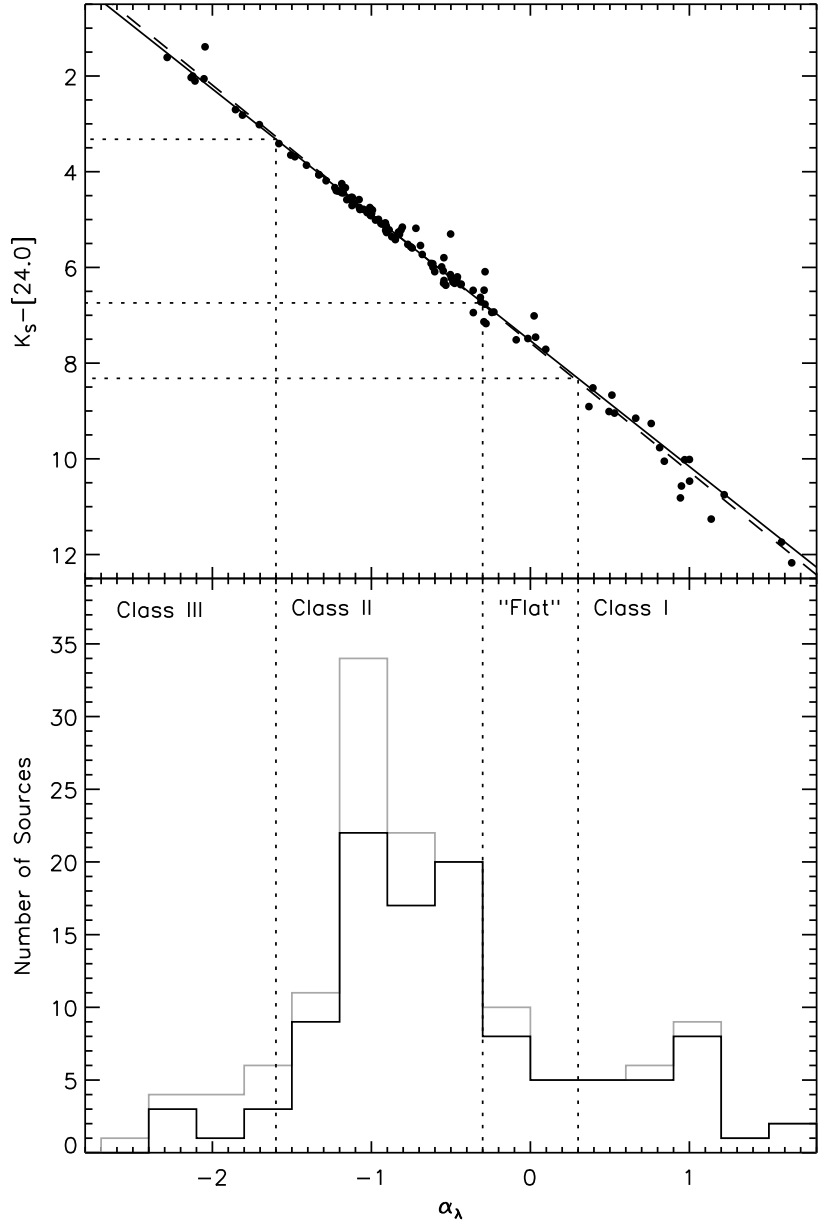}
	}
	\caption{\label{fig:alpha} Upper: Plot showing the equivalence of $\alpha_{\lambda}$ and the $K_{s}$-[24.0] color for the 5-Band YSOs. The solid line shows the theoretical relationship between $\alpha_{\lambda}$ and $K_{s}$-[24.0]. The dashed line is a linear regression to the data. Lower: Histogram of spectral indices for YSO candidates. The black-line shows those YSOs detected by IRAC. The gray-line shows all YSOs including the 2MASS/MIPS YSOs. The dotted lines show the divisions of the classical YSO spectral classes and are extended to show the equivalent $K_{s}$-[24.0] colors in the top diagram. The majority of YSOs have indices in the Class II regime. Not all of the YSOs included in the solid histogram are plotted in the upper graph as not all IRAC YSOs have 24~\micron\ detections. }
\end{figure}

		The majority of the data points used to calculate $\alpha_{\rm IR}$ are in the IRAC regime. For those YSO candidates without an IRAC detection, $\alpha_{\rm IR}$ is only calculated from 2MASS $K_{s}$ and MIPS 24~\micron\ -- effectively the $K_{s}$-[24] color. To test the robustness of using just the 2MASS/MIPS color, the top of Figure \ref{fig:alpha} shows $K_{s}$-[24] versus $\alpha_{\rm IR}$ for the YSOs candidates detected with IRAC. The tight correlation follows a linear relationship thus proving that the equivalent color can be an acceptable proxy for a spectral index calculated using more intermediate points.
		
		The solid line through the points in Figure \ref{fig:alpha} shows the theoretical relationship between $K_{s}$-[24.0] and  $\alpha_{\lambda}$. For two generic bands, A and B, the color and spectral index are linearly related such that $m_{\rm A}-m_{\rm B} = C_{1} \alpha_{\lambda} + C_{2}$ where $C_{1} = -2.5X$, $C_{2}=-2.5+2.5Y$, $X=\log(\lambda_{\rm A})-\log(\lambda_{\rm B})$ and $Y=\log(S_{\rm 0,A})-\log(S_{\rm 0,B})$. The wavelength of each band is $\lambda_{\rm A}$ and $\lambda_{\rm B}$, the flux zero points are $S_{\rm 0,A}$ and $S_{\rm 0,B}$ and the color is ($m_{\rm A}-m_{\rm B}$). Using the data from Table \ref{tab:bands}, the coefficients for $K_{s}$-[24.0] are $C_{1} = 2.63$ and $C_{2} = 7.55$. A linear regression to the data, shown by the dashed line, gives coefficients of $C_{1} = 2.64\pm0.06$ and $C_{2} = 7.73\pm0.06$.
		
		The color-$\alpha_{\rm IR}$ equation can be used to recompute the class boundaries in terms of a source's $K_{s}$-[24.0] color such that they become $K_{s}$-[24.0] = 8.32, 6.74, and 3.32 for the boundaries between Class I/``flat'', ``flat''/Class II, and Class II/III respectively. These values agree with those used by \citet{2006lee} and \cite{2007rebull}. These recomputed boundaries are shown in Figure \ref{fig:alpha} by the dotted lines. 
		
		The lower panel of Figure \ref{fig:alpha} shows a histogram of the YSO spectral indices. The black line histogram shows sources with IRAC coverage. The range of values of $\alpha_{\lambda}$ is -2.28 to 1.63. The gray histogram shows all sources and includes 2MASS/MIPS YSOs. Both distributions peak around -1 and show an increase in the number of sources from Class I to Class II.  The peak of the Cepheus spectral index histograms matches the peak of the distributions seen in Serpens \citep{2007harvey}, Lupus \citep{2008merin} and IC 5146 \citep{2008harvey}. The range of extreme spectral indices is less than in IC 5146 \citep{2008harvey}.
	
		The number of sources classified in each Class via $\alpha{_\lambda}$ and that number as a percentage of the \spitzer\ identified YSOs is listed in column 2 of Table \ref{tab:propStats}. The row for Class 0 sources is left blank for $\alpha_{\rm IR}$, as they can only be distinguished by their bolometeric temperature. We can compare the relative fraction of $\alpha_{\rm IR}$ classes for the Cepheus Flare against the values for other regions from the c2d survey (Figure 5 of \citealt{2008evans}). The percentage of sources in each class closely matches the numbers for the entire c2d survey (shown in column 4 of Table \ref{tab:propStats}). Star formation in the Cepheus Flare is comprised of a series of small YSO groups, isolated cores, and a single (relatively) large YSO group (see Section \ref{extended}). That this mixture gives the Cepheus Flare a relative number of sources equal to the c2d survey may just be a coincidence or it could be that the mixture of star formation modes in Cepheus mirrors the balance of modes across the wider c2d sample. 
				
		Of the individual c2d regions, Cepheus most closely matches the $\alpha_{\rm IR}$ class profile of Serpens \citep{2007harvey}, which is also very similar to that for the full c2d survey \citep{2008evans}. Cepheus also has the same relative number of Class I sources as Serpens \citep{2007harvey}. The relative number of Class I sources in Cepheus is $\sim10$ percentage points higher than that seen in the Chamaeleon II and Lupus clouds although it is $\sim$5 percentage points lower that that seen in the IC5146 and Perseus clouds \citep{2008harvey,2008evans}. 

		\citet{2008kun} include in their review of the Cepheus star formation region a list of  Classical and Weak-Lined T Tauri stars whose pre-main sequence nature has been confirmed by spectroscopic observations. There are 31 \spitzer\ identified YSOs from our sample that appear in that list (these have a reference to the Kun review in the last column of Table \ref{tab:ysofluxes}). Twenty-five of these sources (80\%) are classified as Class II by their $\alpha_{\rm IR}$ value. This match agrees with the usual interpretation of a YSO with a Class II infrared spectrum as most probably being a T Tauri protostar. 
		
		There are 4 sources on the \citeauthor{2008kun} list that are among the previously known YSOs listed in Table \ref{tab:simbadYSOs} that were coincident with our mapped area, but which were not identified as YSO candidates by the \spitzer\ schemes. Despite not being identified as YSOs 3 of these 4 sources were coincident with entries in the \spitzer\ Cepheus catalog. Therefore, 34 out of 35 of the \citeauthor{2008kun} stars had infrared emission that was detectable by \spitzer. It should be noted, however, that the \citeauthor{2008kun} T Tauri list is from a project that is still underway and is not necessarily complete.
		
		Our Class III to Class II ratio is 1/8, whereas WTTS dominate CTTS in spatially complete surveys. For example the WTTS/CTTS ratio in Taurus is ~8 \citep{1995newhaeuser} and 3/2 in IC 348 \citep{2003luhman}. Infrared selection schemes will naturally be less sensitive towards objects with smaller infrared excesses, i.e., Class III YSOs. Estimates of completeness can only be made if there is a previous census of WTTS to compare against. In a comparable study of the Lupus III region, \citet{2008merin} estimated their completeness for Class III sources with infrared excesses against objects with no infrared excess as $\sim50\%$. If the low rate of T Tauri sources in the Cepheus cores without detectable infrared excesses is real and similar to Lupus III, it would further reinforce the idea that the YSO population is comparatively young. 
				
		Figure \ref{fig:non5band-cc} shows that no 2MASS/MIPS YSO candidates are detected with a $K_{s}$-[24] color greater than $\sim5$ (or 6). From Figure \ref{fig:alpha}, we see that this means that the 2MASS/MIPS method is significantly biased towards the detection of Class II and Class III YSOs. Deeper $K_{s}$ data are needed to make a more complete survey of YSOs in the MIPS-only regions. 
	
	\subsection{Color-Color Diagrams}
	\label{ccdiagrams}

\begin{figure*}
	\centering{
		\includegraphics[width=0.32\textwidth]{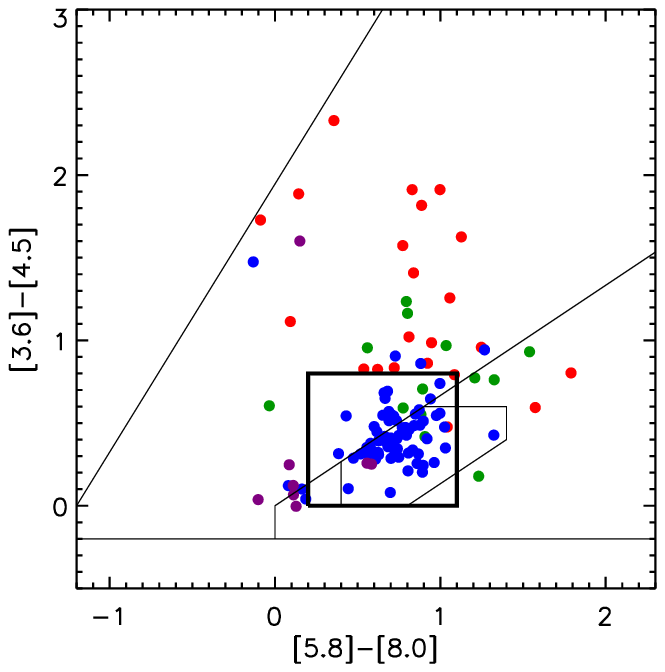}
		\includegraphics[width=0.32\textwidth]{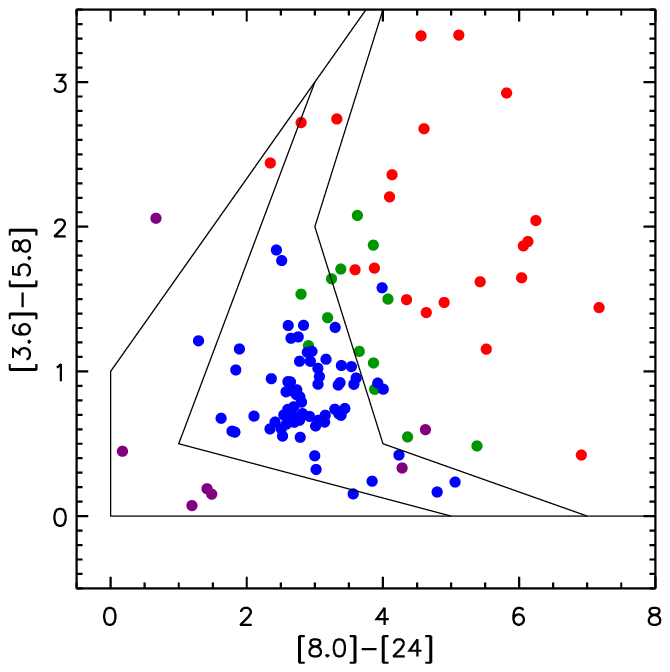}
		\includegraphics[width=0.32\textwidth]{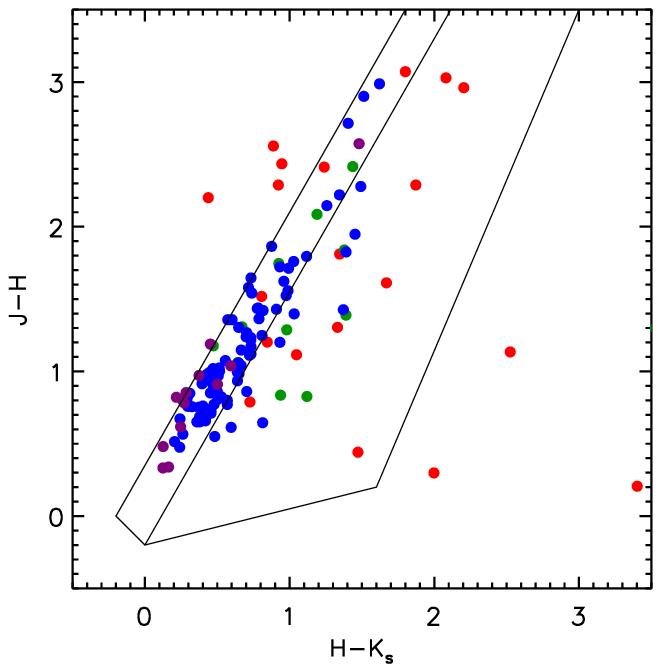}
	}
	\caption{\label{fig:cc} Color-color diagrams showing the correspondence between the colors of the YSO candidates and various color-regions predicted by models of YSO spectra. The points are color coded to the YSOs $\alpha_{IR}$ classification -- red are Class I sources, green are flat spectrum sources, blue are Class II sources, and purple are Class III sources. Approximate regions that contain the Robitaille models are shown on each panel by thin lines \citep{2006robitaille}. The box with the thick outline plotted on the [5.8]-[8.0] vs. [3.6]-[4.5] plot is the region identified by \citet{2004allen} as being the approximate domain of Class II sources. The three Robitaille regions on this plot show the region where any Stage model can be found (bottom right), where the majority of Stage II models are found (small six-sided region), and where the majority of Stage I models can be found (top). The regions plotted on the [8.0]-[24] vs [3.6]-[5.8] plot approximate regions occupied by (from the left) Stage III, II, and I models. The box enclosing the majority of the sources in the $H-K_{s}$ vs. $J-H$ plot shows the region occupied by reddened stellar photospheres. The region to the right of this denotes the area where any evolution stage can occur. }
\end{figure*}

		Figure \ref{fig:cc} shows color-color diagrams of the 143 Cepheus YSOs listed in Table \ref{tab:ysofluxes}. A source is included in a given plot if it has been detected at each of the wavelengths shown in that plot. Red markers show Class I sources, green markers show flat spectrum sources, blue markers show Class II sources and purple markers show Class III sources. \citet{2006robitaille} generated a database of 20,000 YSO models and used them to delineate a series of regions in color-color space that correspond to three different phases of YSO evolution. They termed these evolutionary phases ``Stages'' to differentiate them from the equivalent ``Classes'' (a purely observational parameter which describes a source's infrared signature and may be influenced by reddening, e.g., as a result of varying disk orientation). The approximate boundaries of the regions containing the different evolutionary states are shown on each of the three panels in Figure \ref{fig:cc} (see Figure 23 of \citealt{2006robitaille}).
		
		\citet{2004allen} compared the theoretical \spitzer\ colors of YSOs with data from four embedded clusters. The region they identified as corresponding to the approximate domain of Class II sources contains almost all of our Class II candidates as shown by the thick lined box on the first panel of Figure \ref{fig:cc}.  The small six-sided region coincident with the Allen box shows the region where most of the Stage II Robitaille models lie. Our Class II sources cluster around this region, but there are a number that lie just above it. These could be Class II sources that still retain some degree of reddening. The Class III sources are clustered around the zero point on each axis. Two large regions in the first panel are divided by a line that runs through the Stage II region and Class II box. The region above this line contains the majority of the Robitaille Stage I models whereas the region below it is where any Stage may be present. Our Class I sources agree quite well with the Stage I area.
		
		The middle-panel of Figure \ref{fig:cc} shows IRAC and MIPS colors with the expected color-space limits of the three Robitaille Stages (Stage III, II, and I running left-to-right). The points show the Cepheus YSO candidates with the same color code as the first panel. In this color-space, there is also excellent agreement between the Stage spaces and the Classes. The green points show flat spectrum sources. These do not correspond to a specific Stage, but are found along the Stage I/II boundary, as expected. \citet{2008evans} found a similar result for the entire c2d survey and showed that  extinction correction of the individual fluxes only marginally improved the agreement between the Classes and the Stages. The Stage regions take account of different angles of inclination and this is partially why the Class I sources scatter over such a large area \citep{2006robitaille}. The angle of inclination angle of a disk system to the observer is important as it can have a strong effect on the source's infrared signature and lead to the source being misclassified \citep{2008crapsi}.  
		
		The last panel of Figure \ref{fig:cc} shows the $H-K_{s}$ versus $J-H$ 2MASS colors of the Cepheus YSO candidates. The YSOs are not as separated as in the other two color-spaces, but there is a trend for the Class I and Flat sources to scatter away from the main locus. The box around this locus shows the expected domain of reddened stellar photospheres from the Robitaille models while the enclosed region to the right shows the region where any Stage evolutionary model can be present \citep{2006robitaille}.

\section{Cepheus YSO Properties}
\label{photometric}	
	
	\subsection{Additional Photometry}

		\label{photometry}
	
		\subsubsection{Guide Star Catalog}
		
		A search was made of the {\it HST} Guide Star Catalog-II (GSC-II) for visual companions within 5 arcseconds of the \spitzer/2MASS sources. The GSC-II is an all-sky compilation of astrometric and photometric data from a series of different catalogs including the Palomar Sky Survey-II (POSS-II), the Palomar Quick-V survey, and Tycho Catalogs \citep{2000mclean}. Of the 133 YSOs identified by \spitzer, 93 were also identified in the GSC-II. All of these had either ``star'' or ``non-star'' as their GSC-II spectral  classification; none were classified as ``galaxy.''   
		
		Table \ref{tab:otherflux} lists, where available, the GSC-II derived Johnson-Cousins $BVRI$ photometry for sources listed in Table \ref{tab:ysofluxes}. Column 1 lists the YSO candidates index from Table \ref{tab:ysofluxes}, column 2 lists the GSC-II identifier of the associated source, and columns 3-6 list the equivalent flux photometry in the $BVRI$ bands. The majority of the GSC-II identifications were from the POSS-II survey and had $B_{\rm J}$, $R_{\rm F}$, and $I_{\rm N}$ photometry. $R_{\rm F}$ and $I_{\rm N}$ are from the Johnson-Cousins photometric system \citep{1991reid}, $B_{\rm J}$ was converted to the same system using the \citet{1982blair} color transforms. $V$ data from the Palomar Quick-V survey were assumed to be in the standard Johnson system. The bright $\sim$7 mag $B_{\rm T}$ and $V_{\rm T}$ magnitudes for HD 200775 (YSO \#136) were from the Tycho survey and were converted to the Cousins-Johnson System \citep{1997esa}. All $BVRI$ magnitudes were converted to fluxes using the Cousins photometric zero points \citep{1979bessell}. These zero points are listed in column 4 of Table \ref{tab:bands} and the transforms are summarized in Table \ref{tab:transforms}. The errors on the optical photometry were typically 30\% and could be as high as 50\% in some cases. 
		
		\subsubsection{IRAS}
		
		A search was also made for coincident sources from the IRAS Faint Source Catalog within 15 arcseconds of the \spitzer\ positions \citep{1990moshir}. This catalog was used as it gave more matches than the Point Source Catalog and may be more reliable in regions of strong nebulosity \citep{2007rebull}. The lower resolution of the IRAS survey meant that it was possible to associate multiple \spitzer\ sources with a single IRAS source. In these cases, the IRAS source was assigned to the \spitzer\ source whose MIPS 24~\micron\ flux was nearest to the IRAS source's 25~\micron\ flux. No color corrections were applied to the IRAS fluxes. Column 7 of Table \ref{tab:otherflux} lists the associated IRAS identifier and columns 8-11 list the 12.5, 25, 60, and 100~\micron\ fluxes and upper limits. 
	
		Twenty cross identifications (15\%) were made between the 133 YSO candidates and the IRAS faint source catalogue. Of these 9, are coincident with sources in the \citet{1998kun} list of IRAS-based YSO candidates. A further 2 sources are coincident with the \citet{1998kun} list, but are not cross-identified with IRAS in Table \ref{tab:otherflux} because we restricted ourselves to just the Faint Source Catalogue and did not include the Point Source Catalogue. The Kun IRAS Sources are listed with literature names beginning with K98b (Kun 98 Second Table) in Column 3 of Table \ref{tab:ysofluxes}.

		\subsubsection{Submillimeter} 

		While more evolved protostars may be discerned by a negative spectral slope or a high bolometric temperature, it is more difficult to determine the evolutionary state of young YSOs without examining emission from their envelopes at longer wavelengths. Table \ref{tab:submm} lists available far-infrared and submillimeter photometry for \spitzer\ YSO candidates. The first column lists the YSO index and the second column lists the dark cloud associated with the YSO. Columns 3 to 6 list the actual fluxes quoted to a common precision. The last column lists, where relevant, a literature reference for the quoted flux. The 160~\micron\ fluxes were measured using a 40\arcsec\ radius aperture and 450, 850, and 1200~\micron\ fluxes were measured in a 20 or 25\arcsec\ radius aperture depending on the quoted reference.   
		
		Column 3 of Table \ref{tab:submm} lists MIPS 160~\micron\ fluxes. The fluxes were measured from the unfiltered BCD maps. Backgrounds were subtracted using a sky annuli of 1.0 to 1.87 times the radius of the aperture. Several of the apertures contained more than one YSO candidate. For these sources, the measured flux was divided evenly between the candidates within the aperture. Several bright sources, including PV Cep and L1157, were found to be completely saturated in the 160~\micron\ BCD images. These sources and those within regions of large scale extended emission where flux cannot be reasonably assigned to a compact YSO candidate have been excluded. The MIPS 160~\micron\ maps shown later are constructed from BCD images and have been filtered by a five-pixel diameter median-filter to remove artifacts and to replace pixels excluded due to saturation. 
								
		Columns 4 and 5 of Table \ref{tab:submm} list SCUBA 450 and 850~\micron\ fluxes. The Submillimeter Common User Bolometer Array (SCUBA) on the James Clerk Maxwell Telescope (JCMT) was a submillimeter camera that could map the sky simultaneously at 850 and 450~\micron. Fully sampled maps were produced by either scanning the hexagonal bolometer layout or offsetting the pointing-center in a 64-point jiggle pattern \citep{1999holland}.
		
		Several of the c2d studies of the individual regions have existing SCUBA data for their sources. \citet{2006young} undertook a targeted SCUBA campaign to observe cores in the c2d program. The cores they observed included L1157, L1221, L1228, L1251, and L1155C. The YSOs in L1251B are tightly clustered, making assignment of the lower-resolution SCUBA flux difficult. In their analysis of L1251B, \citet{2006lee} deconvolved the submillimeter emission of the embedded YSO candidates L1251B IRS 1,2 and 4 by using the brightness ratio of the resolved 24~\micron\ sources. Their estimates are listed in Table \ref{tab:submm}. Data from \citet{2006young} were used by the \citet{2008young} study of L1221.
		
		An additional search was made of the SCUBA archive for unpublished data coincident with other regions in this survey. Scan-maps of NGC 7023 taken on 1999 Oct 16-17 and jiggle maps of PV Cep taken on 1997 Oct 16 were downloaded. The data were reduced in the normal manner using the SCUBA User Reduction Facility \citep{1998jenness}. The scan-maps were restored using the Emerson 2 technique \citep{1995emerson}. The submillimeter zenith opacity at 850 and 450~\micron\ was determined using the `skydip' method and by comparison with the 1.3-mm sky opacity. Calibration was performed using observations of the planet Uranus taken during each shift. We estimate that the absolute calibration uncertainty is $\pm$10\% at 850~\micron\ and $\pm$25\% at 450~\micron, based on the consistency and reproducibility of the calibration. Secondary-beam corrected photometry was preformed in a 40\arcsec-diameter aperture at the location of each YSO candidate. 
				
		\citet{2008kauffmann} surveyed a selection of the c2d small cores with the 1.2mm MAMBO bolometer array on the IRAM 30-m telescope \citep{1999kreysa}. Seven of the regions they surveyed are coincident with the Cepheus Flare cores that contain YSO candidates. Photometry for these YSOs is listed in column 6 of Table \ref{tab:submm}.

	\subsection{Spectral Energy Distributions}
		\label{seds}

\begin{figure*}
	\centering{
		\includegraphics[width=0.8\textwidth]{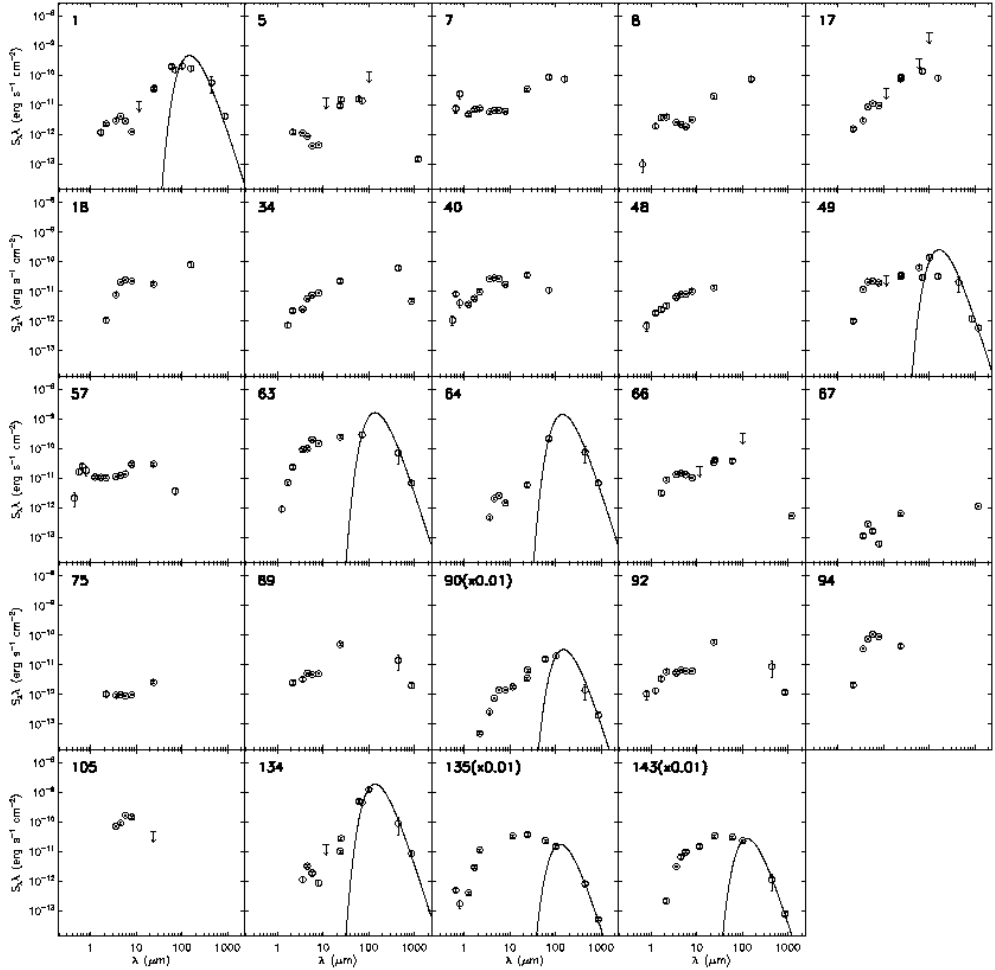}
	}
	\caption{\label{fig:seds1} SEDs for Class I YSO candidates towards the Cepheus Flare. The open circles with error bars show the data points from GSC-II, 2MASS, IRAC, MIPS, IRAS, and SCUBA where available. Upper-limits are also shown where available. The solid curve shows a simple greybody fit to the long wavelength data (see text for details). The YSO Id is shown in the top left hand corner of each box. If this is followed by ``x0.01'' it indicates that the SED has been scaled downwards by two dex in order to place it on the same grid as the other SEDs.}
\end{figure*}

\begin{figure*}
	\centering{
		\includegraphics[width=0.8\textwidth]{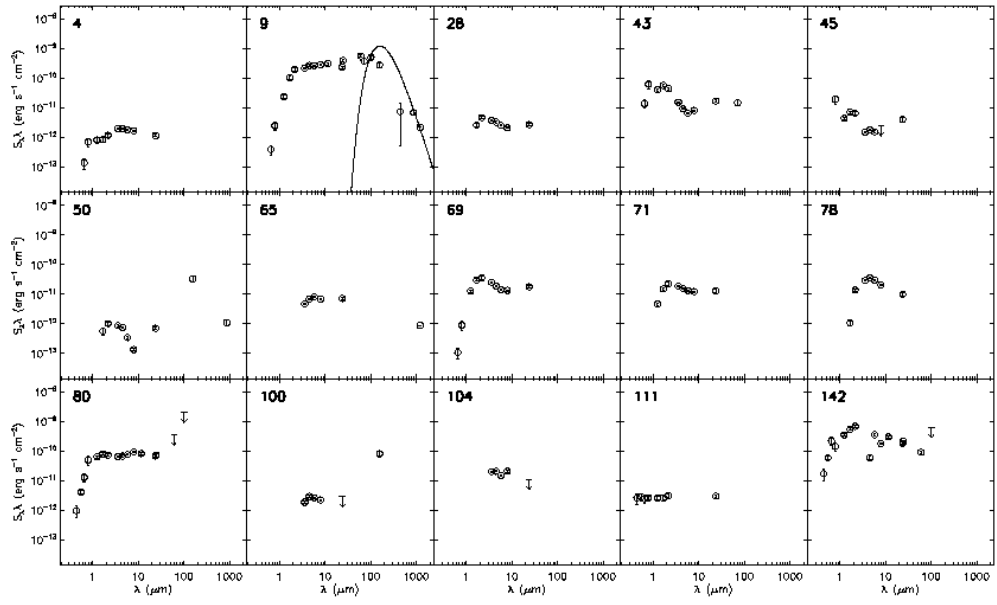}
	}
	\caption{\label{fig:sedsf} SEDs for flat spectral YSO candidates towards the Cepheus Flare. Details as described for Figure  \ref{fig:seds1}. }
\end{figure*}

		Figures \ref{fig:seds1}, \ref{fig:sedsf}, \ref{fig:seds2} and \ref{fig:seds3} show plots of the spectral energy distributions (SEDs) of sources classified as Class I, Flat, Class II and Class III respectively. The open circles show, where available, the photometry from GSC-II, 2MASS, IRAC, MIPS, {\it IRAS}, and SCUBA. The arrows show the position of flux upper limits. The top-left hand corner of each plot is labeled with the index of the YSO. If the source index is followed by ``x0.01'' it indicates that the SED has been scaled downwards by two dex in order to place it  on the same grid as the other SEDs. In Figures \ref{fig:seds2} and \ref{fig:seds3}, the gray lines show two comparison SEDs that have been normalized near the peak of the dereddened SED (usually the 2MASS $J$ band). The solid gray line is a NEXTGEN profile for a K7 star \citep{1999hauschildt} and the dashed gray line with error bars is the median SED for a T Tauri star in Taurus \citep{2005hartmann}. These SEDs are discussed in more detail in Section \ref{sedmodelling}.
		
		For the Class I YSOs in Figure \ref{fig:seds1} with 3 or more detections longwards of 65~\micron, we fit simple greybody SEDs assuming $\beta=2$ (see \citealt{2007kirk} for details) to characterize their submillimeter luminosities. Table \ref{tab:sed1} lists the derived parameters from these fits. Column 1 lists the YSO index, column 2 lists the fitted dust temperature and column 3 lists the submillimeter luminosity integrated under the fitted greybody between 350~\micron\ and 2000~\micron.  
	
		An alternative method to the spectral index for deriving the evolutionary classification of a YSO is to estimate the source's bolometric temperature, $T_{\rm bol}$, which is the temperature of a blackbody that has the same mean frequency as a source's observed SED \citep{1993myers}. $T_{\rm bol}$ can classify embedded protostars more effectively than the spectral index as it uses the entire spectrum and not just the infrared portion used for the spectral index \citep{2008enoch}. Column 5 of Table \ref{tab:ysoprop} lists $T_{\rm bol}$ for each protostar calculated using all available data points. A Class 0 source has $T_{\rm bol}<70$~K, Class I has $T_{\rm bol}=70-650$~K, Class II has $T_{\rm bol}=650-2880$~K and Class III has $T_{\rm bol}>2880$~K \citep{1995chen}. The $T_{\rm bol}$ boundaries for the equivalent of the intermediate flat spectrum class were not set by \citet{1995chen}, but \citet{2008evans} recently suggested boundaries of 350~K and 950~K. Column 6 of Table \ref{tab:ysoprop} lists the Class for each YSO as derived from the \citet{1995chen} boundaries with the \citet{2008evans} modification for flat spectrum sources.
		
		The number of YSOs in each Class as classified by $T_{\rm bol}$ is listed in Column 3 of Table \ref{tab:propStats}. The spectral index cannot classify a source as a Class 0 source so no total is listed the $\alpha_{\rm IR}$ column for such objects. While there is general agreement between the relative number of each Class of YSOs between both schemes (i.e., a few Class III sources compared to a lot of Class II sources), the bolometric method tends to skew a source's classification towards an earlier class (if the class changes at all). The spectral index is only calculated on flux data up to 24~\micron\ so flux data longwards of 24~\micron, as would be expected to dominate the SED of an embedded protostar, will not factor into a source's classification. For example, YSO \#68 has a spectral index of -0.51 which classifies it as a Class II source whereas it has a bolometric temperature of 9.3~K which places it firmly in the Class 0 regime. 
		
		\citet{2008enoch} examined the effect on $T_{\rm bol}$ of excluding a 160~\micron\ data-point. They found that $T_{\rm bol}$ will be over-estimated and the ratio of Class 0 to Class I sources will be skewed towards Class I sources. For the Cepheus data, we found that the exclusion of all data longwards of 150~\micron\ (i.e., the 160~\micron\ and submillimeter data listed in Table \ref{tab:submm}) did not significantly affect the numbers of sources classified as Flat or Class II and III. It did, however, affect the relative number of Class 0 and I sources. The number of Class 0 and I  sources classified by $T_{\rm bol}$ calculated using the submillimeter data is 18 and 9 respectively (as shown in Table \ref{tab:propStats}). Whereas, the number of Class 0 and I sources classified by $T_{\rm bol}$ calculated without the submillimeter data is 4 and 16 respectively in agreement with the \citet{2008enoch} finding.
		
		We retain the submillimeter data points when calculating the value of $T_{\rm bol}$ quoted in Table \ref{tab:ysoprop} and used to calculate the statistics in Table \ref{tab:propStats}. The bolometric method undoubtedly gives a better assessment on the evolutionary status of an embedded protostar and we discuss the individual values for potential Class 0 protostars in Appendix \ref{regions}. The number of sources with high quality submillimeter data, however, is smaller than our total infrared sample size. Since this could bias our source classifications, we retain the $\alpha_{\rm IR}$ classifications as our main scheme. This is also done for consistency with earlier studies.
				
	\subsection{Luminosity Function} 

		The bolometric luminosities, $L_{\rm bol}$, of the YSOs are estimated by integrating under all available data-points in each YSO's SED using a simple trapezoidal method. The distance to each YSO was taken as the distance towards its respective associated dark cloud, as discussed in Section \ref{distance}. The resulting luminosities are listed in column 7 of Table \ref{tab:ysoprop}. \citet{1997yonekura} estimated the IRAS luminosity function of YSO sources associated with the molecular cores in their CO survey of Cepheus. They parametrized the number of sources $dN$ in the luminosity interval $L_{*}+dL_{*}$ as
		\begin{equation}
		\frac{dN}{dL_{*}} = N_{0} \left( \frac{L_{*}}{L_{\odot}} \right)^{-p}
		\end{equation}
		where $N_{0}$ is a normalization factor and $p$ is the power-law index of the function. \citet{1997yonekura} fit $N_{0}$ = 8.9 and $p=1.40\pm0.32$ above their completeness limit of 1$L_{\odot}$ for their close group of IRAS sources. Figure \ref{fig:lumin} shows in its upper panel the function for the 133 \spitzer\ identified YSOs. The slopes were fitted by a least-squares fit above the estimated break of $\log(L_{*}/L_{\odot}) = -1.5$. The slope of the luminosity function fitted to all sources is $p=1.61\pm0.13$. This value agrees with upper range of the \citet{1997yonekura} error bars. The individual regions show  a range of luminosities plus a single source that is approximately one dex more luminous that the rest of the YSO population.
		
		The lower panel of Figure \ref{fig:lumin} shows the same histogram data as in the upper panel but plotted with $dN$ rather than $\log(dN/d(L/L_{\odot}))$ on the abscissa. The luminosity distribution peaks at the point where the upper plot diverges from the single-power law, i.e., at a luminosity of 0.06 $L_{\odot}$. This \spitzer\ completeness limit is similar to that found for other c2d and Gould Belt regions \citep[e.g., ][]{2007harvey} and is 1.5 dex lower than the {\it IRAS} completeness limits \citep{1997yonekura}.
		
		A lot of sources will have differing photometric coverage depending on DSS, SCUBA, and IRAS detections. Therefore, we also calculate the more homogeneous $L_{\rm IR}$, the luminosity integrated between 1--30~\micron\ for all sources with a 3.6~\micron\ detection. The distribution of $L_{\rm IR}$ is shown in Figure \ref{fig:lumin} as the dashed histogram and the individual values are listed in column 8 of Table \ref{tab:ysoprop}. There is good agreement in the peak position and width of the two histograms, but it is noticeable that the $L_{\rm IR}$ lacks $L_{\rm bol}$'s higher luminosity tail. When compared to the equivalent $L_{\rm IR}$ distribution for Serpens and Lupus, the Cepheus $L_{\rm IR}$ distribution peaks in the same place, but the breadth of the peak is wider. All show the same peak just before 0.1 $L_{\odot}$ followed by a sharp drop to zero around 0.01 $L_{\odot}$ \citep{2007harvey, 2008merin}.
	
	\subsection{SED Modeling}
		\label{sedmodelling}

		A dusty circumstellar disk and envelope at a temperature lower than the central YSO will contribute to the combined SED as excess emission at infrared and millimeter wavelengths. The magnitude of this infrared excess can be recovered if the SED of the star is subtracted from the combined spectra. The spectral type of the YSO needs to be known so that a template SED can be used, however. In general the spectral type will not be known without a targeted campaign of spectrographic observations. \citet{2008merin} fitted NEXTGEN stellar models to continuum optical and \spitzer\ data from Lupus to calculate their sources' spectral types and to deredden their SEDs.  \citet{2007harvey} and \citet{2008harvey} made the simplifying assumption that the underlying spectral type of their YSO population was a low-mass K7 T Tauri star or in the case of more luminous stars an A0 type. We will use the same K7 assumption to calculate the infrared excesses for the Cepheus YSOs. 
		
		For the Class II and Class III sources, the effect of reddening by line-of-sight extinction was removed from the SEDs by using the visual extinction towards each source calculated from its 2MASS $J-K_{s}$ color under the assumptions of a K7 underlying spectral type and a $R=5.5$ interstellar extinction law \citep{2001weingartner}. The dereddened data are shown in Figures \ref{fig:seds2} and \ref{fig:seds3} as solid markers. A NEXTGEN K7 profile \citep{1999hauschildt}  normalized against the dereddened $J$-Band flux is plotted for comparison. For a subset of sources, the dereddened optical photometry was significantly higher than the K7 profile that had been normalized to the dereddened $J$-Band flux. These sources include the known variable stars FT Cep, FU Cep, EH Cep and FV Cep. For these sources, it was necessary to modulate the estimated $A_{\rm V}$ down by a factor of 2-4 to make the K7 profile and the dereddened fluxes coincident. 
		
		The dashed line in Figures \ref{fig:seds2} and \ref{fig:seds3} is the median SED for a T Tauri star in Taurus \citep{2005hartmann} that has been normalized to the dereddened 2MASS $J$ flux. The profile represents a prototypical optically thick accretion disk surrounding a T Tauri star. The SEDs of some YSO candidates (e.g. YSO \#3) follow this profile quite closely, but others are closer to the K7 profile and only show a small amount of infrared excess at 24~\micron\ (e.g. YSO \#22). Approximately 30 YSOs (25\% of the disk population) have infrared excesses well below the median SED of the T Tauri stars in Taurus. Implying that these sources have evolved or settled disks. This low fraction also implies that the remaining disks in Cepheus show characteristics of being actively accreting and optically thick.

		Tables \ref{tab:class2model} and \ref{tab:class3model} list the SED modeling parameters and results for the Class II and Class III YSO candidates. Column 1 lists the YSO index. Column 2 lists the $A_{\rm V}$ that was used to deredden the SED and column 3 lists the waveband that was used to normalize the stellar profile. The luminosity of the YSO, $L_{\rm star}$, was estimated by integrating the normalized stellar profile after it had been interpolated to the observed wavelengths. The luminosity of the circumstellar disk, $L_{\rm disk}$, was estimated by integrating the difference between the interpolated normalized stellar SED and the dereddened SED. $L_{\rm star}$ and $\log(L_{\rm disk}/L_{\rm star})$ are listed in columns 4 and 5 of Tables \ref{tab:class2model} and \ref{tab:class3model}. 
		
		A histogram of $\log(L_{\rm disk}/L_{\rm star})$ is shown in Figure \ref{fig:disclumin}. This ratio can be used to characterize a disk as either accretion, passive reprocessing, or debris-like \citep{1987kenyon,2008hillenbrand}. The dividing-line between debris and passive disks suggested by \citet{2008currie} and used by earlier c2d papers was $\log(L_{\rm disk}/L_{\rm star})=-1.7$, however the  \citet{2008hillenbrand} study of young debris disks showed that their luminosities were well below the -1.7 value. We use a division of -3 based on the \citet{2008hillenbrand} result. These divisions are marked in Figure \ref{fig:disclumin}. 
		
		The dashed histogram shown in Figure \ref{fig:disclumin} only includes sources with complete IRAC spectroscopy (i.e., it excludes the 2MASS/MIPS identified sources). The difference between the two histograms shows that the majority of the sources identified as passive disks have been identified using the 2MASS/MIPS scheme. This pattern could be explained if $L_{\rm disk}$ was underestimated due to the lack of IRAC photometry. Assuming that the estimate of $L_{\rm star}$ did not change, the addition of IRAC photometry could cause $L_{\rm disk}$ to increase and would preferentially shift a source towards the accreting disk region. Thus, the observed frequency of passive disks could be lower than actually shown.
		
		It is clear that the majority of the circumstellar disks modeled by this method are accretion disks with a peak in the distribution of $\log(L_{\rm disk}/L_{\rm star})=0.4$. This is in agreement with the disk fractional luminosity for Lupus \citep{2008merin} and Chamaeleon \citep{2008alcala}. We do not, however, find as many passive and debris disks as was found in Serpens \citep{2007harvey}. Differences in stellar ages and masses between clouds could account for different distributions, but a detailed analysis of this possibility is beyond the scope of this paper. The large percentage of accreting disks is consistent with our finding that a relatively small number of sources have infrared excesses below the median T Tauri SED in Taurus. Together these points suggest that the disk average evolutionary status in Cepheus is very close to primordial and that there is little evidence for disk evolution. 
				
		\citet{2007cieza} introduced the second order parameters $\lambda_{\rm turnoff}$ and $\alpha_{\rm excess}$ to characterize the disk infrared excess. The first parameter, $\lambda_{\rm turnoff}$, is the longest wavelength without significant infrared excess. We compute this as the band shortward of the last band where the ratio of disk flux to observed flux is greater than 80\%. When no band has a ratio greater than 80\% we set $\lambda_{\rm turnoff}$ equal to the longest wavelength band. The second parameter, $\alpha_{\rm excess}$, is the spectral index of data points longwards and inclusive of $\lambda_{\rm turnoff}$ computed in the same manner as $\alpha_{\rm IR}$. The calculated values of $\lambda_{\rm turnoff}$ and $\alpha_{\rm excess}$ for the Cepheus YSOs are listed in columns 6 and 7 of Tables \ref{tab:class2model} and \ref{tab:class3model}. Figure \ref{fig:discexcess} shows a plot of $\lambda_{\rm turnoff}$ versus $\alpha_{\rm excess}$. Class II YSOs are shown as open circles and Class III YSOs are shown as asterisks.    
		
		\citet{2007cieza} found an evolutionary sequence along $\lambda_{\rm excess}$ with Classical T Tauri stars having a $\lambda_{\rm excess}$ shortward of 2MASS $K_{s}$. They also showed that the majority of $\alpha_{\rm excess}$ values cluster around -1 irrespective of $\lambda_{\rm turnoff}$, but that the spread in $\alpha_{\rm excess}$ increased with $\lambda_{\rm turnoff}$. Figure \ref{fig:discexcess} shows the same trend for clustering around $\alpha_{\rm excess} \sim -1$, but as most stars have had their stellar profiles normalized to the 2MASS J we would not expect to see values of $\lambda_{\rm turnoff}$ equal to it. Surprisingly, there are no objects with  $\lambda_{\rm turnoff}=4.5$~\micron. The calculation of $\alpha_{\rm excess}$ and $\lambda_{\rm turnoff}$ is affected by the normalization of the stellar profile and to the assumed spectral type. Thus, the empty 4.5~\micron\ band and the scatter of excesses at shorter wavelengths could possibly be due to imperfect knowledge of each source's spectral type.  
	
		\citet{2007harvey} plotted a similar diagram for Serpens and showed that the Serpens Class III sources predominately had a $\lambda_{\rm turnoff} \geq 5.8$~\micron. We find a similar result for the Class III sources in Cepheus as shown by the values of $\lambda_{\rm turnoff}$ in Column 6 of Table \ref{tab:class3model} and the few sources that appear at IRAC 3 and 4 in Figure \ref{fig:discexcess}. We are unable to calculate $\alpha_{\rm excess}$ for the three sources with $\lambda_{\rm turnoff}=24$~\micron\ as none of these sources have longer wavelength data points against which to calculate an index. All three were Class III sources, however, and would have followed the trend for those sources appearing to the right in Figure \ref{fig:discexcess}. 

		\cite{2008merin} showed that an object in the top right of the diagram was likely to have a transitional disk - an optically thick disk with a central cavity larger than several AU. These objects appear as a photospheric SED with an infrared excess only at the longest wavelengths. The object in the top right of Figure \ref{fig:discexcess} is YSO \#83, a Class II YSO in L1251B associated with IRAS F22367+7448. It has been detected in x-rays and was classified as a Classical T Tauri based on its large H$\alpha$ equivalent width and spectroscopic follow-up \citep{2006simon,1993kun, 2008kun}. A single transitional disk source may also be indicative that most of the disks in this region are at an early evolutionary stage. By comparison the \spitzer\ study of three clusters around Cep OB2 found that 10\% of detected disks were transitional in nature \citep{2006aguilar}. Several sources have strong 70~\micron\ excesses, e.g. YSO \#16, 35, 122. Similar objects are discussed in the \spitzer\ studies of Lupus \citep{2008merin} and Chamaeleon \citep{2008alcala} and may represent a very young population of debris disks \citep{2008hillenbrand}.
		
		In Cepheus, there is a population of isolated T Tauri stars ($l\sim117-122^{\circ}$) which are unassociated with molecular material and are outside of the regions surveyed in this paper ($l<115^{\circ}$). The isolation of these objects compared to their evolutionary status suggests that they formed in situ and that their natal clouds have been removed by interaction with the Cepheus Flare Shell \citep{2005tachihara}. If cloud disruption is as effective as argued by \citet{2005tachihara} then it could explain why we are preferentially seeing young disks in our YSO sample. Our observations were specifically targeted towards dense clouds and would therefore be missing older YSOs whose clouds have already been dispersed. In Appendix \ref{l1228}, comparison of the YSO distribution to the pattern of extinction and 160~\micron\ emission shows that the L1228 South YSO group is on the very edge of the L1228 cloud. It is perhaps reasonable to suggest that the L1228 cloud is currently being disrupted and that we are witnessing the unveiling of a cluster of T Tauri stars similar to the isolated group observed by Tachihara.

\section{Extended Structure and YSO Distribution}
\label{extended}

	Figure \ref{fig:distribution} shows the distribution of YSOs and visual extinction towards the regions in the Cepheus survey. Figure \ref{fig:mips160} shows the pattern of MIPS 160~\micron\ emission towards exactly the same regions. All regions are shown at the same angular scale.

	\subsection{Comparison of $A_{\rm V}$ and 160~\micron\ maps}
	
		Figure \ref{fig:distribution} shows $A_{\rm V}$ calculated from the \spitzer\ catalog. The line-of-sight extinction towards sources classified as reddened stellar photospheres between wavelengths of 2 and 24~\micron\ was calculated assuming an extinction law of $R=5.5$. These points were then filtered based on the equivalent 2MASS extinction to correct for contamination from isolated clumps of extinction. These irregularly spaced, filtered-data were then averaged onto a regularly spaced grid using a Gaussian kernel \citep{2007evans}. The resolution of the maps depends on the surface density of available sources. It was found that the highest resolution consistent between all regions in Cepheus was 150 arcseconds.
		
		The approximate areas where we have IRAC and MIPS overlap are shown by the irregular boxes on Figure \ref{fig:distribution}. These are the regions where there is enough photometric coverage to calculate the \spitzer\ $A_{\rm V}$. The greyscale within the overlap regions show the local range of the 2.5 arcminute resolution \spitzer\ $A_{\rm V}$ on a linear stretch. Outside of the overlap area, the greyscale shows the 6 arcminute resolution DSS extinction on a linear stretch between 0.5 and 3 mag \citep{2005dobashi}. Dashed black contours outline the region where the DSS extinction is greater than $A_{\rm V}=1$ and the \spitzer\ extinction is greater than $A_{\rm V}=5$. The \spitzer\ extinction maps show higher values of $A_{\rm V}$ than the Dobashi maps because the infrared stars are visible at higher column densities. 
		
		Figure \ref{fig:mips160} shows 160~\micron\ emission observed by MIPS towards towards the five regions in Figure \ref{fig:distribution}. The greyscale is shown on a log stretch between the local minimum (approximately 20 MJy/sr) and 316 MJy/sr for NGC 7023 or 100 MJy/sr for the rest. The $A_{\rm V}$ contours and overlap box from Figure \ref{fig:distribution} are repeated for reference. There is a strong correlation between the distribution of the visual extinction and the 160~\micron\ emission. This pattern of the 160~\micron\ emission following the $A_{\rm V}$ map has also been seen in many other regions \citep[e.g. ][]{2007rebull,2007chapman}.
		
		In general, the 160~\micron\ emission above $\sim40$~MJy/sr is confined within the $A_{\rm V}=1$ DSS contour. Some regions, particularly L1241 and L1247, also show a correlation at smaller angular scales. There are also subtle differences, however. For example, in L1228S the extinction peaks approximately 4 arcminutes further north than the 160~\micron\ emission. The center of NGC 7023 saturates the MIPS detectors and is left blank in the map.  
		
		Table \ref{tab:cores} lists parameters for the clouds and YSO clusters/groups shown in Figure \ref{fig:distribution}. We identify dense cores based on a comparison of the the extinction and 160~\micron\ maps. The $A_{\rm V}=5$ contour was chosen to define the cores as it was the level where most of them separated from each other. These are labeled on Figures \ref{fig:distribution} and \ref{fig:mips160}. Due to the partial mapping of L1147 and small size of L1155E these regions are merged with L1148 and L1155C when calculating derived properties. L1174A and L1174B with the unlabeled L1174C are listed together as NGC 7023. For each of the remaining fourteen cores, we measured the peak and mean $A_{\rm V}$ within the $A_{\rm V}=5$ contour, the area contained within that contour, and the equivalent mass calculated from the $A_{\rm V}/N_H$ relationship of \citet{1978bohlin}. The c2d and Gould Belt $A_{\rm V}$ values are calculated assuming an extinction law of $R=5.5$. To use the \citet{1978bohlin} relation of
		\begin{equation}
			\frac{A_{\rm V}}{N_{\rm H_{2}}} = 1.1\times10^{-21} \textrm{cm}^{-2} \textrm{ mag},
		\end{equation}
		which was calculated for a value of $R=3.1$, we include a correction of 0.716. These values are listed in columns 11-14 of Table \ref{tab:cores}. The equivalent values for the YSO clusters/groups are calculated using a YSO density contour that defines the edge of the group (see Section \ref{clustering}).

	\subsection{YSO distribution and completeness}
		\label{knownYSOs}
		
		The distribution of YSO candidates is over-plotted on the five regions in Figure \ref{fig:distribution}. The markers are color coded to the class of the candidate - red for Class 0/I, green for Flat spectrum, blue for Class II, and purple for Class III. Columns 2--6 of Table \ref{tab:cores} list the number and type of YSO candidates that were detected towards each core within the $A_{\rm V}=1$ contour or YSO group/cluster boundary. Column 7 of Table \ref{tab:cores} lists the ratio of Class I YSO (column 2) to the total number of YSOs (column 6). In general, the protostellar cores had such ratios in excess of $\sim20\%$ while the cores containing clusters had values less than $\sim20\%$. 
		
		A YSO was considered to be ``off-cloud'' if it was outside of the $A_{\rm V}=1$ contour and was not a member of a formal YSO cluster or group (see Section \ref{clustering}). Results for these YSO candidates are listed as a separate group at the bottom of Table \ref{tab:cores}. The isolated population of Weak-Line T Tauri YSOs in Cepheus have higher YSO-to-cloud separation ($\sim10$~pc) than found in other star formation regions \citep{2005tachihara}. While our observations do not get that far ``off-cloud'', we can use the number of off-cloud YSOs, most of which are Class II and Class III sources, to estimate the density of off-cloud YSOs that are serendipitously located close to dense regions. The area at an extinction of $A_{\rm V}>1$ is 2.14 square degrees and the total mapped area is 8.33 square degrees. We detect 19 \spitzer\ identified YSOs at an extinction of $A_{\rm V}<1$. Assuming that these are not background galaxies, we estimate an off-cloud YSO density of $19/(8.33-2.14)=3.1$ YSOs per square degree. Conversely, we detect $133-19=114$ \spitzer\ identified YSOs at an extinction of $A_{\rm V}>1$.  This gives an on-cloud YSO surface density of $114/2.14 = 53$ YSOs per square degree. 
				
		The SIMBAD object types (the ``otype'' field) of the sources listed in the first three sections of Table \ref{tab:ysofluxes} include YSOs and YSO candidates (Y*0 and Y*?), emission-line stars (EM*), variable star of various classes (V*, Pu*, Mi*, Or*), reflection nebulae (RNe), and T Tauri stars (TT*). A number of sources have multiple object types. A search was made for additional sources with similar object types in a region coincident with our mapped area. The result included 14 variable stars located around NGC 7023 and towards the L1148+L1155 ring. Only HD 200775, however, had a match against a source in the \spitzer\ catalog. \citet{2008kun} reviewed surveys of H$\alpha$ emission line stars towards the Cepheus Flare and produced a list of emission line stars that had been confirmed spectroscopically as T Tauri stars. There are 19 emission-line stars in the list of additional SIMBAD sources. The emission-line stars were preferentially found towards L1228 as this area was the subject of a H$\alpha$ survey by \citet{1990ogura}. Thirty-one of the \citet{2008kun} T Tauri stars are coincident with \spitzer\ YSO candidates, but none of the additional SIMBAD emission line stars are confirmed as a T Tauri star.  
		
		The SIMBAD search was narrowed to only include objects previously identified as YSOs and pre-main sequence stars or candidates for either. A total of 47 YSOs were found in the SIMBAD database of which 36 were cross-identified with \spitzer\ YSO candidates. Half of these cross-identified YSOs are in the L1251B region and are from the \citet{2006lee} c2d paper. Excluding these, 18 out of 30 SIMBAD YSOs were cross-identified with the \spitzer\ YSO candidates. The 12 SIMBAD YSOs that were not identified by the \spitzer\ schemes plus 3 T Tauri stars from \citet{2008kun} that were not identified by the \spitzer\ schemes are listed in Table \ref{tab:simbadYSOs}. Column 4 lists the catalog that the object was taken from (S for SIMBAD, K for \citealt{2008kun}). Column 5 lists the core or region coincident with the object's position. Column 6 lists whether the object position was coincident with an entry in the Cepheus \spitzer\ catalog even if that object was not classified as a YSO candidate or formally detected. Ten of the additional YSOs were coincident with entries in the SGBS Cepheus catalog (PV Cep had only 2MASS data). Each of these non-\spitzer\ identified YSO candidates has been assigned a YSO Id (as listed in column 7) and their photometry is listed in the fourth section of Table \ref{tab:ysofluxes}. 
		
		The embedded Class 0 protostar L1157 was outside of our IRAC area and has no 2MASS detection. L1157 was studied with \spitzer\ by \citet{2007looney}. They presented striking 8~\micron\ images of L1157 that showed its CO outflow in emission and a flattened circumstellar envelope in absorption against the bright 8~\micron\ background. Its IRAC fluxes have been added to our catalog based on 4.5-arcsec radius aperture-photometry conducted on post-BCD data downloaded from the \spitzer\ data archive. \spitzer\ data for PV Cep exists in the data archive\footnote{AOR No. 18955008}, but it is saturated at all four IRAC wavelengths. The 2MASS/MIPS YSO \#118 is also in the PV Cep IRAC field so we add its IRAC fluxes to the catalog in the same manner as for L1157. 
		
		HD 200775 would have been picked up by the IRAC YSO scheme if we had reduced the S/N cut to 2. It was detected in the filters used for the 5-Band scheme, but it was excluded because the 3.6~\micron\ flux was below 2~$\sigma$. The source is very luminous ($L_{\rm bol}=100$~{L}$_{\odot}$) and is saturated in all MIPS bands. It is illuminating the surrounding gas giving it a larger flux uncertainty than an isolated star would have. The position of NGC 7023 S T \citep{1983sellgren} was identified with IRAS F21023+6754 by \citet{1998kun}. Kun also identified the sources IRAS F21025+6801 and F20597+6800 as IRAS YSO candidates within the NGC 7023 region. Of these latter three sources, only IRAS F21025+6801 is coincident with an entry in the Cepheus catalog and then it only has a 24~\micron\ upper limit. None of the \citet{2005tachihara} x-ray T Tauri candidates were coincident with the area we surveyed. 
	
		A total of 143 YSOs were identified from the \spitzer\ catalog and the SIMBAD database. Of these 133 (93\%) were identified by the color-color schemes discussed in Section \ref{candidates} and a further 10 (7\%) were coincident with entries in the catalogue. The number of previously known YSOs that were not identified by \spitzer, but had entries in the Cepheus catalogue is listed in Column 8 of Table \ref{tab:cores}. The total number of known YSOs that were not identified by \spitzer\ is listed in Column 9 of Table \ref{tab:cores}. The number of \spitzer\ identified YSO candidates within $A_{\rm V}=5$ contours discussed in the previous section is Column 10 of Table \ref{tab:cores}.
		
	\subsection{Pre-stellar cores, protostellar cores, and cores with YSO groups/clusters}
		\label{clustering}
		
		The dense cores seen in extinction in Cepheus can be split into three groups. The first division are pre-stellar cores. These are cores that are believed to be taking part in the star formation process but have not yet formed an embedded YSO (i.e., $N_{\rm YSO}=0$). We then define the cores that have $N_{\rm YSO}>0$ within the $A_{\rm V}=5$ contour as protostellar cores. Concentrations of YSOs in protostellar cores with at least 5 and 35 members are termed groups and clusters respectively \citep{2003lada}.
		
		For pre-stellar cores, the YSO count within the $A_{\rm V}=5$ contour is used because the area subtended by the $A_{\rm V}=1$ contour can be large enough to give a significant chance of including a field YSO. For example, the area of L1241 at $A_{\rm V}>1$ is approximately 0.6 square degrees. Based on the average off-cloud YSO surface density, we would expect 1-2 YSOs to be coincident with this region. There is one Class III YSO just inside the $A_{\rm V}=1$ border of L1241. Given its class and the likelihood of finding a coincident YSO within that border, we conclude that L1241 is starless. The parameters for the Cepheus pre-stellar cores are listed in the first section of Table \ref{tab:cores}. The respective masses and areas are calculated from the $A_{\rm V}=5$ contour.   
		
		The cores that have $N_{\rm YSO}>0$ are termed protostellar cores. The distribution of YSOs in Figure \ref{fig:distribution} shows a number of protostellar cores. There have been multiple metrics devised for classifying the composition and concentration of YSOs in protostellar cores. For example, \citet{2004cartwright} used minimal spanning trees to derive a quantity $Q$ that could differentiate between a small, large-scale density gradient and multi-scale sub-clustering. Alternatively, the spatial density of sources can be estimated by calculating the angular distance $r_{\rm N}$ to the Nth nearest neighbor for each point on a regularly spaced grid \citep{1998christopher,1999gladwin, 2005gutermuth}.     
		
		For consistency with previous c2d/SGBS papers, we follow the nearest-neighbor clustering scheme described by \citet{2005gutermuth}. When calculating $r_{\rm N}$, the appropriate correction for spherical coordinates was used (this is vital for Cepheus due to its high declination). This angular distance is then used to calculate a volume number density assuming a spherically symmetric distribution of cluster members \citep{2005gutermuth}. Following \citet{2005gutermuth}, the volume density of stars is given by 
		\begin{equation}
			\rho_{*} = M_{*} \frac{N-1}{\frac{4}{3}\pi r_{\rm N}^{3} } \label{eqn:rhostar}
		\end{equation}
		where $M_{*}$ is the average stellar mass and $N$ is the chosen index for the neighbor. We use $N=5$ and assume an IMF with an average stellar mass of 0.5 M$_{\odot}$ for consistency with other c2d and SGBS papers \citep[e.g. ][]{2008jorgensen,2008harvey}. The value of $\rho_{*}$ was calculated across the regions in our survey. A limitation to this technique is that it can only calculate $\rho_{*}$ for groups with 5 or more members.
		
		The value of $\rho_{*}$ above which a cluster is stable against disruption from passing interstellar clouds is $\rho_{*} \ge 1.0 \textrm{ M}_{\odot}\textrm{ pc}^{-3}$ \citep{1958spitzer, 2003lada}. Cores with peak values of $\rho_{*}$ less than the disruption density or with fewer than $N$ members are listed as purely protostellar and are listed in the middle section of Table \ref{tab:cores}. 
			
		The cores with peak values of $\rho_{*}$ greater than the disruption density are listed in the third section of Table \ref{tab:cores}. Five regions had stellar mass volume densities in excess of the 1.0~M$_{\odot}$~pc$^{-3}$ disruption density. These are NGC 7023/L1172, L1251A, L1251B, L1228S, and L1228N. For this analysis, we have only included the YSO candidates towards NGC 7023 and not the main-sequence members listed in the literature. The pattern of $\rho_{*}$ towards each of these groups is shown by the colored contours on the upper panels in Figure \ref{fig:distribution}. The blue contour traces the disruption density while green contours trace values of 0.125, 0.25, 0.5, 2.0, and 4.0 times the disruption density. 
		
		Figure \ref{fig:distribution} shows that the disruption contours are approximately elliptical, likely because these relatively small and isolated YSO groups are themselves elongated. \citet{2008jorgensen} defined an empirical limit of 25 times the disruption density as a level which appears to more adequately trace the local YSO distribution. This level is shown as a yellow contour in Figure \ref{fig:distribution}. The red contour over NGC 7023 shows the equivalent of the yellow contour after known non-YSO cluster members are included. The position of the 21 non-YSO cluster members is shown by yellow crosses. 
		
		Each group can be classified by the number of members within a particular contour. YSOs within a minimum contour equal to the disruption contour (blue contour) are described as a ``loose'' association and YSOs within a minimum contour equal to 25 times the disruption density (yellow contour) are described as a ``tight'' association \citep{2008jorgensen}. A cluster is defined as an association with at least 35 members \citep{2003lada}. Based on these definitions, no association in Cepheus can be considered a true cluster although NGC 7023 is close. It is a loose association of 32 members which contains two tight groups, A and B, with 20 and 4 members each. The other four cores contain individual YSO groups that have tight and loose members, although the tight regions do not fragment as in NGC 7023.
	
		We have defined four different contours that can be used to define the extent of the cores. These are the $A_{\rm V}=1$ and $A_{\rm V}=5$ contours used for the cores themselves and the $\rho_{*}=1$ and $25 M_{\odot}$~pc$^{-3}$ used for the YSO groups/clusters. The third section of Table \ref{tab:cores} lists the YSO statistics and physical parameters for the cores with YSO groups. Each core is listed three times, first for the core itself as defined by the $A_{\rm V}=1$ contour, then for the loose and tight YSO groups as defined by the blue and green contours in Figure \ref{fig:distribution}. The relevant contour is used for the calculation of the group/cluster statistics. The tight groups for NGC 7023 A and B are listed separately. 
		
		In general, the statistics for the entire core and for the loose groups are approximately equal, but there are a few noticeable differences. The statistics calculated from the ``loose'' contour characterize the cores with groups better than the $A_{\rm V}=5$ contour. The NGC 7023 loose cluster subtends an area that is significantly larger than the area of the core as measured by its $A_{\rm V}=5$ contour. The L1228 YSO group straddles both $A_{\rm V}$ contours so using those contours to count the YSOs gives an artificially low number.
		
		The final column on Table \ref{tab:cores} lists the star formation efficiency (SFE) for each core as given by,
		\begin{equation}
			\textrm{SFE} = \frac{M_{\rm stars}}{M_{\rm cloud}+M_{\rm stars}}
		\end{equation}
		where $M_{\rm cloud}$ is the mass of the gas and dust in the core (as listed in column 14 of Table \ref{tab:cores}) calculated from the mean $A_{\rm V}$ within the core's boundary contour. The mass of the stars, $M_{\rm stars}$, is given by $N_{\rm YSO} M_{*}$ where $M_{*}$ is the same average stellar mass as assumed for the calculation of $\rho_{*}$. 
		
		There is no clear trend between the masses of the cores and the number of YSOs they contain. Thus, a larger stellar content automatically translates into a higher star formation efficiency. The protostellar cores without YSO groups/clusters have SFEs $\sim1\%$ while the overall SFE for the cores with YSO groups/clusters is $\sim8\%$. The SFE for the tight component of the groups/clusters is $>3\times$  higher than for the loose component. This can be interpreted as either an increase in the in situ star formation within the dense parts of the core or a drift of YSOs into the central part of the core from other parts of the cloud. In the former, the SFE is real and, in the latter, it is over-estimated because the mass of gas that those stars formed from is under-estimated. 
	
		The bottom line of Table \ref{tab:cores} shows the aggregated statistics for the entire mapped region. Columns 2-5 reiterate the YSO class statistics from column 2 of Table \ref{tab:propStats}. Table \ref{tab:regionstats} lists additional star formation statistics in the same format as table 3 of \citet{2008evans}. Columns 1-4 of Table \ref{tab:regionstats} list the boundary used for the analysis, the number of YSOs within that boundary $N_{\rm YSO}$ , the solid angle of the region $\Omega$, and the integrated area of the region $A$ assuming an average distance of 300~pc. Columns 5 and 6 respectively list the number of YSOs per square degree and square parsec. Column 7 lists the star formation rate (SFR$= N_{\rm YSO} M_{*}/\tau_{\rm sf}$ ). Where we have used the aforementioned assumption that $M_{*}=0.5$~M$_\odot$ (the same used in the clustering analysis) and that the timescale for star formation is $\tau_{\rm sf}=2$~Myr \citep{2008evans}. Column 8 lists the star formation rate per square parsec (SFR/$A$). Column 9 and 10 list the total mass of the cloud ($M_{\rm cloud}$) and the surface density of the cloud ($M_{\rm cloud}/A$). 
		
	\section{Discussion: Star Formation in the Cepheus Flare}
		
		Star formation in Cepheus is mostly concentrated in small, isolated groups of YSOs. A total of 78 out of 133 \spitzer-identified YSOs, i.e., 59\%, were determined to be members of loose YSO groups, leaving 41\% of the YSOs distributed elsewhere. By comparison, the average distributed YSO fraction in c2d clouds was only $\sim10$\% \citep{2008evans}. Figure \ref{fig:coverageYSOs} shows the distribution of YSO and YSO candidates towards the Cepheus Flare plotted over the distribution of H$\alpha$ \citep[log scaled greyscale; ][]{2003finkbeiner} and integrated CO emission (black contours; the same data as the white contours from Figure \ref{fig:coverage}, \citealt{2001dame}). The black foot prints show the limit of our survey and the red crosses show the positions of \spitzer\ YSO candidates from this paper. The positions of spectroscopically confirmed T Tauri stars are shown by open blue squares \citep{2008kun} and the position of X-ray WTTS are shown by filled blue squares \citep{2005tachihara}. The green circles show an unvetted list of objects described in the SIMBAD database as YSOs irrespective of their survey of origin. 
		
		The masses of cores with YSO groups ($N_{\rm YSO}\geq5$) and cores with only a few YSOs ($N_{\rm YSO}<5$) were found to be broadly similar, which suggests a factor other than core mass determines how many stars form. If we examine the location of the YSO groups (L1251A, L1251B, L1228S, and L1228N) in Figure \ref{fig:coverageYSOs}, we see that they are all located in the region nearest the Cepheus Flare Shell (CFS), whereas the majority of the cores with less than 5 YSOs (L1148, L1157, etc) are located on the opposite site of the region. There appears to be two different modes of star formation occurring here. The first mode of groups of YSOs has a SFE of $\sim8\%$ and has produced a YSO population that is dominated by Class II sources. The second unperturbed, almost quiescent mode has a SFE of $\sim1\%$ and has produced YSOs with a fairly uniform spread of infrared classes. 
						
		There is mounting evidence that star formation in the Cepheus Flare is triggered/influenced by the passage of the CFS \citep{1989grenier,2005tachihara,2008kun}. \citet{1998kun} noted that the surface distribution of IRAS identified YSOs in the Cepheus Flare peaked towards the edges of clouds and inferred that their formation was triggered by external shocks. Likewise, Figure \ref{fig:coverageYSOs} shows that the locations of the small YSO groups in L1228 and L1251 are on the edge of the CO contours facing the Cepheus Void. The cometary shape of C$^{18}$O emission in L1251 and an unusual velocity gradient across the cloud appears to have been caused by interaction with a passing wind/shock wave, but the failure to detect SiO emission from the cloud may indicate that the shock phase, at least in L1251, has passed \citep{1994sato}. It is the cometary head of L1251 that contains the L1251A and B YSO groups. 
		
		The L1228S YSO group not only lies on the edge of the CO contours but, as shown in Figure \ref{fig:distribution}, it also lies on the edge of the DSS $A_{\rm V}=1$ and \spitzer\ $A_{\rm V}=5$ extinction contours. The peak of the L1228S \spitzer\ $A_{\rm V}$ map is offset from the peak of the 160~\micron\ emission. The difference in position could be due to heating of the dust caused as the core is dispersed by interaction with the CFS. There are small collections of T Tauri stars (open blue boxes) on the Void side of both the L1228 and L1251 groups, suggesting that they have become exposed as material has been stripped away from the leading edge of the molecular clouds. For the group of T Tauri stars near L1251, \citet{1995toth} estimated that it would have taken 0.1~Myr after the passage of the shock for an offset of 1$^\circ$ to form, assuming an expansion velocity of 10~kms$^{-1}$.     
		
		There is a collection of SIMBAD YSOs and T Tauri stars close to the end of the CO filament at $l=112^\circ$, $b=+14^\circ$. It is outside of our selection criteria (it has peak $A_{\rm V}<3$), but it is coincident with the L1235 dark nebula. Given the position of the YSOs, the CO contour, and its orientation to the Cepheus Void it is possible that L1235 is another small YSO group, but one that is intermediate in evolution between L1228S and the Cepheus Void WTTS Group. Two other concentrations of YSOs are shown in Figure \ref{fig:coverageYSOs}, L1199 at a distance of 500~pc and a group associated with L1217 at a distance of 400~pc \citep{1998kun}. Both L1217 and L1235 are coincident with the G109+11 infrared loop \citep{2006kiss}. The small, isolated protostellar core L1221 was included in our survey and is also coincident with the interior of the infrared loop, but its assumed distance of 250~pc places it foreground of L1217 and L1235.
		
		The age of the supernova at the center of the Cepheus Flare Shell is 40,000~yr \citep{1989grenier}, but the bubble that the shell encloses appears to be significantly older, 7~Myr, and was probably a pre-existing wind driven cavity created by a star that subsequently exploded as a supernova \citep[][ and discussion therein]{2006olano}. Ionization pressure from \ion{H}{2} regions can enhance gravitational collapse in surrounding material, e.g., the Horsehead Nebula \citep{2006wardthompson}, and trigger star formation in the heads of pillar-like filaments, e.g., M16's ``Pillars of Creation'' \citep{1996hester}. We conjecture that this mode of star formation may have been responsible for some of the WTTS currently currently situated inside the Cepheus Void. The Cepheus Flare Shell is an expanding \ion{H}{1} ring \citep{2006olano}, however, and shows negligible H$\alpha$ emission when compared with the Ceph OB2 \ion{H}{2} region (see Figure \ref{fig:coverageYSOs}). The lack of H$\alpha$ emission means that any hot gas associated with the CFS has had sufficient time to cool or escape.   
				
		Following the \citet{2006olano} analysis of the dynamics of the CFS we estimate that its radius was 30~pc at 4~Myr ago and 40~pc 2~Myr ago. These two radii are shown by the concentric, dotted circles in Figure \ref{fig:coverageYSOs}. The current radius of 50~pc is shown by the solid circle. These circles are only sketches of the Shell's extent and ignore differences in the local density of material the Shell was propagating into and the effects of a 2 dimensional shape imposed on the projected distribution of a 3 dimensional structure. Nevertheless, the isolated Void WTTS are clearly located within the central circle and most of them are within a radius of 4$^\circ$ (the CFS radius after only 1-2 Myr). The Void WTTS have ages comparable to the age of the bubble \citep{2005tachihara}. 
		
		The ratio of more evolved YSOs to younger YSOs (Class I+F/Class II+III), i.e., the ``class ratio'', in protostellar cores ($N_{\rm YSO}<5$) and loose groups ($N_{\rm YSO} \ge 5$) is 11/9=1.22 and 28/53=0.52 respectively. The class ratio for the entire region is 35/98=0.36. Taken at face value, this implies that the protostellar cores represent a younger YSO population than the loose groups and that the loose groups have a relatively narrow range of YSO ages. The class ratio for distributed YSOs and for loose YSO groups averaged across the c2d clouds is 0.24 and 0.89 \citep{2008evans}. The c2d average for distributed YSOs is similar to the class ratio for the entire Cepheus Flare and further supports the idea that relatively distributed star formation is the dominant mode of star formation in this region. 
		
		Our SED analysis has shown that the disk population in Cepheus is skewed towards young, accreting disks. The age of the disks would be expected if the YSOs in the groups formed simultaneously due to a fairly recent triggering event, such as the passage of a shock wave. The current best estimate for the lifetime of low mass star formation prior to the Class III phase is 2-3~Myr \citep[see ][and discussion therein]{2008evans}. The four loose YSO groups are in the zone between the current position of the CFS and its estimated position 4~Myr ago. Given that L1251 and L1228 are believed to be on opposite sides of the Cepheus Void \citep{2008kun}, projection effects could mean that L1251 is closer to the Shell wall than it actually appears. Therefore, the general empirical ages of Class II dominated groups are consistent with their formation at a time when the CFS was coincident with their position.
		
		The Void WTTS \citep{2005tachihara} occur at the same galactic latitude ($b\sim16^\circ$) as L1241, but they are distributed towards the projected interior of the Cepheus Void at $115<l<123^\circ$. The region on the other side of L1251 and L1228, in the longitude range $107<l<110^\circ$, appears in Figure \ref{fig:coverageYSOs} to hold a reservoir of material that has not yet started forming stars in the same manner. L1241 also represents a large mass of material that has not yet started forming stars. It has usually been placed at the same distance as L1251 (see Section \ref{distance}), but the lack of star formation may indicate that it is actually at a different distance and, like the material in the range $107<l<110^\circ$, has yet to encounter the CFS. Alternatively, the CO emission towards L1241 may be the superposition of two or more low density clouds, \citet{1997yonekura} list two different components to $V_{\rm LSR}$ towards L1241.  
		
		Simulations of the impact of ionizing radiation on a turbulent ISM show that it can enhance pre-existing density contrasts and the efficiency of star formation \citep{2007dale,2009gritschneder}. This enhancement could explain why we see more efficient star formation in the cores towards the CFS. The surface density of the \citet{2009gritschneder} simulation at $t=250$~kyr (their Figure 1) resembles the distribution of CO seen in the Cepheus Flare (our Figure \ref{fig:coverageYSOs}). However, the time scale and physical dimensions of their simulation are shorter and smaller than inferred for the CFS (see above). 

		Moving further west there is a zone clear of CO emission at $l\sim106^\circ$ and then we reach the locations of L1172+L1174 at $l\sim104^\circ$ and L1148+L1152+L1155 at $l\sim102^\circ$. Even here, furthest from the CFS, it appears that some degree of triggering is at work. The NGC 7023 cluster in L1172 is being shaped by the powerful illumination of HD 200775 and holds the largest concentration of YSOs in the region we have surveyed. We presume that the formation of these YSOs has been triggered by compression of the material around the reflection nebula caused by HD 200775. L1148+L1152+L1155 are comprised of a series of pre-stellar cores and protostellar cores that contain only a few YSOs each. It would appear that this final region is the most unperturbed, but again there is some evidence of outside influences on the cores \citep{2008nutter}.
		
		The general scenario for star formation in the Cepheus Flare has the initial conditions of a turbulent ISM where distributed, quiescent, low-efficiency star formation is the norm. Approximately 7~Myr ago, a progenitor high-mass star with a strong circumstellar wind formed at the center of what would become the CFS \citep{2006olano}. The wind or radiation from the star compressed the surrounding gas into pillars/filaments in which a series of new YSOs formed. The influence of the progenitor star eventually stripped the natal molecular gas from this generation of YSOs and left them behind as the Cepheus Flare WTTSs \citep{2005tachihara}. The wind from the progenitor continued building up an expanding shell of material \citep{2006olano} until it went supernova approximately 40~kyrs years ago \citep{1989grenier}. A second generation of star formation occurred on surfaces facing the interior of the wind bubble. The passage of the CFS enhanced a core's SFE relative to the original quiescent mode of star formation and produced a series of small YSO groups. The material of the clouds that surrounded the YSO groups was stripped away before the co-evolving group, on average, reached the Class III phase.
		
\section{Summary}
	
	\begin{itemize}
	
		\item We have presented \spitzer\ IRAC and MIPS observations of the L1148+L1152+L1155, L1172+L1174, L1221, L1228, L1241, and L1247+L1251 dark cloud associations in the Cepheus Flare region ($D\sim300$ pc) as part of the \spitzer\ Gould Belt Legacy Survey (SGBS) of star formation within 500~pc.
		
		\item The SGBS delivery catalog (which is available from the SSC as a legacy data product) contains over 71 thousand sources within the area common to all wavelengths and detected in at least one IRAC band. Of these, 6.5 thousand sources have detections at all four IRAC wavelengths and 888 also have detections at MIPS 24~\micron. Across the entire area 392 sources were detected at MIPS 70~\micron.
		
		\item Three color-magnitude schemes based on 2MASS+IRAC+MIPS, 2MASS+MIPS, and IRAC-only photometry were used to reject background galaxies from the catalog. A total of 133 candidate YSOs were identified in this manner. Two-thirds of these were classified by the infrared spectral index as Class II YSOs (generally classical T Tauri YSOs) and one-fifth were classified as Class I YSOs (embedded protostars). Cross-identifications were made for 93 candidate YSOs in the GSC-II and 20 candidate YSOs in the IRAS Faint Source Catalog. Ten additional previously known YSOs that were not identified by \spitzer\ were found to be coincident with entries in the catalog.
		
		\item The Cepheus Flare \spitzer\ luminosity function peaks at $\log(L/L_{\odot})\sim-1.5$. Above this value, the luminosity function has a power law index of 1.6 in agreement with that found for {\it IRAS}-only YSO candidates.
		
		\item SED modeling was conducted, following \citet{2008harvey} and \citet{2007cieza}, to estimate the degree of infrared excess for Class II and Class III YSOs. The majority of the YSOs was found to have accretion-style disks. The values of $A_{\rm V}$ estimated from the 2MASS J-K$_{s}$ color were over-estimated for known variable stars. We suggest that YSO \#83 is a possible transitional disk candidate.
		
		\item Comparison of 2.5\arcmin\ resolution extinction maps to 160~\micron\ emission maps showed 14-18 dense cores (depending on consolidation) split between the six dark cloud associations. Three of these, L1155, L1241, and L1247, are confirmed as starless above the $A_{\rm V}=5$ contour. It was found that the morphology of the 160~\micron\ and higher resolution \spitzer\ $A_{\rm V}$ maps agreed particularly well for quiescent pre-stellar cores, but diverged in less quiescent regions (e.g., the heart of the NGC 7023 reflection nebula). 
		
		\item Five YSO associations with peak stellar mass volume densities greater than a theoretical cluster disruption density were found. L1228N, L1228S, L1251B, and L1251A are small groups with 5-15 members each. The larger NGC 7023 cluster contains 32 YSOs and 21 non-YSO members. The star formation rate for dense cores with and without formal associations of YSOs was found to be $\sim8\%$ and $\sim1\%$ respectively. No difference in the mass of pre-stellar and protostellar cores was found.

	\end{itemize}

\acknowledgments

	We are grateful to Maria Kun and her co-authors for a preprint of their chapter, ``Star Forming Regions in Cepheus'', from the Handbook of Star Forming Regions, Vol 1: The Northern Sky, (2008, ed. B. Reipurth, ASP Monograph Series).

	JMK thanks UK STFC for PDRA funding through the Cardiff Astronomy Rolling Grant. 
	
	This work has made use of data and resources from the {\it Spitzer Space Telescope}, the Two Micron All Sky Survey, the Digitized Sky Survey II, the Guide Star Catalog II, the James Clerk Maxwell Telescope, the Canadian Astronomy Data Centre SCUBA Archive, and the SIMBAD database. The {\it Spitzer Space Telescope} is operated by the Jet Propulsion Laboratory, California Institute of Technology under a contract with NASA. The Two Micron All Sky Survey is a joint project of the University of Massachusetts and the Infrared Processing and Analysis Center/California Institute of Technology, funded by the National Aeronautics and Space Administration and the National Science Foundation. The Digitized Sky Survey-II is based on photographic data obtained using The UK Schmidt Telescope. The UK Schmidt Telescope was operated by the Royal Observatory Edinburgh, with funding from the UK Science and Engineering Research Council, until 1988 June, and thereafter by the Anglo-Australian Observatory. Original plate material is copyright (c) of the Royal Observatory Edinburgh and the Anglo-Australian Observatory. The plates were processed into the present compressed digital form with their permission. The Digitized Sky Survey was produced at the Space Telescope Science Institute under US Government grant NAG W-2166. The Guide Star Catalog-II is a joint project of the Space Telescope Science Institute and the Osservatorio Astronomico di Torino. Space Telescope Science Institute is operated by the Association of Universities for Research in Astronomy, for the National Aeronautics and Space Administration under contract NAS5-26555. The participation of the Osservatorio Astronomico di Torino is supported by the Italian Council for Research in Astronomy. Additional support is provided by  European Southern Observatory, Space Telescope European Coordinating Facility, the International GEMINI project and the European Space Agency Astrophysics Division. The James Clerk Maxwell Telescope is operated by The Joint Astronomy Centre on behalf of the Science and Technology Facilities Council of the United Kingdom, the Netherlands Organisation for Scientific Research, and the National Research Council of Canada. SCUBA was built at the Royal Observatory, Edinburgh. The Canadian Astronomy Data Centre is operated by the National Research Council of Canada with the support of the Canadian Space Agency. The SIMBAD database is operated at CDS, Strasbourg, France. Marvelous fellows, all of them.

{\it Facilities:} \facility{Spitzer}, \facility{CTIO:2MASS ()}, \facility{JCMT}

\appendix

\section{Discussion of Individual Regions}
	\label{regions}
	
	The SCUBA Legacy Catalog \citep{2008difrancesco} was used to identify regions of the Cepheus Flare that have been observed with SCUBA. The majority of the cores in the Cepheus Flare have been observed with the SCUBA scan-map method at one time or another and with varying levels of signal-to-noise. The exceptions are L1251W and L1228S which were not observed, and PV Cep which has only been observed with the jiggle-map method. The SCUBA Legacy maps have been filtered to remove large scale angular variations which may or may not be real for any given map \citep{2008difrancesco}. The Cepheus Flare scan-maps were rereduced by us to compare possible extended structure against the MIPS 160~\micron\ emission. The entire set of raw data for the Cepheus Flare scan maps and the jiggle maps for PV Cep and L1172 (which is higher signal-to-noise than the L1172 scan-map) were downloaded from the archive and reduced as described in Section \ref{photometry}. A median baseline was used during the reduction of the scan-maps. The data have been smoothed to a FWHM resolution of 18\arcsec\ to improve the image quality.
	
	Figure \ref{fig:scuba} shows IRAC 3.6~\micron\ images towards twelve Cepheus Flare cores. The IRAC images are shown with a log stretch to emphasize faint nebulosity. The white contours show the SCUBA 850~\micron\ data. The contours are at intervals of 2~$\sigma$ (3~$\sigma$ for L1228N and 10~$\sigma$ for PV Cep) and start at 3~$\sigma$. The local 1~$\sigma$ rms was measured as the point-to-point variation towards off-source regions and was typically 10-30 mJy per 18\arcsec\ beam. The dashed contours show $A_{\rm V}$ from Figure \ref{fig:distribution} at levels of 5, 9, 13, 17, and 21 mag. The positions of YSO candidates are shown by star markers and each of them is labeled with their index number. Additional objects of interest are shown by white crosses and are discussed in the following sections. Selected outflows are shown by dashed lines. Table \ref{tab:outflows} lists the driving sources, Herbig-Haro numbers (where relevant), and references for the outflows shown in Figure \ref{fig:scuba}. The relation of the cores to each other is shown in Figure \ref{fig:coverage}.

	\subsection{The L1148+L1152+L1155+L1157 Dark Cloud Association}

		The $A_{\rm V}$ and 160~\micron\ maps both show a ring- or loop-like structure (see Figures \ref{fig:distribution} and \ref{fig:mips160}) that includes the dense cores L1148, L1155, L1152 and L1157. This ring is remarkable because three out of these four cores contain relatively isolated Class 0 protostars suggesting that star formation is still ongoing within this structure. Two bright point sources are visible at 160~\micron, the Class 0 protostar L1157 and the Herbig AeBe star PV Cep.
		
		\subsubsection{PV Cep}
		
		PV Cep is a bright Herbig Ae/Be star \citep{1994li}. Figure \ref{fig:scuba}a shows the region around PV Cep. The SCUBA 850~\micron\ emission is point-like, but there is a slight vein of low-level emission that points to the south. This could be CO contamination from the outflow in the 850~\micron\ waveband \citep[e.g, see ][]{2001chini}. The MIPS 160~\micron\ emission is also point-like, but is saturated in the BCD images. The IRAC image shown in Figure \ref{fig:scuba}a is a BCD image from the SSC archive and still has some of the artifacts that were removed from other images by the c2d/SGBS processing. Inspection of the SSC images shows that PV Cep is also saturated at all four IRAC bands. Nevertheless, the breadth of additional photometry does allow for a fairly complete reconstruction of its SED as shown in Figure \ref{fig:seds1}. The SED has had to be scaled downwards to fit on the same axes as the other SEDs as it is so bright.
		
		The white crosses in Figure \ref{fig:scuba}a show the positions of the HH 315 outflow. The closest two HH objects to PV Cep have been connected to show the orientation of the outflow.  Lines connecting paired HH objects on either side of PV Cep do not cross in the same location and were taken by \citet{2004goodman} as evidence that the source was moving to the west. Taken with other evidence, they concluded that PV Cep had an unusually high local motion and has escaped from the NGC 7023 cluster.
		
		\subsubsection{L1148}
		
		\citet{2005kauffmann} studied L1148 and reported the discovery of a very low luminosity object they called L1148-IRS. To the south of L1148 in Figure \ref{fig:distribution} the $A_{\rm V}=5$ contour breaks in two between a northern component containing L1148-IRS and a southern component that runs off the bottom of the high resolution $A_{\rm V}$ box. This southern core is L1147. 
		
		Figure \ref{fig:scuba}b shows the region around L1148 at IRAC 3.6~\micron. Care must be taken in interpreting the SCUBA 850~\micron\ data as much of the brighter emission is towards the edge of the map and could easily be Fourier artifacts introduced by the SCUBA scan-map restoration. Comparison with the MAMBO filaments \citep{2005kauffmann,2008kauffmann}, the $A_{\rm V}$ contours, the MIPS 160~\micron\ emission and {\it Akari} far-IR emission \citep{2008nutter}, however, show common features that supports the probability that SCUBA is tracing extended structure. Taken together, they appear to show a loop or square of material/emission with L1148-IRS on the northeastern side. The two MAMBO dust filaments that \citet{2005kauffmann} observed towards L1148 are also part of this loop. The filament that contains L1148-IRS (YSO \#5 in Figure \ref{fig:scuba}b) follows the line of the extinction, but the brighter MAMBO filament is 3\arcmin\ to the east and is coincident with a 160~\micron\ feature. The SCUBA contours also show these two filaments aligned NE-SW with L1148 IRS at the very top of one of them.
		
		The extinction peaks around L1148-IRS and to the north-west. The 160~\micron\ emission also peaks to the north-west, but is relatively weak around L1148-IRS and is stronger towards the south. \citet{2008nutter} observed the L1148+L1155 filament at 90, 140, and 160~\micron\ with the {\it Akari} far-infrared satellite. They showed that the extended far-infrared emission towards L1148 and L1155 followed particular edges of regions of high visual extinction and attributed the difference in the two distributions in L1155 to the effect of an exterior source heating the dust along one side of the core. They also detect the source L1148-IRS strongly at 90~\micron. There is good agreement between the 160~\micron\ emission as observed by MIPS and {\it Akari}.  
		
		\subsubsection{L1152}
				
		\citet{2008chapman} observed L1152 with \spitzer\ as part of their study of the extinction law in four dense cores. They found three YSO candidates in this field - the two Class II sources YSO \#2 and \#3 and the Class 0 source YSO \#1. The IRAS sources F20353+6742 and F20358+6746 are associated with YSOs \#1 and \#3. Figure \ref{fig:scuba}c shows the region around L1152 at IRAC 3.6~\micron. An alternative reduction of this SCUBA data was originally presented in \citet{2006young}. 
		
		The Herbig Haro (HH) 376 jet cuts through this region - its orientation is shown by the dashed line and the position of HH 376A is indicated \citep{1997reipurth}. It is immediately noticeable that the HH 376 outflow is aligned tightly with the objects of interest in this field. The northern end of the outflow is coincident with the cometary nebula GM 3-12 (RNO 124 = YSO \#3), this is a conical nebula with two helical arms \citep{2004movsessian}. The HH-object 376A has a bow-shock structure suggesting that it is moving to the S-W away from YSO \#3 \citep{1997reipurth,2004movsessian}. Given the proximity of the isolated group of 850~\micron\ contours to HH 376A, it is possible that we are seeing the bow-shaped HH376A \citep{2004movsessian} plowing into the clump of material that is ahead of it.
		
		YSO \#1 is associated with the peak of the SCUBA emission in the map. It has a $\sim$1\arcmin\ long bipolar molecular (CO) outflow that is offset from the direction of the HH 376 outflow \citep[shown as the shorter dotted-line in Figure \ref{fig:scuba}c]{1996bontemps}. The HH 376 outflow could be powered by YSO \#1 \cite[e.g., see ][]{1997reipurth}, but given the offset between it and YSO this would appear unlikely. The projected line of the HH outflow, however, is close enough to the dense material seen at 850~\micron\ that the outflow could possibly have influenced the evolution of YSO \#1 or even triggered its formation. There is a region of 3.6~\micron\ nebulosity that stretches away from YSO \#1 parallel to the direction of the HH 376 outflow. YSO \#1 does show faint nebulosity in JHK which is clipped to the NE \citep{1990heyer,2007connelley}. \citet{2008chapman} identify the nebulosity as an outflow that is visible from 2MASS $K_s$ to IRAC 8.0~\micron\ that is altering the dust grain properties within surrounding material. This nebulosity is coincident with the 850~\micron\ contours.  
 		
		\citet{1991clark} suggested that YSO \#1 is a very young candidate protostar. The spectral index and $T_{\rm bol}$ estimates of the classification of YSO \#3 and \#1 agree that the former is more evolved than the latter. YSO \#1 has a value of $T_{\rm bol}=33$~K suggesting that it is a Class 0 source. It also has $L_{\rm submm}/L_{\rm bol}=5.0\%$ which is the canonical value dividing Class 0 and Class I sources. YSO \#3 has neither a continuum detection at millimeter wavelenghts \citep[e.g., a disk; ][]{1993terebey} nor a maser detection  \citep{2003furuya}. 
		
		\subsubsection{L1155}
						
		The 160~\micron\ emission in L1155 is fragmented within the $A_{\rm V} = 1$ contour. There is a clear east-west break in the 160~\micron\ emission between the smaller eastern core and the western core. The larger western core has two peaks which correspond to the position of L1155C (north-peak) and L1155H (southern peak) \citep{2005kirk}. The smaller eastern core has been listed as either L1155E or L1158 \citep{2008kun}. The L1155D core is the faint peak of emission slightly to the SE of L1158 \citep{2005kirk}.
		
		There is general agreement between the distribution of 160~\micron\ and 850~\micron\ emission towards L1155, as seen with other pre-stellar cores \citep{2007kirk}, although the SCUBA emission does peak slightly further to the north. The strengths of the L1155C and L1155H 160~\micron\ peaks are approximately equal. Both the 160~\micron\ and 850~\micron\ emission run along the side of the low-resolution \cite{2005dobashi} $A_{\rm V}$ map. \citet{2008nutter} used a comparison of {\it Akari}, SCUBA, and ISO data to show that there was a monotonic spatial shift of the emission peak with wavelength that was caused by an edge-to-center negative temperature gradient of 2~K created by the external illumination of L1155 by the nearby A star BD+671263. This star was one of those used by \citet{1992straizys} to derive his distance estimate to these cores (see Section \ref{distance}).

		\subsubsection{L1157}

		The L1157-MM embedded protostar is usually given the same name as the dark cloud. It was not observed by our campaign, but was observed with \spitzer\ by \citet{2007looney} who detected a flattened pseudo-disk in absorption at 8~\micron\ that matched a similar structure see in N$_2$H$^+$ and DCO$^+$. The orientation of the disk was perpendicular to the orientation of the YSO's outflow. It was added to our catalog as YSO \#134. It has a bolometric temperature of 37~K, a $L_{\rm submm}/L_{\rm bol}$ ratio of 2.8\%, and is listed as a Class 0 source in the literature \citep{2008kun}.
		
		Figure \ref{fig:scuba}e shows the region around L1157 at 3.6~\micron\ and SCUBA 850~\micron. Both tracers show the same basic pattern, a cross-like structure formed by the N-S outflow and EW nebulosity. The outflow from L1157 has been extensively studied and is considered the ``prototype for chemically active outflows'' \citep{2008kun}. \citet{2001chini} compared 850~\micron\ emission to CO (1-0) emission towards L1157. They saw a similar alignment of the outflow to the 850~\micron\ emission and concluded that it must contain significant CO line contamination.
		
		The 160~\micron\ emission across L1157 is dominated by a bright point-source centered upon the YSO. Fainter extended emission appears to the SE in the approximate position of the contours in the bottom-left corner of Figure \ref{fig:scuba}. There is no such emission to the north of the YSO seen at either 160~\micron\ or 850~\micron\ and the YSO sits on the northern edge of the core as mapped by the Dobashi $A_{\rm V}$ extinction. This suggests that the northern portion of the outflow is emerging into a lower density environment than the southern outflow.  
	
	\subsection{The L1172+L1174 (NGC 7023) Dark Cloud Association}
	
		The L1172+L1174 (NGC 7023) complex is a clustered region of star formation. The extinction map in Figure \ref{fig:distribution} shows the same T-shaped cloud as Figure \ref{fig:ngc7023fc} with regions of high extinction towards the extremities of the T and a void near the center. The NGC 7023 reflection nebula, otherwise known as the Iris Nebula, sits at the center of the T and is powered by the seventh magnitude Herbig AeBe star HD 200775. The 160~\micron\ emission shows the same basic T-shape, but is dominated by bright emission from the center of the nebula that completely saturates the MIPS detectors. The dark nebula associated with the dense material east and west of NGC 7023 is called L1174. We use the designations L1174A, L1174B, and L1174C to refer to the material to the east, to the west, and in the center of NGC 7023 respectively, based on decreasing order of peak $A_{\rm V}$. The dark nebula that forms the T-stem is L1172.
		
		Figure \ref{fig:scuba}f shows a close up of the region around the NGC 7023 nebula and cluster. The asymmetric bi-conical cavity to either side of HD 200775 is clearly visible. The position of HD 200775 is shown by the central source YSO \#138. There is excellent agreement between the lower SCUBA contour and the $A_{\rm V}$ contours. The SCUBA and extinction contours show the dense material that forms L1174, crossbar of the T, with the reflection nebula immediately to the south.  
		
		The eastern lobe of the nebula appears in bright 3.6~\micron\ emission while only the partial rim on the western lobe appears. The eastern lobe is constrained by a greater level of extinction than the western lobe. The 3.6~\micron\ emission traces the southern and western part of the western lobe, but the northern part is much fainter. This same pattern can be seen in the 160~\micron\ map where the south-west part of the lobe appears in strong emission. Where the broken-western lobe encounters the edge of L1174B at 21\lath 00\latm 20\lats\ +68\degr 13\arcmin 00\arcsec, there is a small group of YSO candidates.  This is the NGC 7023 Tight B group as identified in the clustering analysis. An enlargement of these YSOs is shown in the smaller image in Figure \ref{fig:scuba}g where they appear to form a small ring-like nebula visible at 3.6~\micron. 
		
		L1174C, the central extinction peak just to the north of HD 200775, appears as a short filament with three separate 850~\micron\ peaks. The middle peak is coincident with the embedded YSO \#34. It has insufficient data points to calculate $L_{\rm submm}$, but it has a bolometric temperature of just 14~K making it a probable Class 0 source. The southern edge of the filament follows the line of a 3.6~\micron\ filament. This filament has been mapped with multiple wavelengths, including earlier \spitzer\ observations, and traces the dissociation front at the end of the dense molecular gas \citep{2003an, 2004werner2}. Conditions along the NGC 7023 dissociation front have made it an ideal location to test dust-associated photo-luminescence models \citep{2006witt} and emission properties of PAH features \citep{2006flagey}. \citet{2004werner2} reported the existence of a small ring of IRAC 4.5~\micron\ emission between HD 200775 and the bright ridge just north of it, also seen at 3.3~\micron\ by \citet{2003an}. The feature is also present in our maps, but is not highlighted by the stretch we have used to display the images. The top of the bright emission from the nebula forms a sharp dark line that is coincident with a strong linear HI filament \citep{1998fuente}. The strongest part of the filament is also coincident with the 850~\micron\ emission from L1174C.

		Figure \ref{fig:scuba}f and the extinction map show that the eastern lobe is truncated at 21\lath 02\latm 10\lats\ by the edge of the extinction from L1174A. It is a large block of extinction that runs eastwards for about 10\arcmin. It is noticeable that star formation is not occurring within this extinction and that all YSO candidate markers are positioned towards its edge or at lower $A_{\rm V}$.  
		
		The molecular gas in NGC 7023 has been extensively studied, but the associated cluster has received relatively little attention \citep{2008kun}. \citet{1983sellgren} identified thirty cluster members of various spectral types. We identify 9 of Sellgren's cluster members as YSO candidates (these are labeled as such in Table \ref{tab:ysofluxes}). Non-YSO cluster members are shown on Figure \ref{fig:scuba}f as white crosses and in Figure \ref{fig:distribution} as yellow crosses. Sellgren identified seven probable pre-main sequence stars based on their variability, infrared excess, and hydrogen-line emission. Of those seven, we identify all but the variable stars SX Cep and HZ Cep as YSO candidates. 

		The majority of the YSO candidates in this region are part of the Tight A Group and the majority of these are situated northwards of the two lobes of the nebula in the region coincident with the area of the highest SCUBA emission. The smaller eastern lobe is coincident with several YSOs and stellar cluster members, but the western lobe is noticeably less populated. The majority of the stellar cluster members which do not have YSO candidate cross-identifications are situated to the south and east of HD 200775. The clustering analysis was repeated to include the non-YSO cluster members. The revised $25 M_{\odot}$~pc$^{-3}$ contour (see equation \ref{eqn:rhostar}) is shown in Figure \ref{fig:distribution} by the red contour. The inclusion of the non-YSO members extends the Tight A group to cover most of the eastern lobe.
		
		Figure \ref{fig:scuba}h shows the L1172D region which is at the northern tip of the southern extinction maximum in Figure \ref{fig:distribution}. It contains a triplet of YSOs in its densest region. The region was originally mapped with SCUBA by \citet{2002visser} who labeled the peaks associated with YSO \#49 and \#50 as L1172-SMM1 and L1172-SMM2. They believed L1172-SMM2 to be starless, but the high sensitivity of our IRAC observations has revealed the presence of a faint YSO at that location. The third submillimeter source, L1172-SMM3, is 3\arcmin\ further to the south and is confirmed starless. SMM1 and SMM2 both have very low values of $T_{\rm bol}$, 42~K and 24~K respectively, indicating that they are Class 0 sources. SMM1 has a broad CO outflow that was detected by \citet{1988myers} and its $L_{\rm submm}/L_{\rm bol}$ ratio is greater than 5\%.  
		
		The Class II YSO \#120  (not shown) is coincident with the listed position of L1171 and a small clump of $A_{\rm V}=1$ extinction. 

	\subsection{L1221}
	
		The protostellar core L1221 lies towards the southern edge of the G109+11 infrared loop (the top of which is shown in Figure \ref{fig:coverage}), closer to the Galactic plane than the other regions in our sample. Figure \ref{fig:scuba}i shows two SCUBA sources towards L1221. The northern submillimeter source contains a pair of YSO candidates. \citet{2008young} first published the \spitzer\ data for L1221 and detected three infrared sources embedded within the SCUBA emission which they labeled IRS 1, 2, and 3. These correspond to the YSO candidates we index as \#63, \#105, and \#64 respectively. YSOs \#63 and \#64 have values of $T_{\rm bol}$ of 66~K and 21~K and $L_{\rm submm}/L_{\rm bol}$ ratios of 4 and 9\% respectively. Based on these values YSO \#64 is a definite Class 0 source while YSO \#63 is a Class I source. The approximate orientation of a broad molecular outflow \citep{2005lee} that has been detected towards L1221 IRS 1 is shown by a dashed line in Figure \ref{fig:scuba}i. The properties of the SCUBA envelopes around IRS1 and 3 are very similar and the relative position of the infrared sources to the envelopes are virtually identical, but the protostars themselves are not as similar \citep{2008young}.

	\subsection{The L1228 Dark Cloud Association}
		\label{l1228}
		
		The extinction map towards L1228 shows two distinct peaks. A northern peak at a declination of +77\degr 35\arcmin\ and a lower $A_{\rm V}$ southern peak at +77\degr 08\arcmin. We call these two cores L1228N and L1228S respectively, although we note that L1228N has also been called L1228A in the literature \citep{2008kun}. L1228N and L1228S each contain small groups of YSO candidates with sufficient stellar densities to be picked up by the clustering analysis. The SIMBAD database shows 10 YSOs in the entire L1228 region of which 4 are just outside of the region mapped with \spitzer\ (all scattered to the west). L1228N has received more attention than L1228S and is the site of several Herbig Haro outflows.
		
		\subsubsection{L1228N}
		
		The extinction map towards L1228N shows a core that is elongated approximately E-W while the distribution of YSO candidates is approximately N-S. The 160~\micron\ map shows a similar morphology, but is dominated by a central bright point source. The source is coincident with YSO \#9 and the position of IRAS 20582+7724. Figure \ref{fig:scuba}j shows IRAC and SCUBA emission towards the L1228N region. It also shows two families of HH objects as white crosses and the approximate orientation of the HH 199 outflow which emanates from YSO \#9 and the HH 200 outflow which emanates from YSO \#7 \citep{1995bally}. Additionally, YSO \#9 has an east-west molecular outflow \citep{2006arce} which appears as faint nebulosity in the IRAC image. An enhanced image generated from the same IRAC data was shown by \citet{2008velusamy} as part of their demonstration of the application of the HiRes image deconvolution technique to \spitzer\ data.  
				
		\cite{2008chapman} examined the mid-infrared extinction in L1228N and concluded that the outflows were altering the extinction law by destroying larger dust grains within the outflow cavities. The four VLA sources were detected in this region at 3.6~cm by \citet{2004reipurth}. VLA 1 and VLA 4 match the positions of YSO candidates \#9 and \#7 from our survey, but no YSO matches are found for VLA 2 and VLA 3 though they may have IRAC 3.6~\micron\ detections. The positions of VLA 2 and VLA 3 are labeled in Figure \ref{fig:scuba}j.
		
		The SCUBA emission towards L1228 is dominated by a strong point source coincident with the position of YSO \#9 and a similar point source seen at 160~\micron. Surrounding this is low-level emission that is elongated in a N-S direction, perpendicular to the direction of the HH 199 outflow. The east-west 3.6~\micron\ nebulosity surrounding YSO \#9 is co-incident with a 2.2~\micron\ jet \citep{1995bally}. The same emission was seen in the $K$ band by \citet{1994hodapp} who classified the sources surrounding it as a small cluster - albeit at the limits of their definition of a cluster. YSO \#9 has a broad SED, meaning that it is spectrally classified as a flat spectrum source rather than a borderline 0/I source as would expected by its bolometric temperature (79 K). YSO \#7 is not detected by SCUBA, but has a bolometric temperature (54K) below the Class 0 cut-off.
		
		\subsubsection{L1228S}
		
		Figure \ref{fig:scuba}k shows L1228S. There are no SCUBA data for this core. The \spitzer\ observations of L1228S were first presented by \citet{2004padgett} who showed a cluster of 9 sources with Class II or III SEDs. L1228S is situated at the furthest southern extent of the Dobashi $A_{\rm V}=1$ contour. As noted previously, the L1228S extinction appears to be separated from the 160~\micron\ peak. The proximity of L1228S to the edge of the cloud and the proximity of L1228 to the Cepheus Flare Shell leads to the possibility that the cloud around these YSOs is being removed by the passage of the Shell, leaving the YSOs free of their natal cloud \citep{2008kun}.
	
	\subsection{L1241} 
		
		L1241 is a relatively massive starless core between the active regions of L1251 and L1228. L1241 is not shown in Figure \ref{fig:scuba} as it has not been observed with SCUBA and is not associated with any sources in our YSO catalog. It is associated with a large mass of CO on the west side of the Cepheus Flare (see Figure \ref{fig:coverage} and Table \ref{tab:assocprop}), but it appears to show little evidence for current star formation. Only a single YSO candidate, a Class III YSO at 21\lath 56\latm 13.3\lats +76\degr 58\arcmin 14.2\arcsec, was found towards this core, but it is well away from the $A_{\rm V}$ peak and this core should still be considered starless. No YSO candidates were founded in the SIMBAD database. The nearest SIMBAD YSO is 2$^\circ$ away ([K98c EM* 61]) and the nearest \citet{2005tachihara} x-ray selected T Tauri is 3$^\circ$ away. Why such a large mass of gas is not undergoing star formation is not immediately obvious as the regions L1251 and L1228 seem to be under going star formation triggered by the passage of the Cepheus Flare Shell. If L1241 is not undergoing star formation then it may be because it has not yet encountered the Cepheus Flare Shell and is at a marginally different distance than originally thought.
				
	\subsection{L1247+L1251} 
	
		The map of $A_{\rm V}$ towards the L1247+L1251 dark cloud association shows a chain of cores proceeding east-west with the peak $A_{\rm V}$ decreasing towards the east. The cores L1251A, B, and W all contain YSO candidates (see Figure \ref{fig:scuba}l-n), but L1247 appears starless (not shown). The YSO candidates in this region form three distinct groups - the L1251A and L1251B groups and a small quartet in L1251W. We follow \citet{2006lee} in naming these cores A and B after the \citet{1989sato} outflow sources and we adopt the name L1251W to refer to the elongated western core. 
		
		The L1247 core has no associated YSOs and is confirmed starless within the sensitivity of our survey. MIPS 160~\micron\ was not taken for the c2d cores region so there is no 160~\micron\ data available for L1251. Data were taken for L1247 however, it shows excellent correlation between the 160~\micron\ emission and the visual extinction. The 160~\micron\ emission and the extinction map peak in the same position and show a narrow extension/filament to the north of the peak. 		
		
		In L1247+L1251 a total of 39 YSO candidates was identified from the \spitzer\ photometry. Of these, 16 are not in the SIMBAD database. The majority of these, however, are from the L1251A core and will be discussed in the paper by \cite{2008lee}. There is a one-to-one correspondence between the YSO candidates \#66, 67, and 68 in L1251A and the ammonia cores T3, T2, and T1 \citep{1996toth}. The outflow HH 189 \citep{1994eiroa} is shown by the white crosses and dashed line in Figure \ref{fig:scuba}m. It emanates from the tight YSO group associated with the small knot of SCUBA emission, but it is not entirely clear which YSO is driving it. The approximate orientation of the HH 149 is shown by the dashed line in Figure \ref{fig:scuba}n. The driving source outflow is coincident with YSO \#143 \citep{1989sato, 1992balazs}.
		
		The L1251B cluster was studied by \citet{2006lee} using the same data presented here. They found that a tight cluster of YSO candidates centered around the bright source L1251B IRS 1. See \citet{2008lee} for further discussion of this region.

\bibliographystyle{apj}
\bibliography{cepheus}

\clearpage
\clearpage
\clearpage
\clearpage
\clearpage
\clearpage	
\clearpage
\clearpage
\clearpage	

\clearpage

\clearpage
\clearpage
\begin{figure*}
	\centering{
		\includegraphics[width=0.8\textwidth]{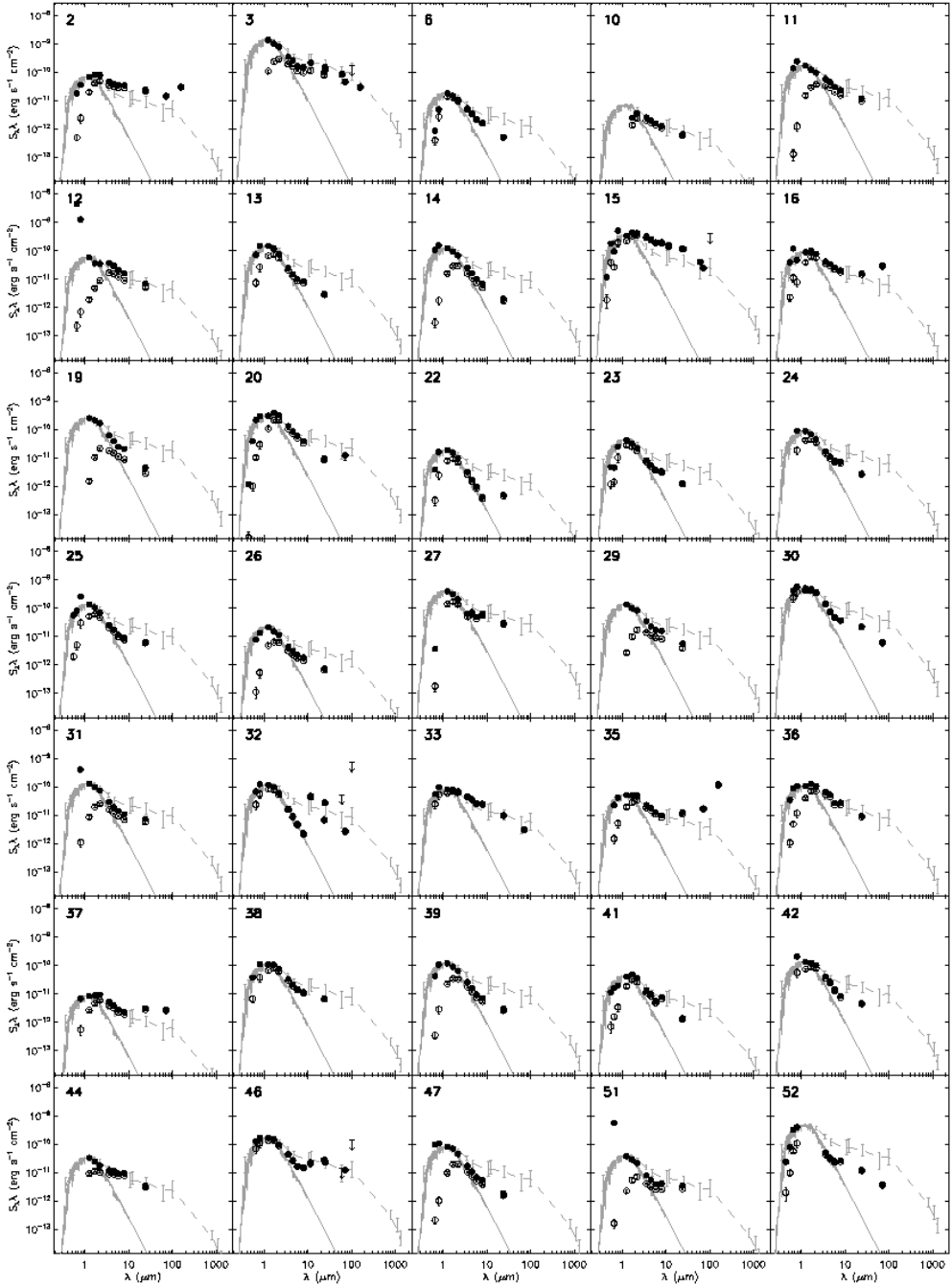}
	}
	\caption{\label{fig:seds2} Class II candidate SEDs. Details as described for Figure \ref{fig:seds1}. Additionally, the filled circles show the dereddened data. The gray lines show two comparison SEDs that have been normalized near the peak of the dereddened SED (usually the 2MASS $J$ band). The solid gray line is a NEXTGEN profile for a K7 star \citep{1999hauschildt} and the dashed gray line with error bars is the median SED for a T Tauri star in Taurus \citep{2005hartmann}.}
\end{figure*}
\clearpage
\begin{figure*}
	\centering{
		\includegraphics[width=0.8\textwidth]{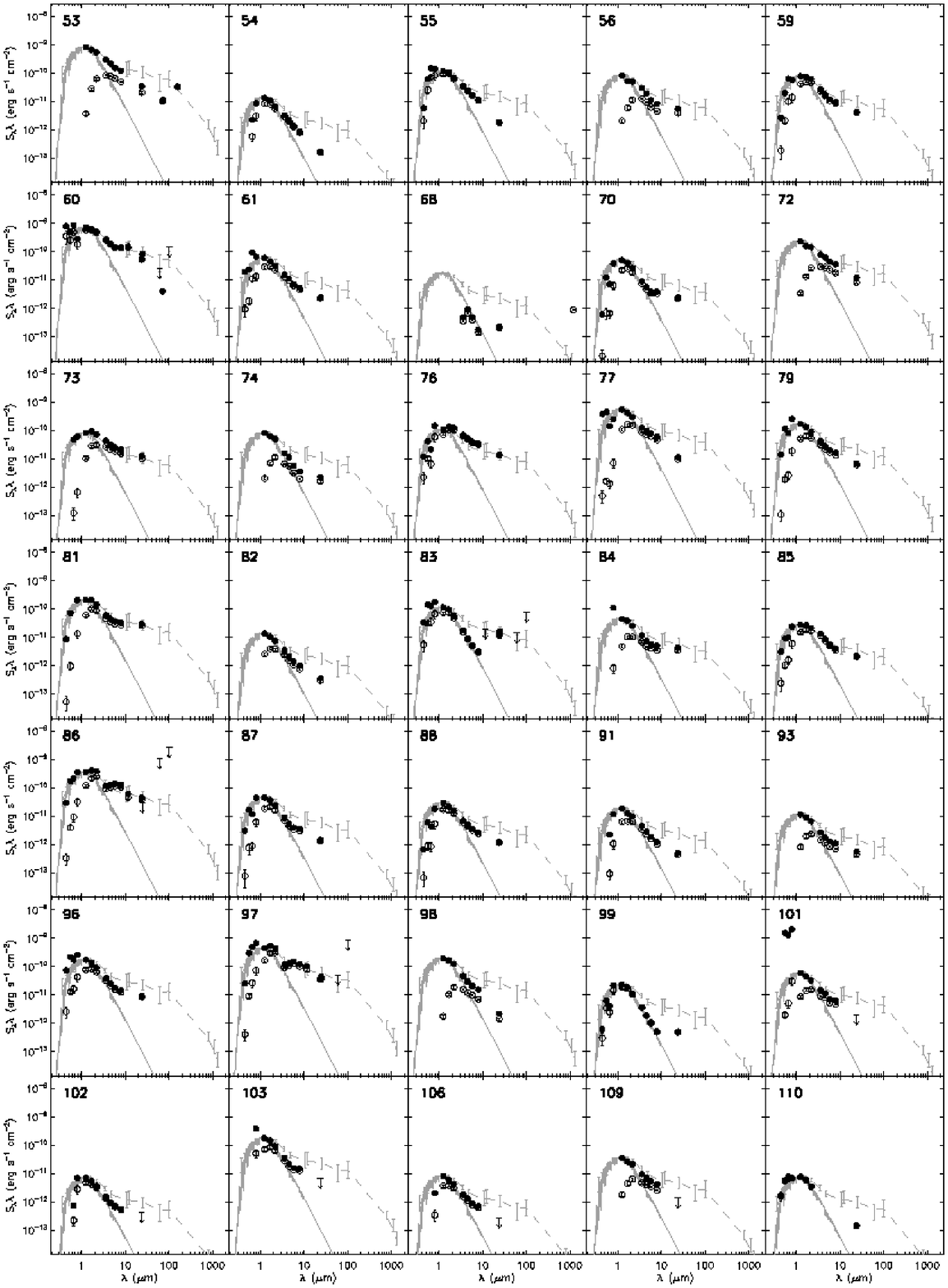}
	}
	\contcaption{(...continued)}
\end{figure*}
\clearpage
\begin{figure*}
	\centering{
		\includegraphics[width=0.8\textwidth]{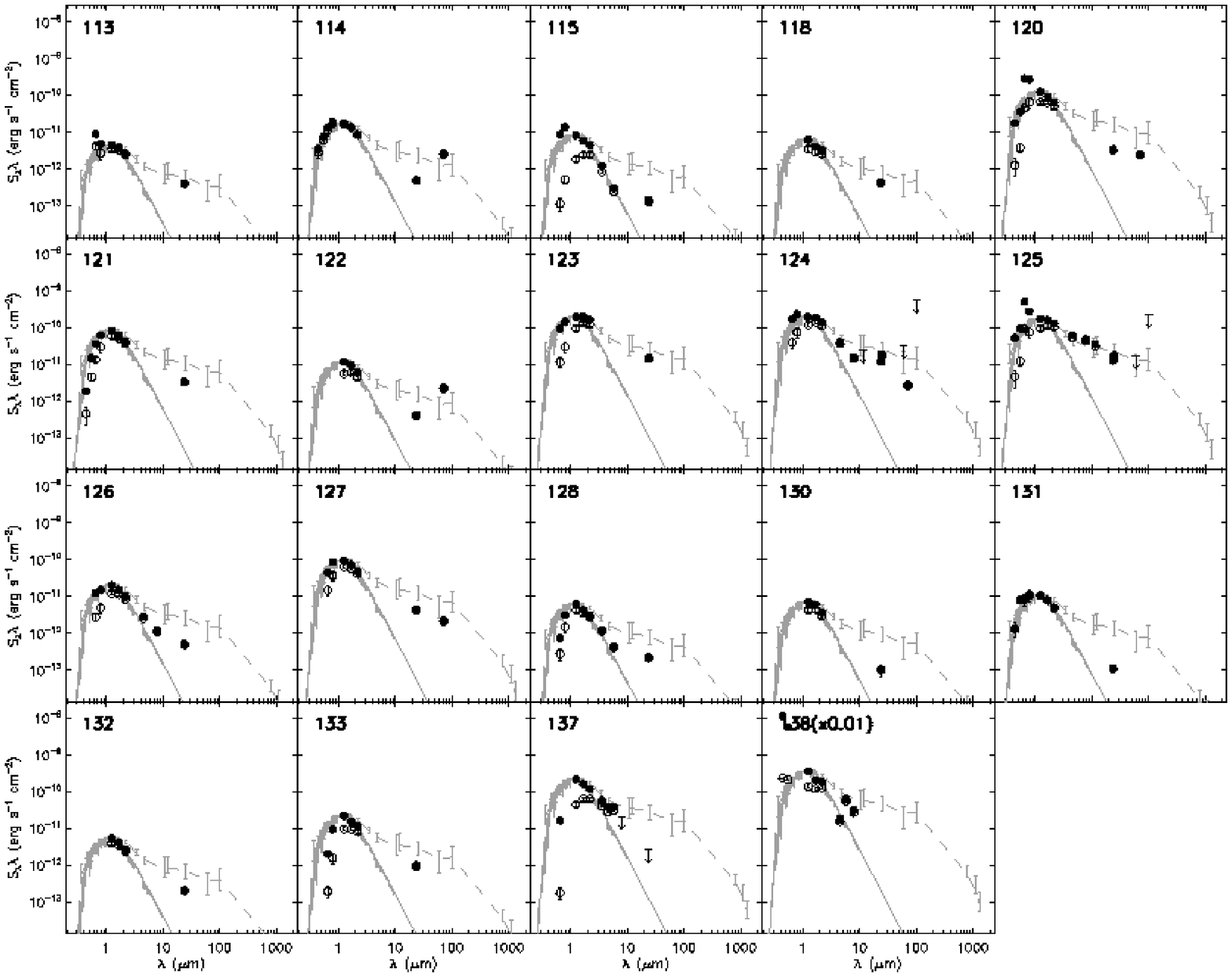}
	}
	\contcaption{(...continued)}
\end{figure*}
\clearpage
\begin{figure*}
	\centering{
		\includegraphics[width=0.8\textwidth]{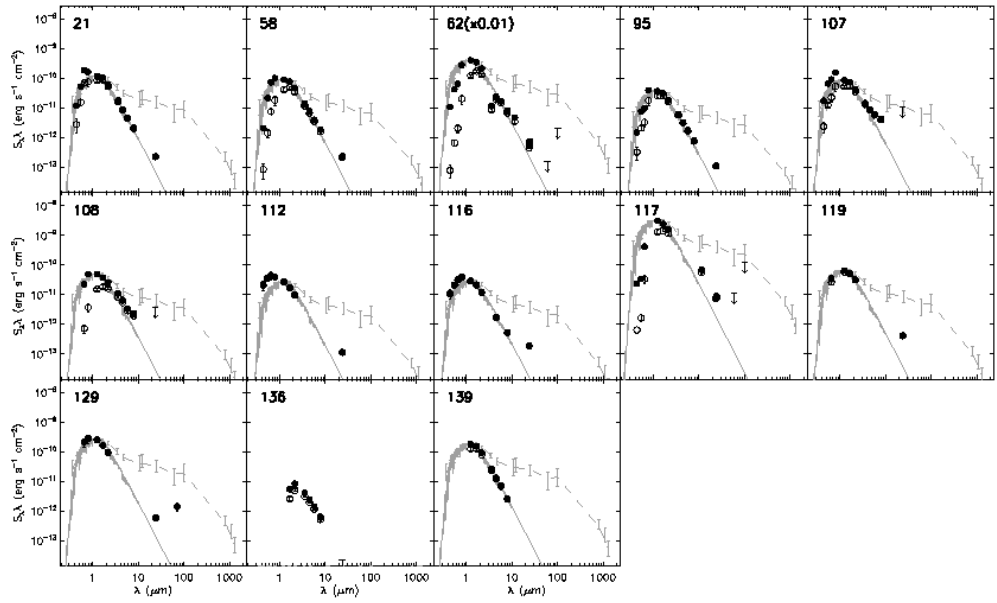}
	}
	\caption{\label{fig:seds3} Class III candidate SEDs. Details as described for Figure \ref{fig:seds2}. }
\end{figure*}		
\clearpage
\begin{figure}
	\centering{
	\includegraphics[width=0.7\columnwidth]{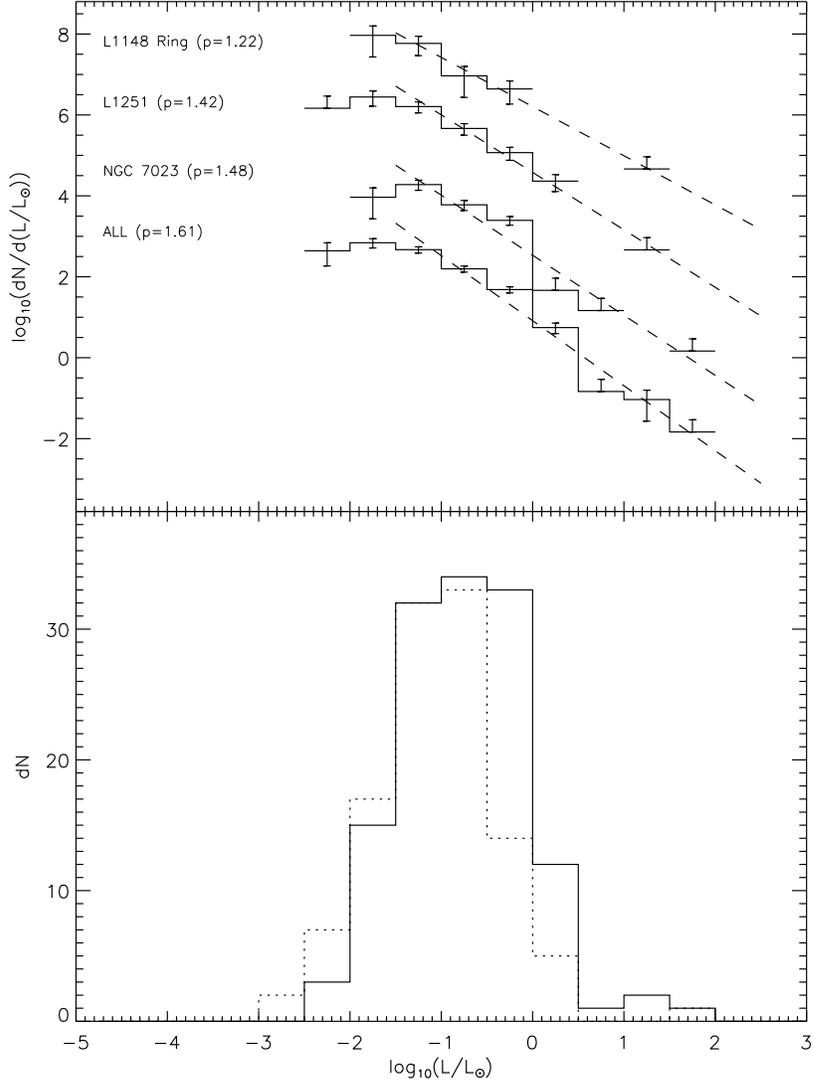}
	}
	\caption{\label{fig:lumin} Top: Bolometric luminosity functions for the 133 \spitzer\ identified YSOs. Each luminosity function is offset by 2.0 dex from the preceding one. The index of the luminosity function is shown for each region. No lower limit is shown for bins with only one source. Lower: Histogram of bolometric luminosity ($L_{\rm bol}$) for all sources (solid line) and infrared luminosity ($L_{\rm IR}$) for sources detected with IRAC (dashed line). Both plots show a break at our limiting luminosity of $L\sim0.06$~L$_{\odot}$. }
\end{figure}
\clearpage
\begin{figure}
	\centering{
		\includegraphics[width=\columnwidth]{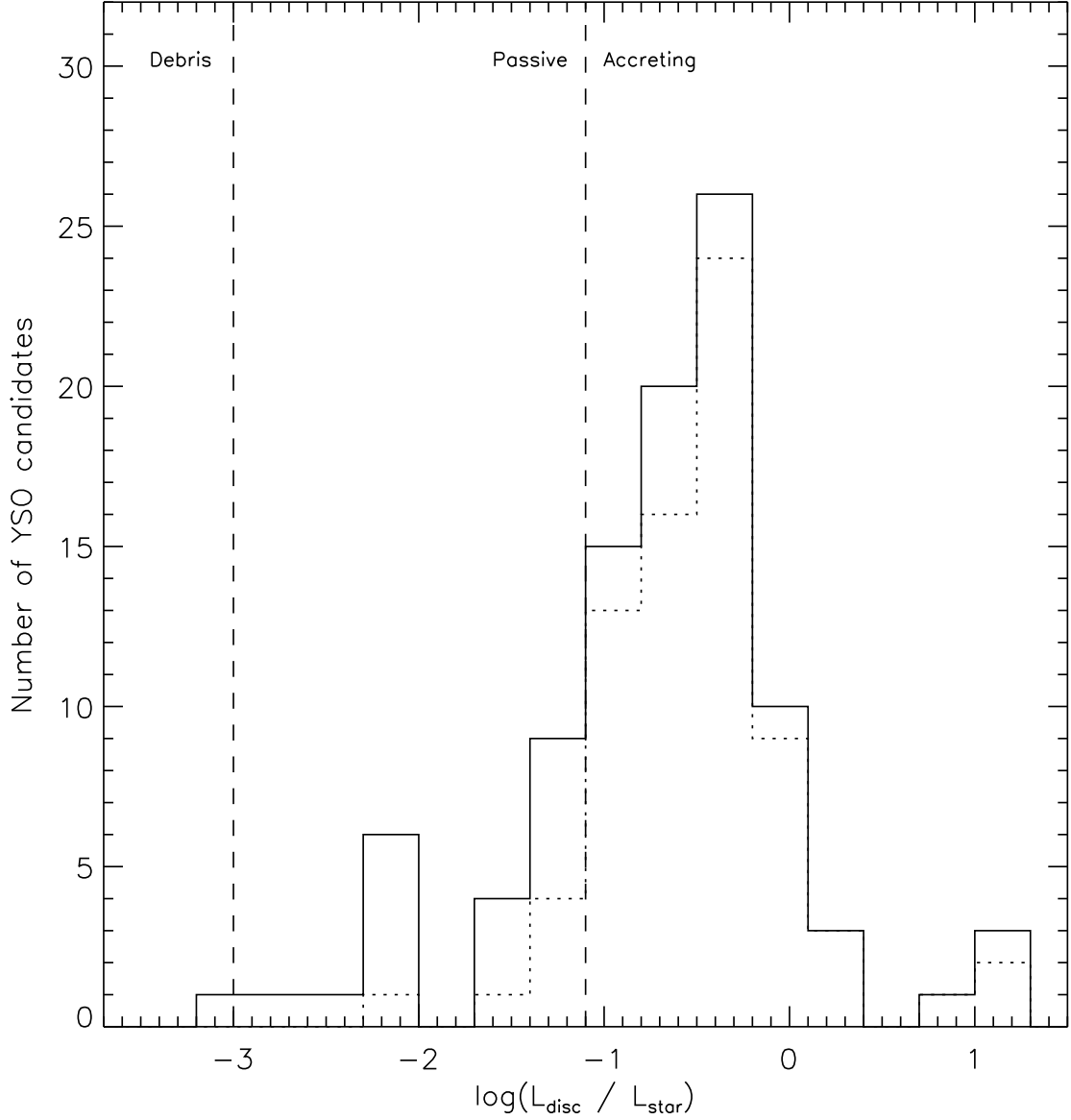}
	}
	\caption{\label{fig:disclumin} Histogram of the ratio of disk to star luminosity ($L_{\rm disk}/L_{\rm star}$) as derived from the SED modeling for Class II and Class III candidate sources. The solid line is for all sources and the dashed line excludes sources identified with the 2MASS/MIPS scheme. The vertical dashed lines show the expected regions for accreting, passive and debris disks. The histogram shows that the majority of the disks are of the accreting type.}
\end{figure}
\clearpage
\begin{figure}
	\centering{
		\includegraphics[width=\columnwidth]{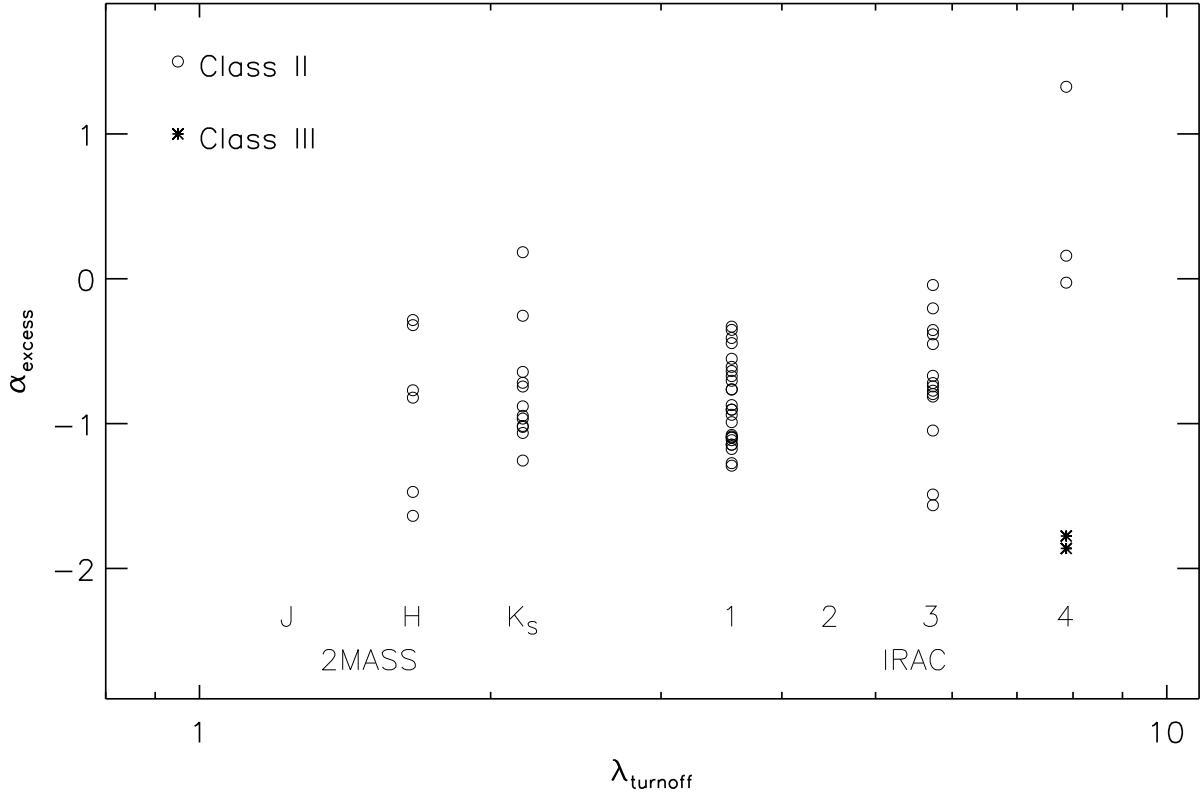}
	}
	\caption{\label{fig:discexcess} Plot of $\lambda_{\rm turnoff}$, the wavelength where the dereddened SED diverges from the normalized stellar photosphere, against $\alpha_{\rm excess}$ the spectral index of SED points longwards of $\lambda_{\rm turnoff}$. Each marker represents a single Class II or Class III candidate as shown by the key. The wavelength bands are labeled with their instrument band. The Class III candidates only show an excess at the longest values of $\lambda_{\rm turnoff}$.}
\end{figure}
\clearpage
\begin{figure*}
	\centering{
		\includegraphics[width=\textwidth]{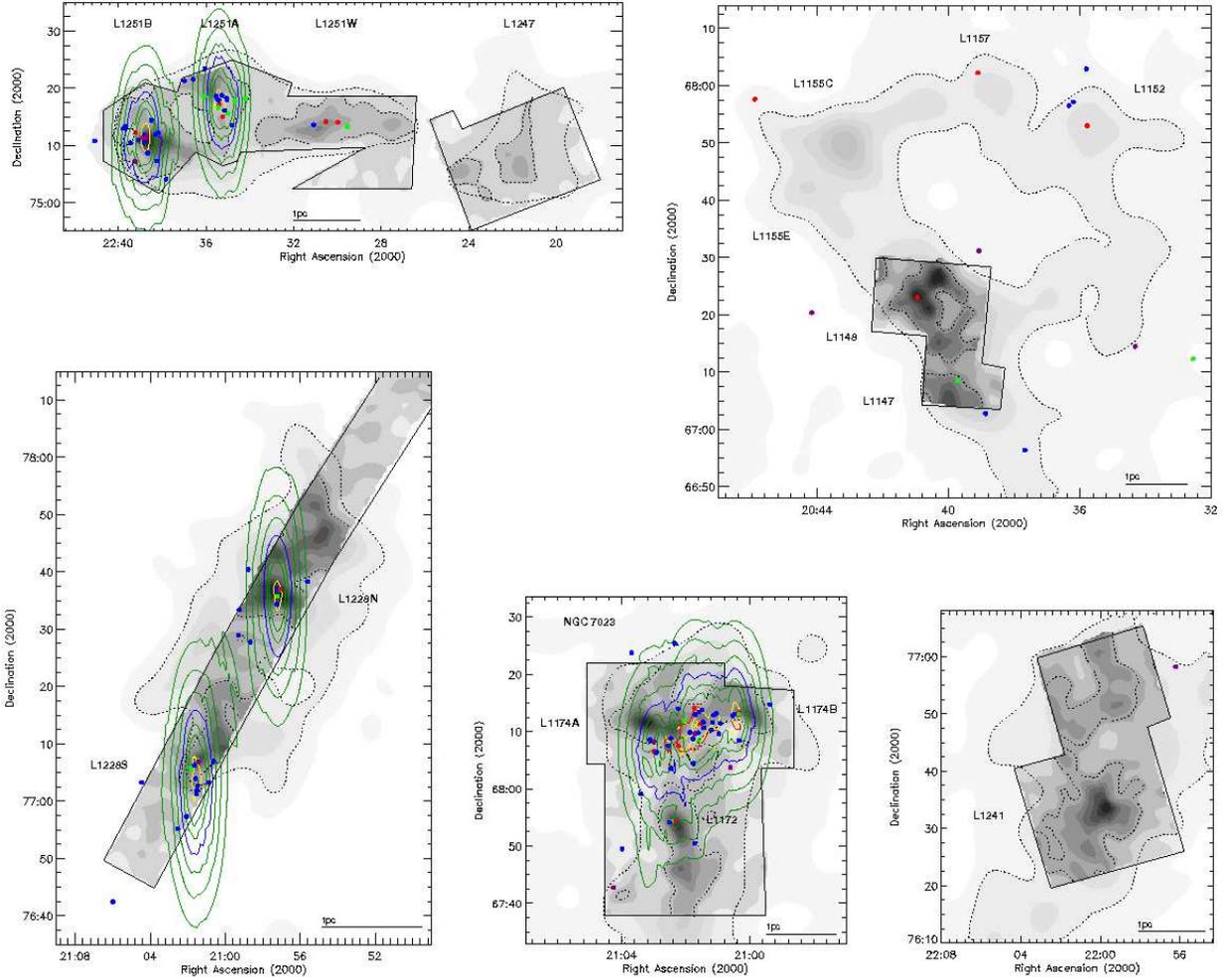} 
	}
	\caption{\label{fig:distribution} Distribution of YSO candidates, clustering results, and extended structure towards the Cepheus dark clouds associations.  The greyscale shows the distribution of visual extinction derived via two different datasets. The extinction as derived from the Digitized Sky Survey \citep{2005dobashi} is shown across the entire map. The higher-resolution \spitzer\ extinction maps are superimposed over this in the regions where there is IRAC/MIPS overlap (as shown by the boxes). Two dashed contours, Dobashi $A_V$=1 and \spitzer\ $A_V$=5, are shown. The colored markers show the location and spectral type of the YSO candidates. The colors are the same as used in Figure \ref{fig:cc} (red/green/blue/purple for Class I/Flat/II/III). The colored contours, green unless noted, show $\rho_{*}$ from the clustering analysis in units of 0.125, 0.25, 0.5, 1.0 (blue), 2.0, 4.0, 25.0 (yellow) $\times$ M$_\odot$ pc$^{-3}$. }
\end{figure*}
\clearpage
\begin{figure*}
	\centering{
		\includegraphics[width=\textwidth]{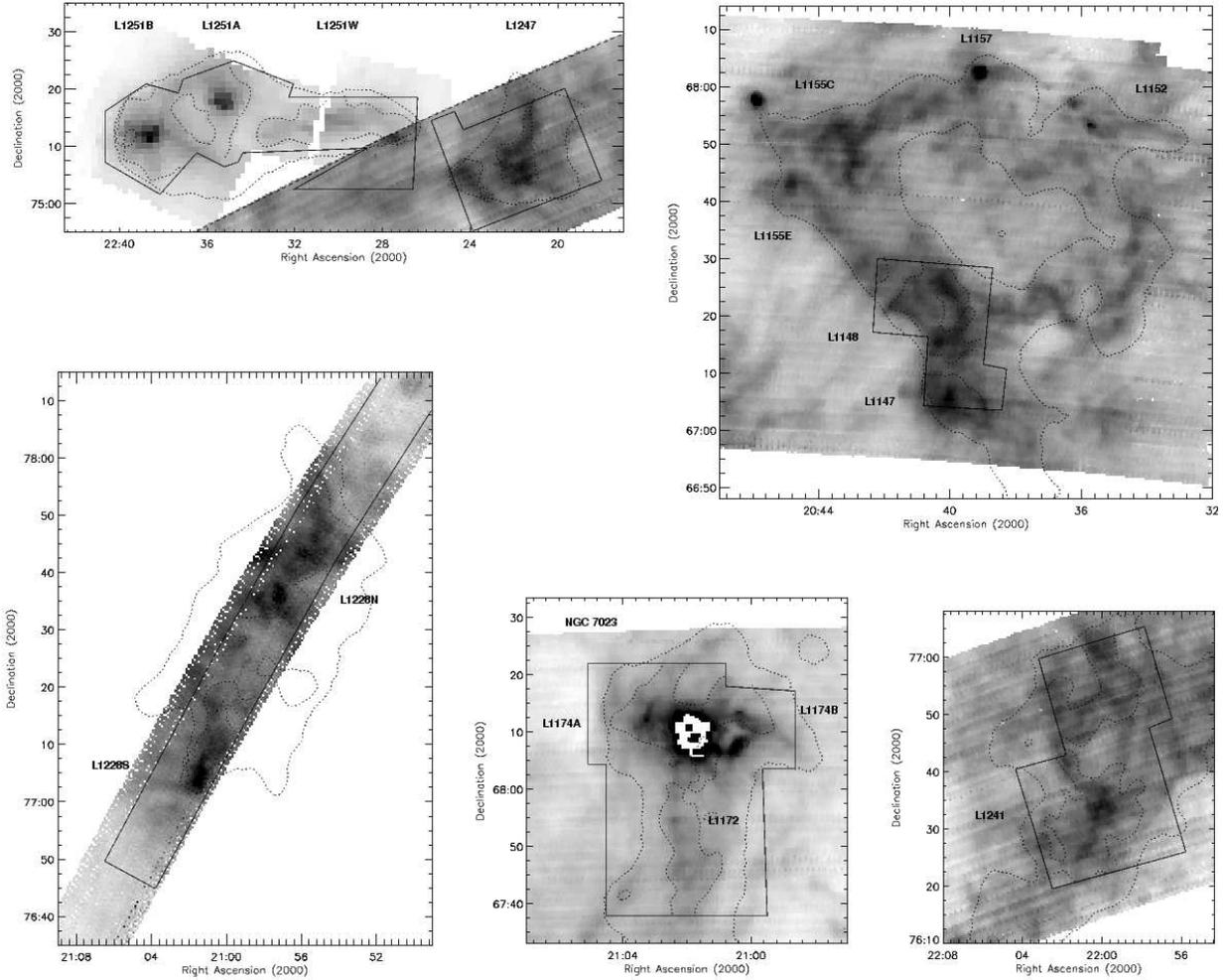} 
	}
	\caption{\label{fig:mips160} Distribution of MIPS 160~\micron\ emission towards the Cepheus dark cloud associations. The greyscale shows a log stretch between the local minimum ($\sim20$~MJy/sr) and 316~MJy/sr (NGC 7023) or 100~MJy/sr (all other regions). The Dobashi $A_V$=1 and \spitzer\ $A_V$=5 contours and the overlap boxes from Figure \ref{fig:distribution} are shown for reference. No MIPS 160~\micron\ data exists for L1251 so an ISOPHOT 200~\micron\ map scaled to the 160~\micron\ intensity in L1247 is shown instead \citep{1996lemke}. The line separating the 200~\micron\ and 160~\micron\ maps is shown by a dashed line. }
\end{figure*}
\clearpage
\begin{figure*}
	\centering{
		\includegraphics[width=\textwidth]{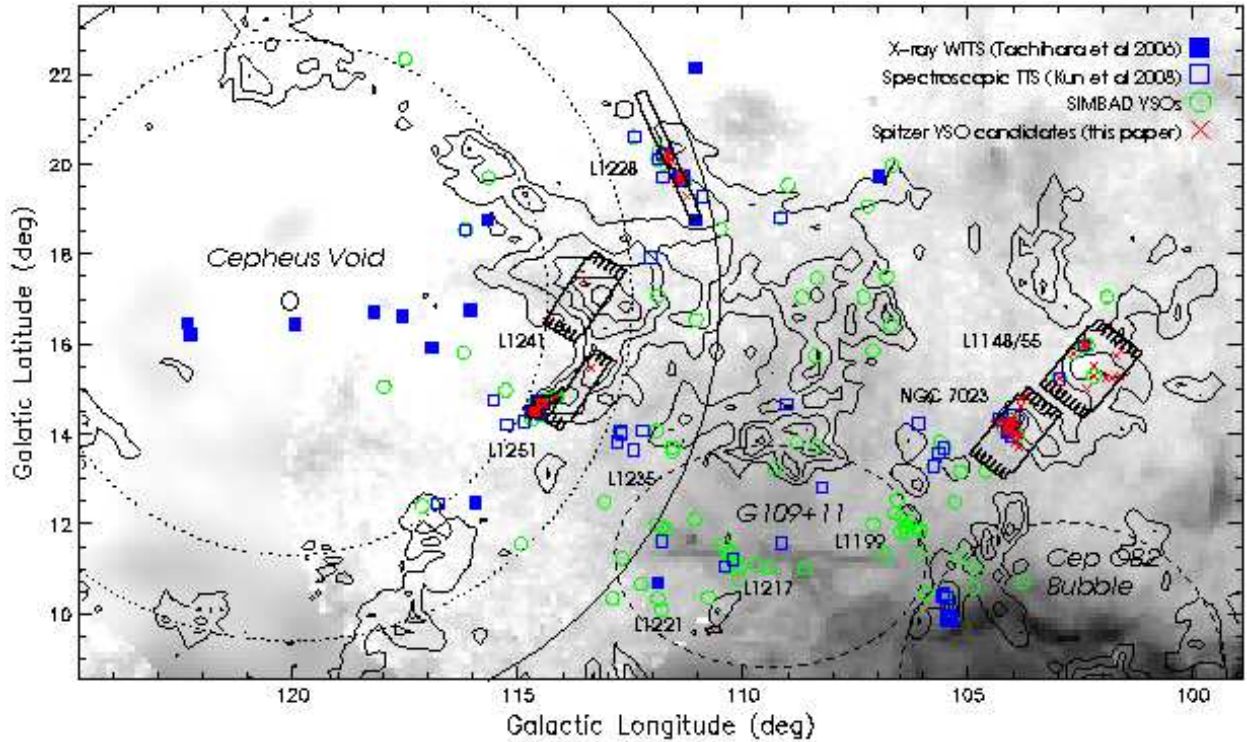} 
	}
	\caption{\label{fig:coverageYSOs} The distribution of YSOs and YSO candidates towards the Cepheus Flare. A key to the different YSO symbol is shown top-right. The areas mapped with \spitzer\ at 24~\micron\ are shown as black footprints. The labels are the names of selected Lynds dark nebulae \citep{1962lynds}. The greyscale shows log scaled H$\alpha$ emission \citep{2003finkbeiner}. The contours shows integrated CO emission at 5, 10, 15, 20 and 25 K kms$^{-1}$ \citep{2001dame}. The solid black line shows the present extent of the Cepheus Flare Shell that surrounds the Cepheus Void \citep{2006olano}. The two dotted circles show the estimated extent of the Shell 2~Myr and 4~Myr ago. The dashed-black lines denote the G109+11 and Cepheus OB2 Bubble. }
\end{figure*}

\clearpage

\begin{figure*}
	\begin{tabular}{ll}
	\vspace{2mm} a) PV Cep & b) L1148  \\ 
	
	\vspace{2mm} \includegraphics[height=3in]{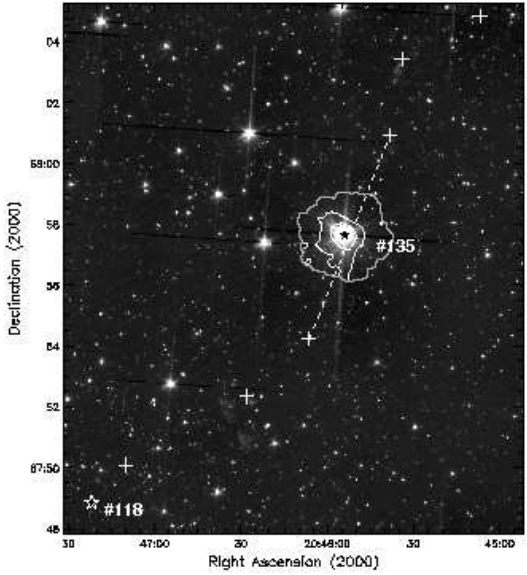} &
		\includegraphics[height=3in]{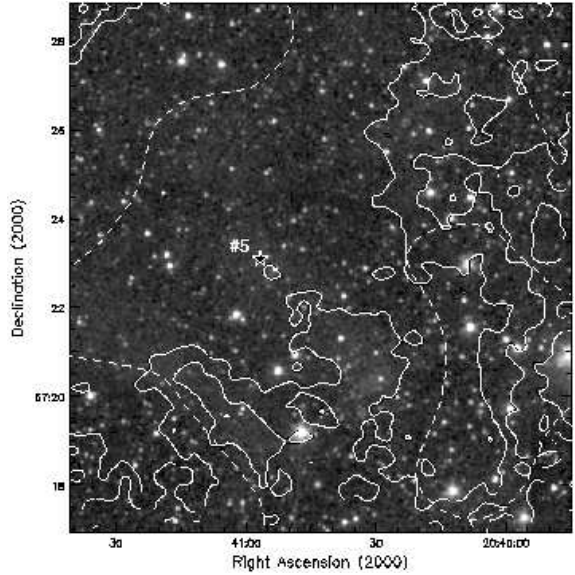} \\
	\vspace{2mm} c) L1152 & d) L1155C \\
	\vspace{2mm} \includegraphics[height=3in]{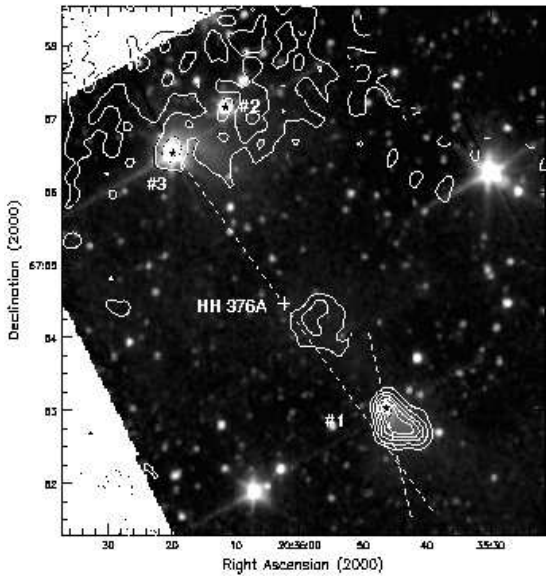} &
		\includegraphics[height=3in]{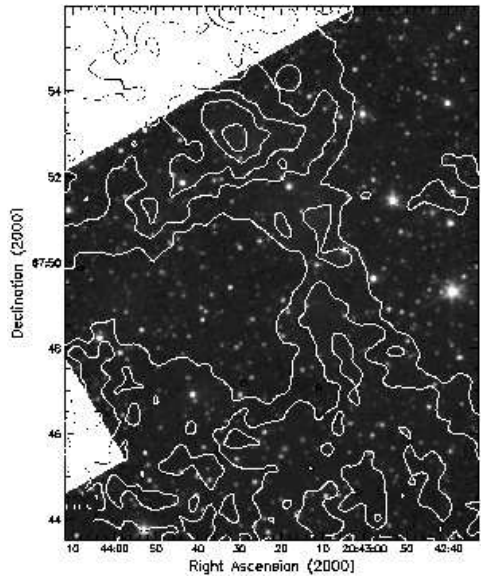} \\
	
	\end{tabular}
	\caption{\label{fig:scuba} Images of IRAC 3.6~\micron\ emission towards the centers of the Cepheus Flare cores. The solid contours show SCUBA 850~\micron\ emission in intervals of 2~$\sigma$ starting at 3~$\sigma$ (intervals of 3~$\sigma$ and 10~$\sigma$ are used for L1228N and PV Cep respectively). The edges of the mapped SCUBA areas are shown by the gray contours. The dashed contours show Spitzer $A_V$ in intervals of 4 mag starting at 5 mag. The straight dashed lines show the approximate orientation of selected outflows. The positions of YSO candidates are shown by the star markers. These are labeled with the YSO index. The positions of other objects discussed in the main text are shown by the white crosses.
	}
\end{figure*}
\clearpage
\begin{figure*}

\begin{tabular}{ll}
	\vspace{2mm} e) L1157 & g) L1174 Enlargement  \\ 
	\vspace{2mm} \includegraphics[height=3in]{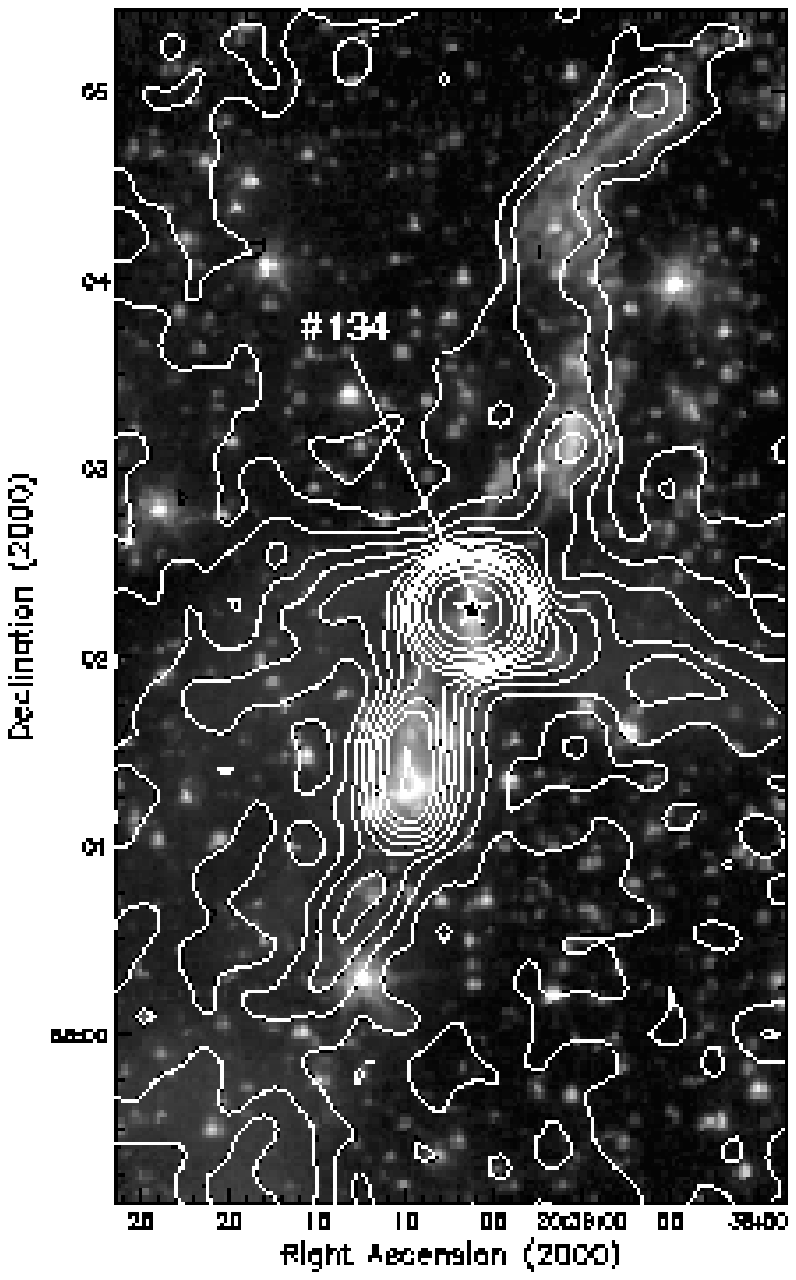} &
		\includegraphics[width=3in]{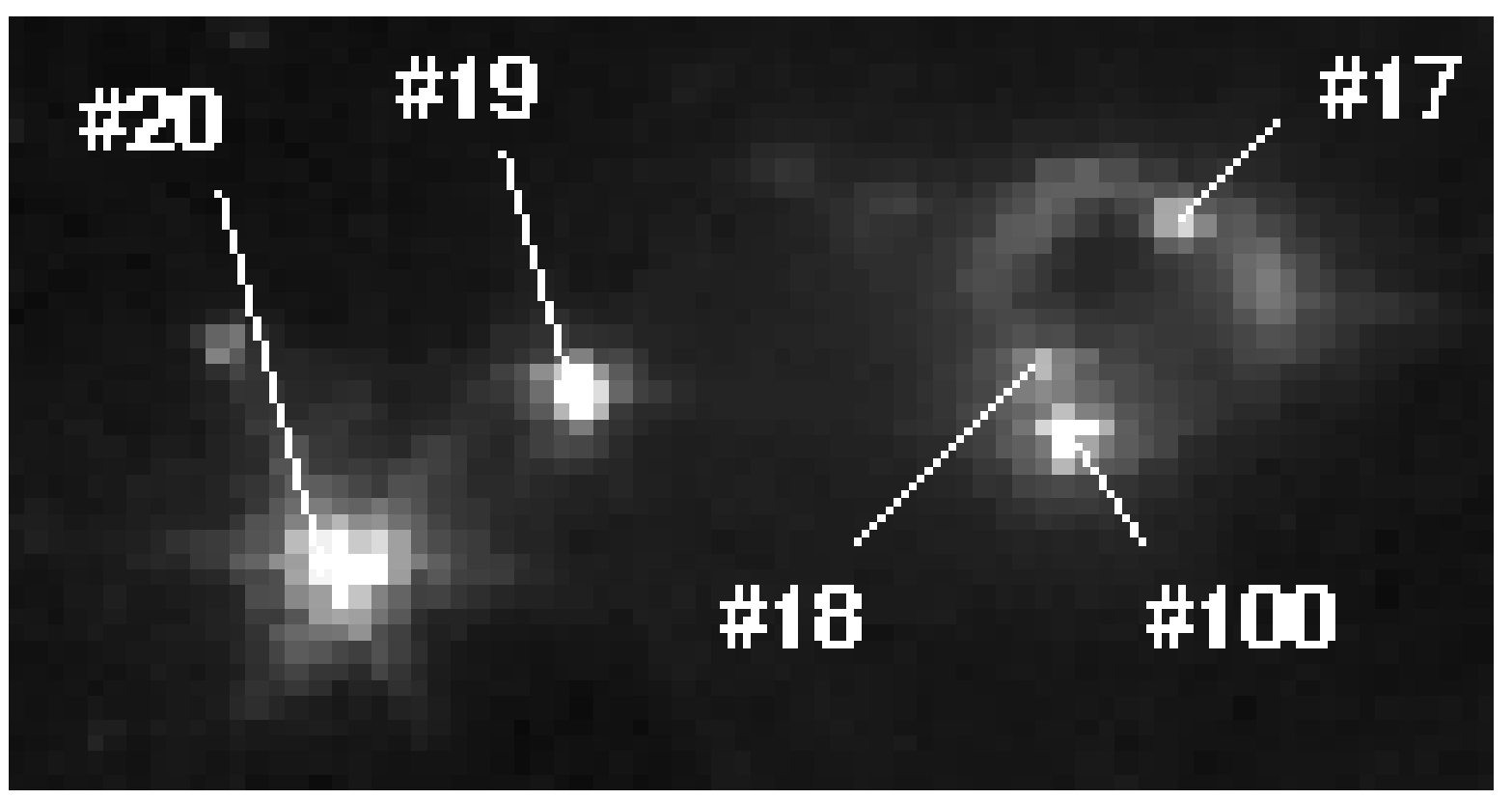} \\
	\multicolumn{2}{l}{f) L1174 (NGC 7023)} \\
	\multicolumn{2}{l}{\includegraphics[height=4in]{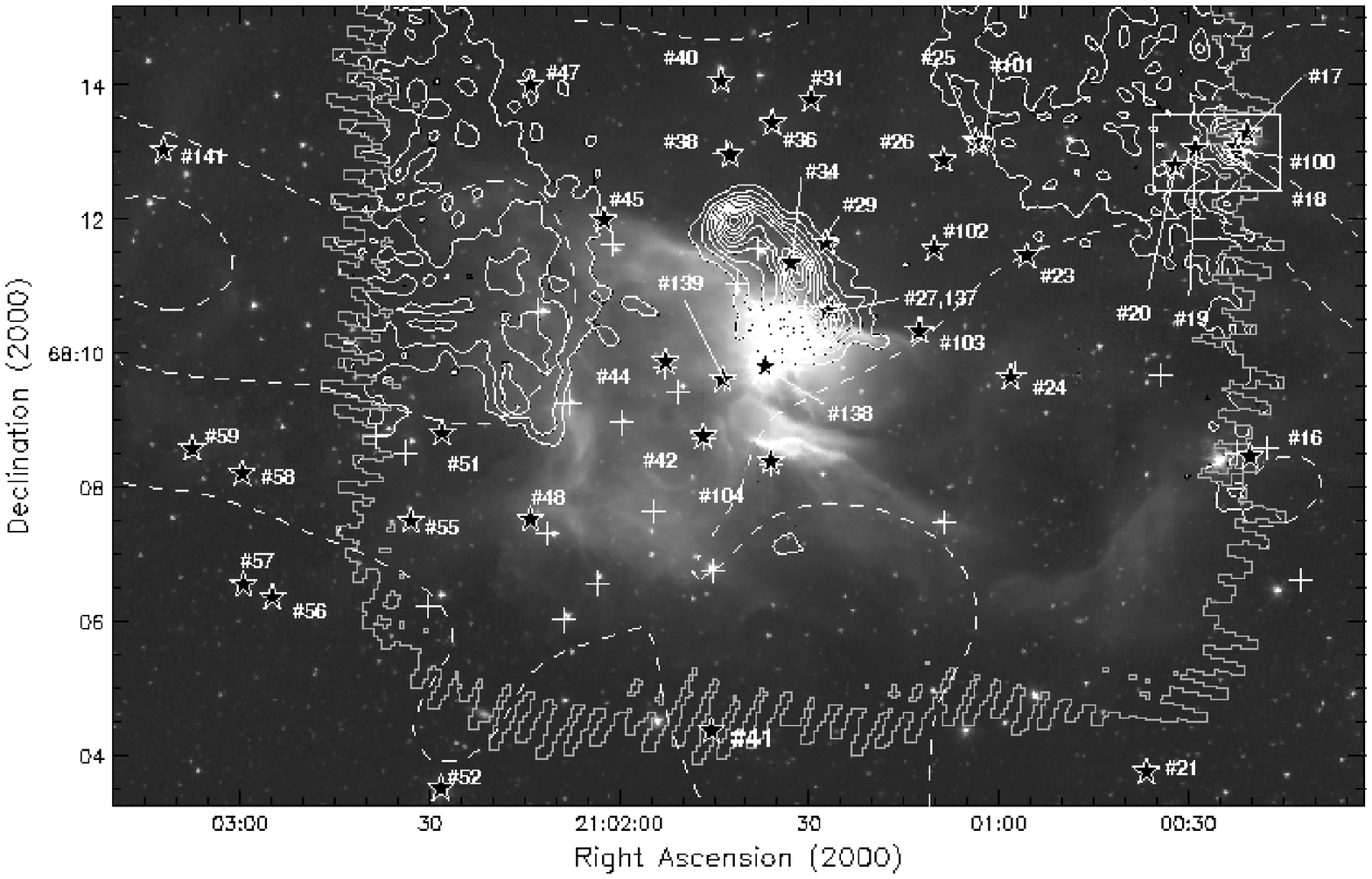}} \\
	\end{tabular}
	
	\contcaption{(...continued)}
\end{figure*}
\clearpage
\begin{figure*}
	\begin{tabular}{ll}
		\vspace{2mm} h) L1172D & i) L1221 \\
		\vspace{2mm} \includegraphics[height=3in]{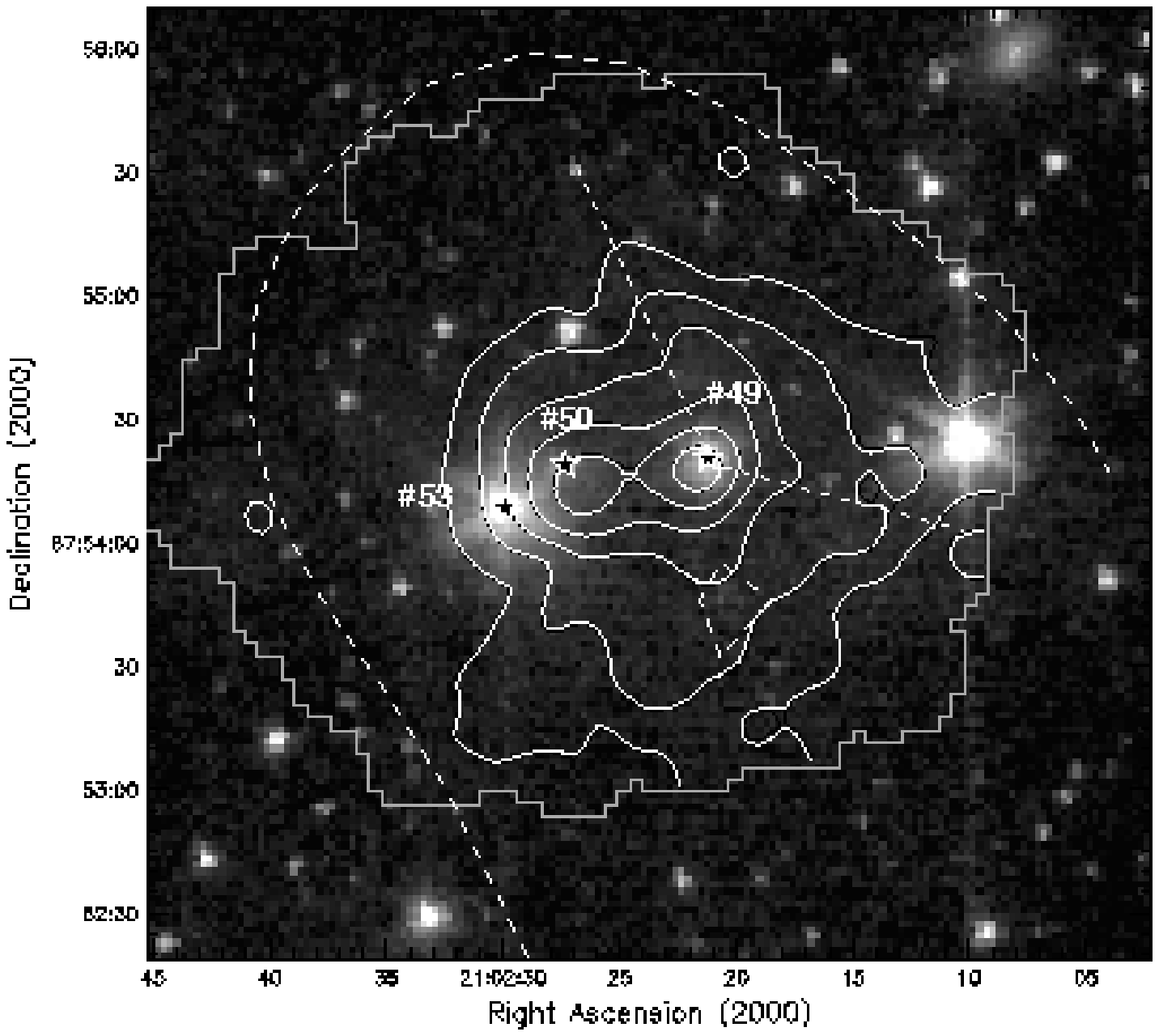} &
			\includegraphics[height=2.5in]{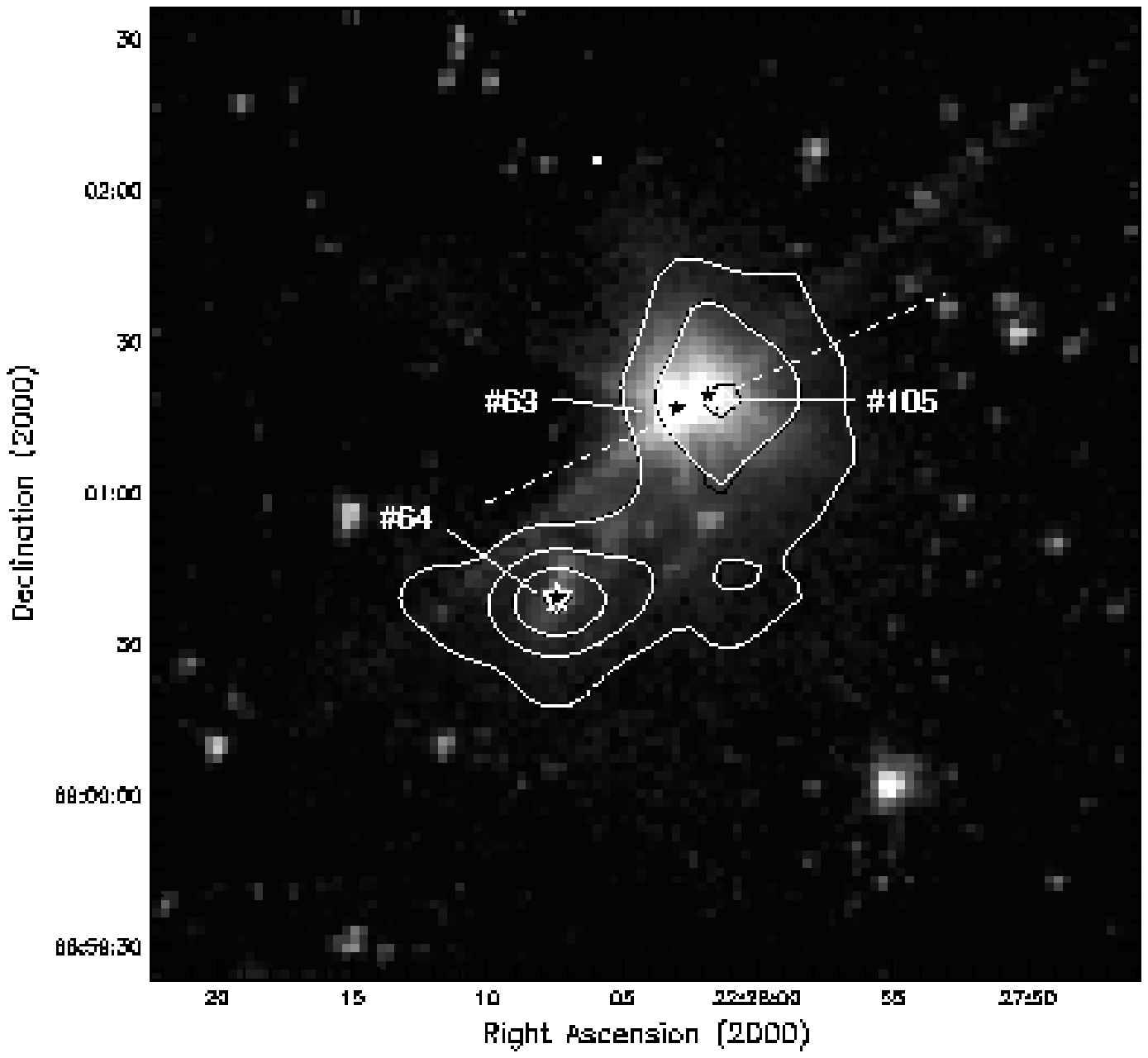} \\
		\vspace{2mm} j) L1228N & k) L1228S \\
		\vspace{2mm} \includegraphics[width=3in]{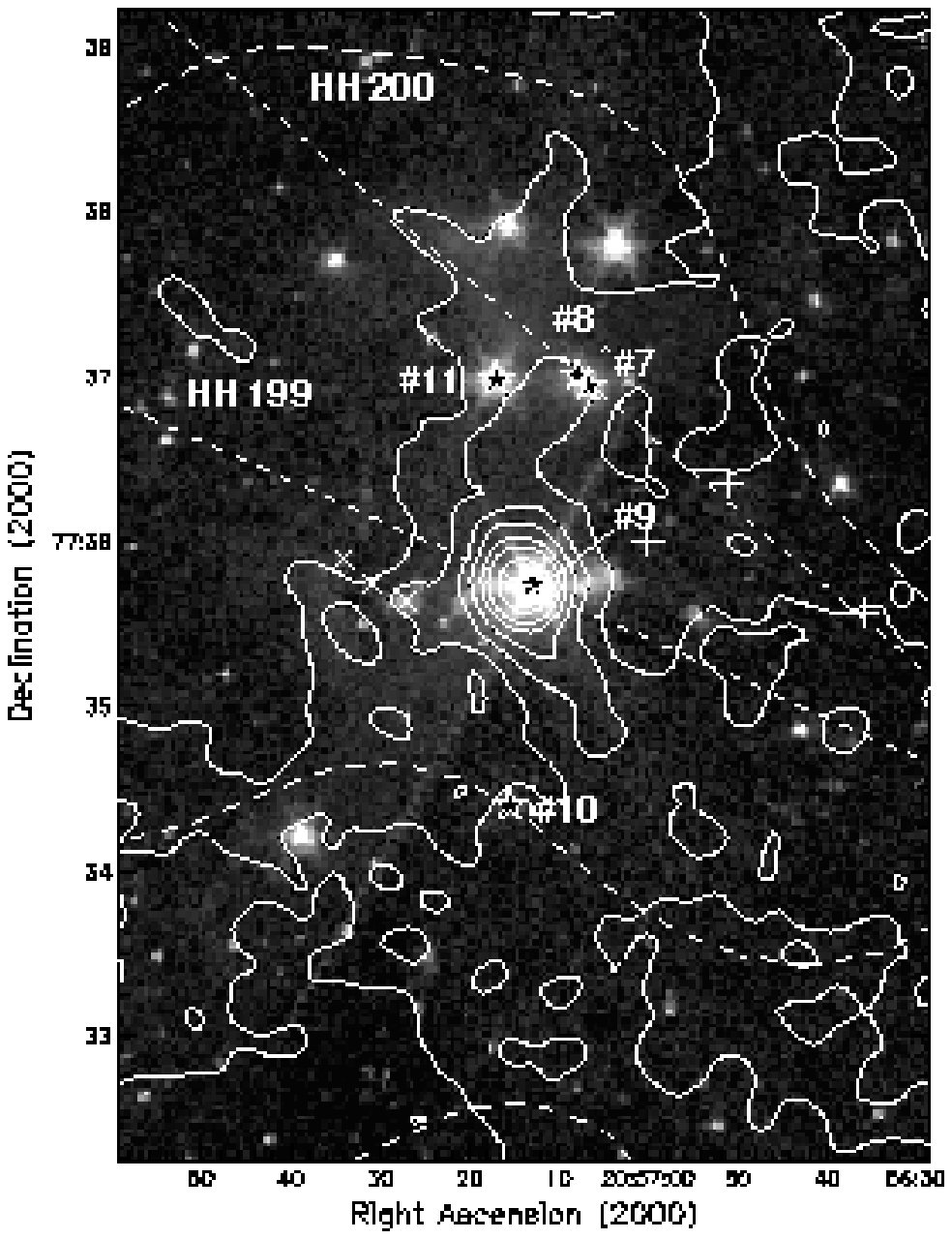} &
			\includegraphics[height=4in]{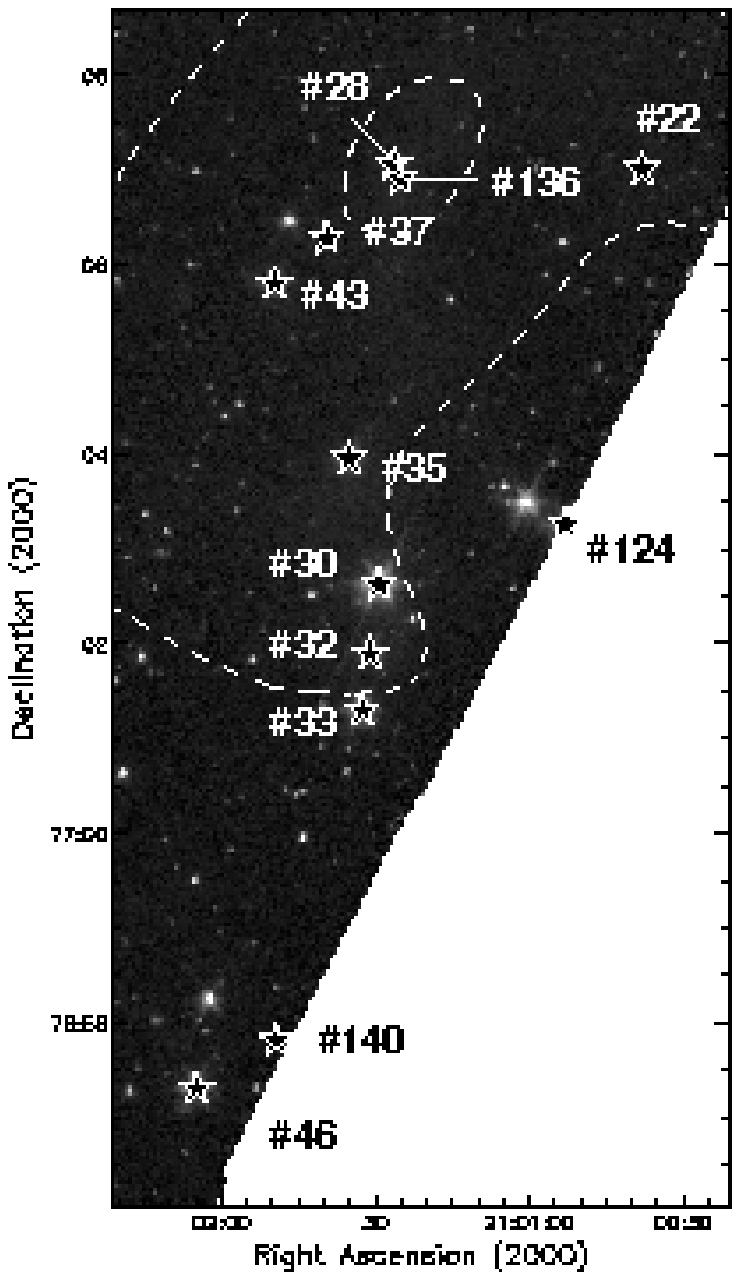} \\
	\end{tabular}
	\contcaption{(...continued)}
\end{figure*}
\clearpage
\begin{figure*}
	\begin{tabular}{ll}
		\vspace{2mm} l) L1251A  & m) L1251B \\
		\vspace{2mm} \includegraphics[height=3in]{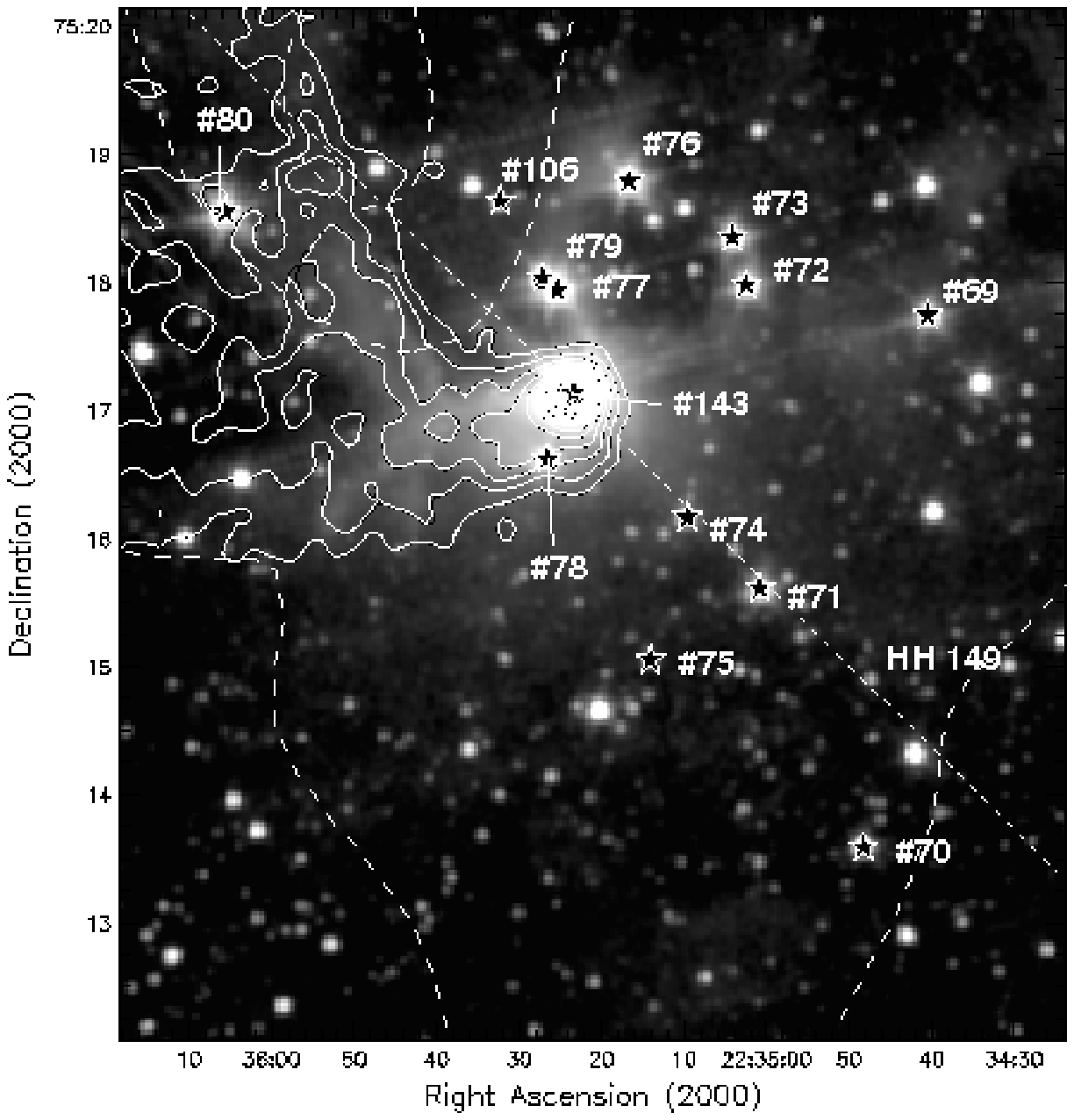} &
		\includegraphics[height=3in]{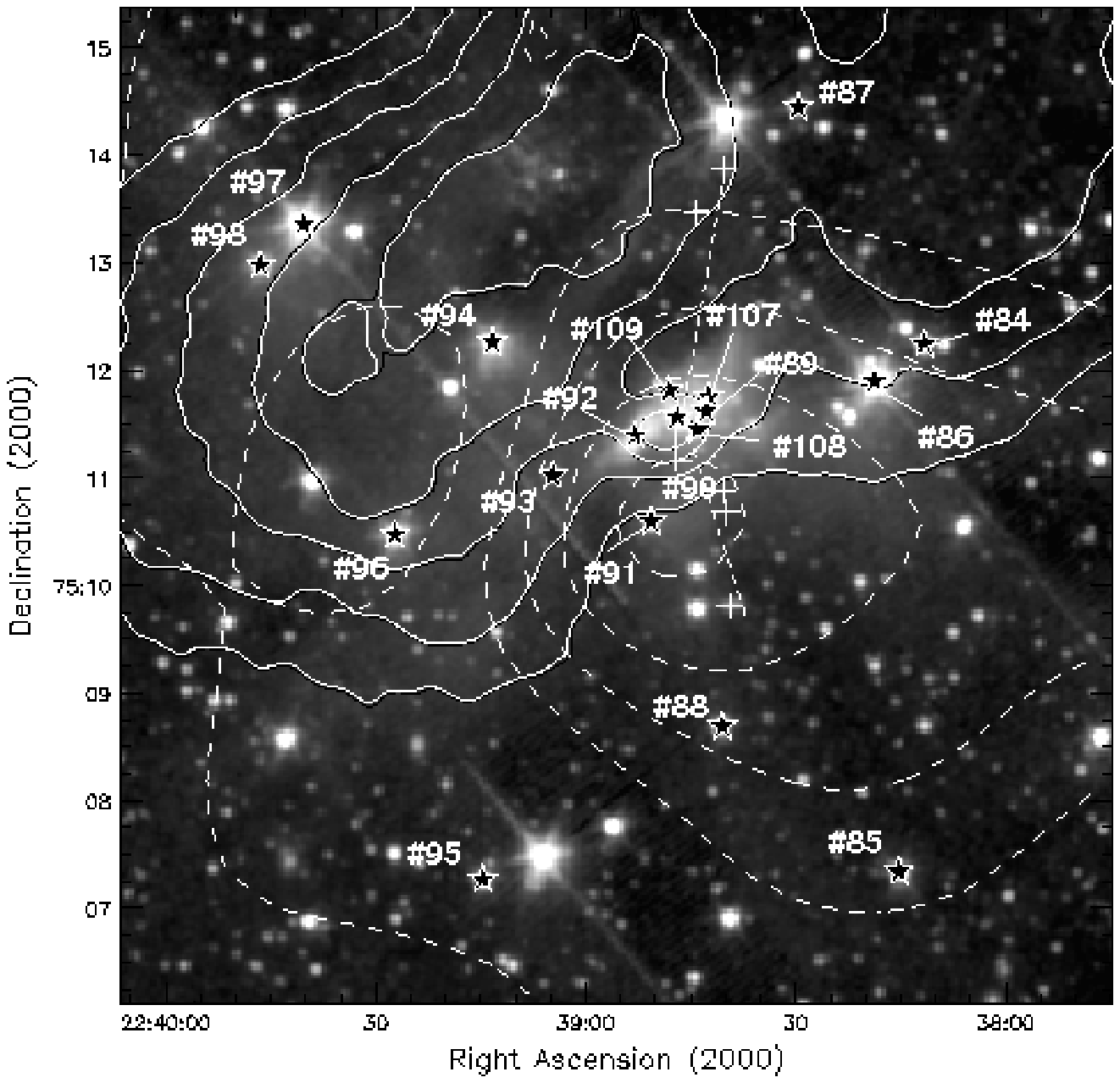} \\
		\vspace{2mm} n) L1251W & \\
		\vspace{2mm} 
		\includegraphics[height=2in]{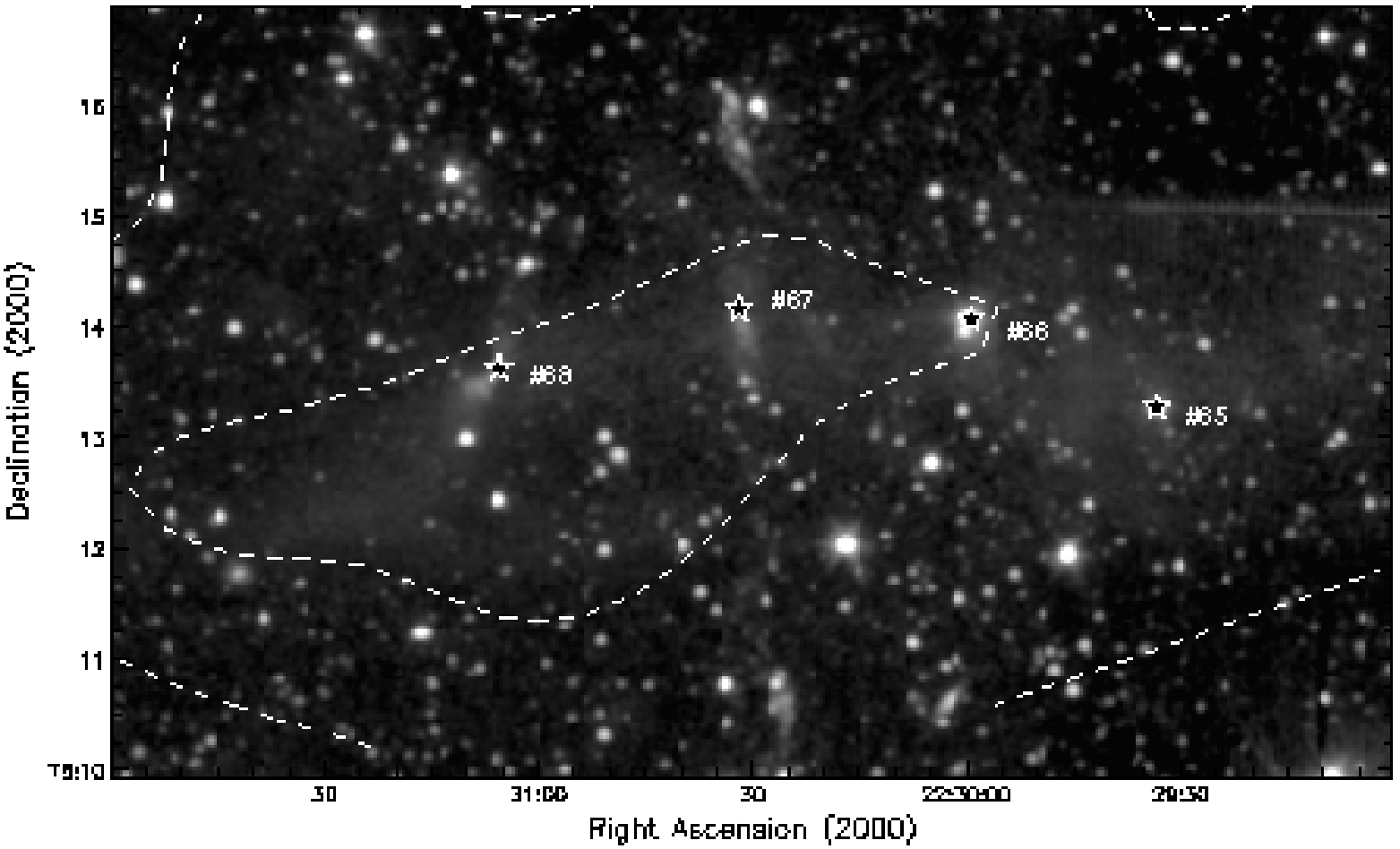} & \\

	\end{tabular}
	\contcaption{(...continued)}
\end{figure*}
	

\clearpage	
\begin{deluxetable}{lcccclrcc}
	\rotate
	\tablecaption{\label{tab:assocprop} Dark Cloud Association Properties}
	\tablecolumns{9}
	\tablewidth{0pt}
	\tablehead{
		\colhead{Association} & \colhead{$l$} & \colhead{$b$} & \colhead{R.A.} & \colhead{Dec.} & \colhead{Distance} & \colhead{$M_{\rm cloud}$} & \colhead{$\Delta V(^{13}\textrm{CO})$} & \colhead{IRAC/MIPS Area} \\
		 & \colhead{[\degr]} & \colhead{[\degr]} & \colhead{(2000)} & \colhead{(2000)} &  \colhead{[pc]} & \colhead{[$M_{\odot}$]} & \colhead{[km\,s$^{-1}$]} & \colhead{[Sqr Deg]} \\
		}
	\startdata
		L1148+L1152+L1155	& 102.4 & 15.5 & 20\lath 40\latm & 67\degr 40\arcmin & 325$\pm$13 & 830 & 1.0 & 0.31/2.46 \\
		L1172+L1174		& 104.0 & 14.2 & 21\lath 02\latm & 68\degr 10\arcmin & 288$\pm$25 & 730 & 1.6 & 0.45/1.93 \\
		L1221		& 110.7 & 9.73 & 22\lath 28\latm & 69\degr 07\arcmin & 250$\pm$50 & 40 & 1.5 & 0.05/0.40 \\
		L1228		& 111.7 & 20.2 & 20\lath 57\latm & 77\degr 35\arcmin & $200^{+100}_{-20}$ & 230 & 1.6 & 0.44/0.77 \\
		L1241		& 113.4 & 16.9 & 22\lath 00\latm & 76\degr 30\arcmin & $300^{+50}_{-10}$ & 1500 & 2.2 & 0.30/1.67 \\
		L1247+L1251	& 114.1 & 14.8 & 22\lath 30\latm & 75\degr 10\arcmin & $300^{+50}_{-10}$ & 790 & 1.9 & 0.56/1.10 \\	
	\enddata
\end{deluxetable}
	
\clearpage
\begin{deluxetable}{lrllllllll}
\rotate
\tablecaption{\label{tab:aors} AORs in the SGBS Cepheus Catalog }
\tablecolumns{10}
\tablewidth{0pt}
\tabletypesize{\footnotesize}
\tablehead{
\colhead{Dark Cloud}	&	\colhead{PID}	&	\multicolumn{4}{c}{IRAC AORs}	&	\multicolumn{4}{c}{MIPS AORs} 	\\
\colhead{Association}	&		&	\colhead{Epoch 1}	&	\colhead{Epoch 2}	&	\colhead{Date}	&	\colhead{Pipeline}	&	\colhead{Epoch 1}	&	\colhead{Epoch 2}	&	\colhead{Date}	&	\colhead{Pipeline}	\\
}
\startdata
L1148	&	30574	&	0019963136	&	0019963392	&	2006-11-26	&	S15.0.5	&	0019962624	&	0019962880	&	2007-02-26	&	S15.3.0	\\
	&	30574	&	0019995136	&	0019995648	&	2006-11-26	&	S15.0.5	&	\nodata	&	\nodata	&	\nodata	&	\nodata	\\
L1152	&	139	&	0005156352	&	\nodata	&	2004-07-03	&	S13.2.0	&	0009438976	&	\nodata	&	2004-09-22	&	S13.2.0	\\
L1155C	&	139	&	0005158656	&	\nodata	&	2004-07-03	&	S13.2.0	&	0009408768	&	\nodata	&	2004-12-02	&	S13.2.0	\\
L1155E	&	139	&	0005159168	&	\nodata	&	2004-07-03	&	S13.2.0	&	0009441792	&	\nodata	&	2004-09-22	&	S13.2.0	\\
L1172	&	30574	&	0019965184	&	0019965696	&	2006-11-26	&	S15.0.5	&	0019994624	&	0019994880	&	2007-02-26	&	S15.3.0	\\
	&	30574	&	0019996928	&	0019997184	&	2006-11-26	&	S15.0.5	&	\nodata	&	\nodata	&	\nodata	&	\nodata	\\
	&	30574	&	0025037312	&	0025037568	&	2008-03-12	&	S17.0.4	&	\nodata	&	\nodata	&	\nodata	&	\nodata	\\
\enddata
\end{deluxetable}

\clearpage
\begin{deluxetable}{ll r@{.}l r@{.}l cc}
\tablecaption{\label{tab:bands} Spectral Bands In This Survey}
\tablecolumns{8}
\tablewidth{0pt}
\tabletypesize{\footnotesize}
\tablehead{
	\colhead{Origin} & \colhead{Filter} & \multicolumn{2}{c}{Wavelength} & \multicolumn{2}{c}{$S_{\rm 0pt}$} & \colhead{Limiting} & \colhead{Ref}   \\
	& & \multicolumn{2}{c}{[\micron]} & \multicolumn{2}{c}{[Jy]} & \colhead{Magnitude\tablenotemark{a}}   \\
}
\startdata
GSC-II\tablenotemark{b}		& $B$		& 0&44	& 4260&		&\nodata&  1 \\
GSC-II\tablenotemark{b}		& $V$		& 0&55	& 3640&		&\nodata&  1 \\
GSC-II\tablenotemark{b}		& $R$		& 0&64	& 3080&		&\nodata&  1 \\
GSC-II\tablenotemark{b}		& $I$		& 0&79	& 2550&		&\nodata&  1 \\
2MASS		& $J$		& 1&24	& 1594&		& 16.3 &  2 \\
2MASS		& $H$		& 1&66	& 1024&		& 15.5 &  2 \\
2MASS		& $K_{s}$	& 2&16	&  666&7	& 15.3 &  2 \\
IRAC		& IRAC1		& 3&6	&  280&9	& 17.2 &  3 \\
\enddata
\tablenotetext{a}{Taken as the turnover in the S/N limited source count distribution.}
\tablenotetext{b}{The GSC-II data have been converted into the Johnson-Cousins photometric system from the natural systems used by its component surveys}   
\tablerefs{ (1) \citealt{1979bessell}; (2) \citealt{2006skrutskie}; (3) \citealt{2007evans}; (4) \citealt{1984neugebauer}; (5) \citealt{1999holland}  }
\end{deluxetable}
\clearpage
\begin{deluxetable}{lr}
\tablecaption{\label{tab:stat} Detection (S/N~$\geq$~3) statistics across the delivered area }
\tablecolumns{2}
\tablewidth{0pt}
\tabletypesize{\footnotesize}
\tablehead{
\colhead{Detected with/in\ldots} & \colhead{Detections} \\
}
\startdata
\ldots at least one IRAC band &    71076 \\
\ldots all four IRAC bands &     6518 \\
\ldots three IRAC bands &     6167 \\
\ldots two IRAC bands &    34841 \\
\ldots only one IRAC band &    23558 \\
\ldots MIPS 1 &     2140 \\
\ldots all four IRAC bands and MIPS 1 &      888 \\
\ldots all four IRAC bands and with 2MASS\tablenotemark{a} &     3839 \\
\ldots all four IRAC bands, but not with 2MASS\tablenotemark{a} &     2679 \\
\ldots 2MASS alone\tablenotemark{a}	 &        1 \\
\ldots IRAC, but not in 2MASS\tablenotemark{a} &    67080 \\
\cutinhead{Excluding Extended Sources} 
\\
\ldots all four IRAC bands &     6183 \\
\ldots all four IRAC bands and with 2MASS\tablenotemark{a} &     3821 \\
\ldots IRAC1+2 and 2MASS\tablenotemark{a} &     3971 \\
\enddata
\tablenotetext{a}{A 2MASS detection is taken as one with S/N~$>$~10 in both $H$ and $K_{s}$}
\end{deluxetable}

\clearpage
\begin{deluxetable}{lrrrrrr}
\tablecaption{\label{tab:filterstats} Detection of sources above S/N Thresholds
for the entire mapped area at each wavelength.}
\tablecolumns{7}
\tablewidth{0pt}
\tablehead{
& \multicolumn{6}{c}{Wavelength (\micron)} \\ 
\colhead{Detection} 	 & \colhead{3.6} & \colhead{4.5} & \colhead{5.8} & \colhead{8.0} & \colhead{24} & \colhead{70}  \\ }
\startdata

S/N~$>$~ 3	&    80010 &    61001 &    14672 &    10221 &     9338 &      392 \\
S/N~$>$~ 5	&    63148 &    44093 &     8391 &     6172 &     2559 &      349 \\
S/N~$>$~ 7	&    48854 &    32061 &     6402 &     4589 &     1021 &      180 \\
S/N~$>$~ 10	&    32037 &    19894 &     4270 &     3079 &      389 &        2 \\
S/N~$>$~ 15	&    14672 &     9379 &     2082 &     1626 &        1 &        1  \\
\enddata
\end{deluxetable}

\clearpage
\begin{deluxetable}{r c p{1.5in} r@{ $\pm$ }l r@{ $\pm$ }l r@{ $\pm$ }l r@{ $\pm$ }l r@{ $\pm$ }l r@{ $\pm$ }l l l}
\rotate
\tablecaption{\label{tab:ysofluxes} Cepheus YSO Candidates}
\tablecolumns{17}
\tabletypesize{\scriptsize}
\tablewidth{0pt}
\tablehead{
\colhead{Id.} & \colhead{Name/Position} & \colhead{Literature  Name\tablenotemark{a}} & \multicolumn{2}{c}{3.6\micron} & \multicolumn{2}{c}{4.5\micron} & \multicolumn{2}{c}{5.8\micron} & \multicolumn{2}{c}{8.0\micron} & \multicolumn{2}{c}{24.0\micron} & \multicolumn{2}{c}{70.0\micron} & \colhead{Flags\tablenotemark{b}} & \colhead{Ref.} \\
& (SSTgbsJ) & & \multicolumn{2}{c}{[mJy]} & \multicolumn{2}{c}{[mJy]} & \multicolumn{2}{c}{[mJy]} & \multicolumn{2}{c}{[mJy]} & \multicolumn{2}{c}{[mJy]} & \multicolumn{2}{c}{[mJy]} & &  \\
}
\startdata
\cutinhead{1. 5-Band YSO Candidates}
 1 & 2035463+6753020 & {\raggedright L1152~1; [K98b]~4} &     3.53&   0.19 &     6.30&   0.33 &     5.45&   0.27 &     3.31&   0.16 &     273.&    25. &    3670.&   341. & F\phm{I}\phm{M} & 1; 8 \\
 2 & 2036116+6757093 & {\raggedright L1152~2} &     41.3&    2.0 &     47.9&    2.3 &     56.2&    2.7 &     74.5&    3.6 &     173.&    16. &     347.&    33. & FIM & 1 \\
 3 & 2036198+6756316 & {\raggedright L1152~3; [K98b]~6; [K98c]~EM*6; GM~3-12} &     228.&    12. &     243.&    16. &     215.&    10. &     262.&    13. &     634.&    59. &    1080.&   102. & FIM & 1; 8; 10; 15 \\
 4 & 2039430+6708301 & {\raggedright } &     2.39&   0.12 &     2.93&   0.14 &     3.46&   0.18 &     4.39&   0.21 &     9.20&   0.87 & \multicolumn{2}{c}{$\cdots$} & FI\phm{M} &  \\
 5 & 2040567+6723050 & {\raggedright L1148-IRS; [K98b]~8} &     1.32&   0.07 &     1.31&   0.06 &    0.797&  0.049 &     1.16&   0.06 &     75.3&    7.0 &     331.&    33. & FI\phm{M} & 2; 8 \\
 6 & 2055371+7738196 & {\raggedright L1228~4} &     5.96&   0.29 &     5.13&   0.25 &     4.25&   0.22 &     4.21&   0.20 &     4.05&   0.39 & \multicolumn{2}{c}{$\cdots$} & FIM & 1 \\
 7 & 2057067+7736561 & {\raggedright L1228~5; HH~200~IRS} &     6.87&   0.38 &     9.72&   0.48 &     12.5&    0.6 &     16.3&    0.8 &     269.&    25. &    2060.&   236. & FIM & 1; 15 \\
 8 & 2057079+7736597 & {\raggedright L1228~6} &     3.03&   0.15 &     3.35&   0.17 &     3.59&   0.18 &     8.53&   0.41 &     153.&    14. & \multicolumn{2}{c}{$\cdots$} & FIm & 1 \\
\enddata
\tablerefs{The SIMBAD source prefix, where relevant, is listed after each reference in square brackets (a lower case letter denotes a particular table in that reference). 1 - \citealt{2008chapman}, 2 - \citealt{2005kauffmann}, 3 - \citealt{2004padgett}, 4 - \citealt{2008young}, 5 - \citealt{2006lee}, 6 - \citealt{2002visser}, 7 - \citealt{1983sellgren}, 8 - \citealt{1998kun} [K98], 9 - \citealt{1993kun} [K93], 10 - \citealt{1977gyul} [GM], 11 - \citealt{1996rodriguez} [RR96], 12 - \citealt{1996toth} [TW96]. 13 - \citealt{1998rodriguez} [RR98], 14 - \citealt{1995rosvick} [RD95], 15 - \citealt{2008kun} }
\tablenotetext{a}{For c2d papers (refs 1--5) the literature name, where one is given, is taken as the core name followed by the index or name given to the source in the actual paper. }
\tablenotetext{b}{YSO identification flag. The source was identified as a YSO candidate by:  F - the 5-Band method with a 24\micron\ detection, f - the 5-Band method without a 24\micron\ detection, I - the IRAC colour-magnitude method, M - the 2MASS/MIPS colour-magnitude method (greater than 1 mag away from the limits), m - the 2MASS/MIPS colour-magnitude method (less than 1 mag away from the limits) }
\end{deluxetable}

\clearpage
\begin{deluxetable}{l l r@{.}l c r@{.}l c r@{.}l r@{.}l }
\tablecaption{\label{tab:ysoprop}YSO Properties}
\tablecolumns{12}
\tabletypesize{\scriptsize}
\tablewidth{0pt}
\tablehead{ 
	\colhead{Index} & \colhead{Region} & \multicolumn{2}{c}{$\alpha_{\rm IR}$} & \colhead{$\alpha_{\rm IR}$} & \multicolumn{2}{c}{$T_{\rm bol}$} & \colhead{$T_{\rm bol}$} & \multicolumn{2}{c}{$L_{\rm bol}$} & \multicolumn{2}{c}{$L_{\rm IR}$} \\ 
	& & \multicolumn{2}{c}{  } & \colhead{Class} & \multicolumn{2}{c}{[K]} & \colhead{Class} & \multicolumn{2}{c}{[$L_{\odot}$]} & \multicolumn{2}{c}{[$L_{\odot}$]} \\  
	& & \multicolumn{2}{c}{(a)} & & \multicolumn{2}{c}{(b)} & \colhead{(c)} & \multicolumn{2}{c}{(d)} & \multicolumn{2}{c}{(e)} \\ }
\startdata
 1 & L1152 &     1&0 & I &      33&  & 0&    1&6 &   0&060 \\
 2 & L1152 &  -0&31 & II &      160&  & I&  0&50 &   0&34 \\
 3 & L1152 &  -0&50 & II &      270&  & I&    2&1 &     1&6 \\
 4 & L1147 &  -0&089 & F &      630&  & F&  0&017 &   0&016 \\
 5 & L1148 &   0&81 & I &      81&  & I&  0&44 &   0&017 \\
 6 & L1228N &    -1&2 & II &      1500&  & II&  0&023 &   0&019 \\
 7 & L1228N &   0&66 & I &      54&  & 0&  0&22 &   0&037 \\
 8 & L1228N &   0&76 & I &      35&  & 0&  0&13 &   0&019 \\
\enddata
\tablenotetext{a}{The infrared spectral index, $\alpha_{\rm IR}$, calculated by a least-squares fit between 2MASS $K_s$ and 24~\micron. Where IRAC data were not available $\alpha_{\rm IR}$ was calculated from the $K_s$-[24] colour as described in Section \ref{classification}}.
\tablenotetext{b}{The bolometric temperature, $T_{\rm bol}$, calculated using all available data from $B$-band to 1200~\micron.} 
\tablenotetext{c}{The equivalent class calculated from $T_{\rm bol}$ (see Section \ref{seds}). } 
\tablenotetext{d}{The bolometric luminosity, $L_{\rm bol}$, calculated using all available data from $B$-band to 1200~\micron. } 
\tablenotetext{e}{The infrared luminosity, $L_{\rm IR}$, calculated using all available data from 1 to 30~\micron. } 
\end{deluxetable}

\clearpage
\begin{deluxetable}{rccc}
\tablecaption{\label{tab:propStats}Class statistics of \spitzer\ identified YSOs}
\tablecolumns{4}
\tablewidth{0pt}
\tablehead{
	\colhead{Class} & \multicolumn{3}{c}{No. of YSOs} \\
	& \colhead{$\alpha_{\rm IR}$} & \colhead{$T_{\rm bol}$} & \colhead{c2d\tablenotemark{a}} \\
}
\startdata
0 	& \nodata 	&  17 ( 13\%) & \nodata \\
I 	& 21 ( 16\%) 	&  10 (  8\%) & 16\% \\
``Flat'' & 14 ( 11\%) 	&  37 ( 28\%) & 12\% \\
II 	& 87 ( 65\%) 	&  64 ( 48\%) & 60\% \\
III 	& 11 (  8\%) 	&   5 (  4\%) & 12\%\\
\enddata
\tablenotetext{a}{Total YSO statistics for all c2d regions \citep{2008evans}.} 
\end{deluxetable}16, 12, 60, and 12

\clearpage
\begin{deluxetable}{l l r@{.}l r@{.}l r@{.}l r@{.}l@{\hspace{1cm}} l r@{.}l r@{.}l r@{.}l r@{.}l}
\tablecaption{\label{tab:otherflux}Additional Fluxes}
\tablecolumns{18}
\tabletypesize{\scriptsize}
\tablewidth{0pt}
\tablehead{
 \colhead{Id.} & \colhead{GSC-II Name} & \multicolumn{2}{c}{$B$} & \multicolumn{2}{c}{$V$} & \multicolumn{2}{c}{$R$} & \multicolumn{2}{c}{$I$} & \colhead{IRAS Name} & \multicolumn{2}{c}{12.5\micron} & \multicolumn{2}{c}{25\micron} & \multicolumn{2}{c}{60\micron} & \multicolumn{2}{c}{100\micron} \\
& & \multicolumn{2}{c}{[mJy]} & \multicolumn{2}{c}{[mJy]} & \multicolumn{2}{c}{[mJy]} & \multicolumn{2}{c}{[mJy]} & & \multicolumn{2}{c}{[Jy]} & \multicolumn{2}{c}{[Jy]} & \multicolumn{2}{c}{[Jy]} & \multicolumn{2}{c}{[Jy]}  \\
}
\startdata
 1 &  & \multicolumn{2}{c}{$\cdots$} & \multicolumn{2}{c}{$\cdots$} & \multicolumn{2}{c}{$\cdots$} & \multicolumn{2}{c}{$\cdots$} & F20353+6742 & $<$ 0&0522 &  0&302 &    4&00 &    7&00 \\
 2 & N1KR010841 & \multicolumn{2}{c}{$\cdots$} & \multicolumn{2}{c}{$\cdots$} &  0&109 &  0&641 &  & \multicolumn{2}{c}{$\cdots$} & \multicolumn{2}{c}{$\cdots$} & \multicolumn{2}{c}{$\cdots$} & \multicolumn{2}{c}{$\cdots$} \\
 3 &  & \multicolumn{2}{c}{$\cdots$} & \multicolumn{2}{c}{$\cdots$} & \multicolumn{2}{c}{$\cdots$} & \multicolumn{2}{c}{$\cdots$} & F20358+6746 &  0&447 &  1&000 &    1&68 & $<$   5&71 \\
 4 & N1KP003099 & \multicolumn{2}{c}{$\cdots$} & \multicolumn{2}{c}{$\cdots$} &  0&0302 &  0&185 &  & \multicolumn{2}{c}{$\cdots$} & \multicolumn{2}{c}{$\cdots$} & \multicolumn{2}{c}{$\cdots$} & \multicolumn{2}{c}{$\cdots$} \\
 5 &  & \multicolumn{2}{c}{$\cdots$} & \multicolumn{2}{c}{$\cdots$} & \multicolumn{2}{c}{$\cdots$} & \multicolumn{2}{c}{$\cdots$} & F20404+6712 & $<$ 0&0636 &  0&126 &  0&320 & $<$   4&39 \\
 6 & N0Y2007286 & \multicolumn{2}{c}{$\cdots$} & \multicolumn{2}{c}{$\cdots$} &  0&0872 &  0&715 &  & \multicolumn{2}{c}{$\cdots$} & \multicolumn{2}{c}{$\cdots$} & \multicolumn{2}{c}{$\cdots$} & \multicolumn{2}{c}{$\cdots$} \\
 7 & N0Y4008344 & \multicolumn{2}{c}{$\cdots$} & \multicolumn{2}{c}{$\cdots$} &    1&62 &    6&17 &  & \multicolumn{2}{c}{$\cdots$} & \multicolumn{2}{c}{$\cdots$} & \multicolumn{2}{c}{$\cdots$} & \multicolumn{2}{c}{$\cdots$} \\
 8 & N0Y4008341 & \multicolumn{2}{c}{$\cdots$} & \multicolumn{2}{c}{$\cdots$} &  0&0215 & \multicolumn{2}{c}{$\cdots$} &  & \multicolumn{2}{c}{$\cdots$} & \multicolumn{2}{c}{$\cdots$} & \multicolumn{2}{c}{$\cdots$} & \multicolumn{2}{c}{$\cdots$} \\
\enddata
\end{deluxetable}

\clearpage
\begin{deluxetable}{l r@{=}l l}
	\tablecaption{\label{tab:transforms} Color-color transforms to the Cousins-Johnson System}
	\tablecolumns{3}
	\tablewidth{0pt}
	\tablehead{ \colhead{System} & \multicolumn{2}{c}{Transforms} & \colhead{Reference} \\}
	\startdata
	Quick	& $V$ & $V_{\textrm{Quick}}$ & \\
	\tableline
	Tycho 	& $V$ & $V_{\rm T} - 0.090 (B_{\rm T}-V_{\rm T})$ &  The Hipparcos and Tycho  \\
		& $B$ & $ 0.850 (B_{\rm T}-V_{\rm T}) - V$ & Catalogs \citep{1997esa} \\
	\tableline
	POSS-II & $B$ & $(B_{\rm J} - 0.28V ) / 0.72$ & \citealt{1982blair} \\
		& $R$ & $R_{\rm F}$ & \\
		& $I$ & $I_{\rm N}$ & \\
	\enddata
\end{deluxetable}

\clearpage
\begin{deluxetable}{ r l r@{ $\pm$ }l r@{ $\pm$ }l r@{ $\pm$ }l r@{ $\pm$ }l l}
	\tablecaption{\label{tab:submm} MIPS 160~\micron, SCUBA 450 \& 850~\micron, and MAMBO 1200~\micron\ photometry of Cepheus YSO candidates}
	\tablewidth{0pt}
	\tablecolumns{11}
	\tabletypesize{\small}
	\tablehead{
		\colhead{YSO \#} & 
		\colhead{Region} & 
		\multicolumn{2}{c}{160\micron} & 
		\multicolumn{2}{c}{450\micron} & 
		\multicolumn{2}{c}{850\micron} &
		\multicolumn{2}{c}{1200\micron} &
		\colhead{Ref.\tablenotemark{a}} \\
		&  & \multicolumn{2}{c}{[Jy]} & \multicolumn{2}{c}{[Jy]} & \multicolumn{2}{c}{[Jy]} & \multicolumn{2}{c}{[Jy]} & \\
	}
	
	\startdata
	1	& L1152    & 8.8 & 0.3 &  8.3 & 4.4 & 1.2 & 0.2  &  \multicolumn{2}{c}{\nodata}  & 1 \\
	2	& L1152	   & 1.6 & 0.1 &   \multicolumn{2}{c}{\nodata}  &  \multicolumn{2}{c}{\nodata} &  \multicolumn{2}{c}{\nodata}  \\
	3	& L1152	   & 1.6 & 0.1 &   \multicolumn{2}{c}{\nodata}  &  \multicolumn{2}{c}{\nodata} &  \multicolumn{2}{c}{\nodata}  \\
	5	& L1148    & \multicolumn{2}{c}{\nodata} &   \multicolumn{2}{c}{\nodata}  &  \multicolumn{2}{c}{\nodata} & 0.059 & 0.007 & 4 \\
	7	& L1228N   & 3.9 & 0.1 &   \multicolumn{2}{c}{\nodata}  &  \multicolumn{2}{c}{\nodata} &  \multicolumn{2}{c}{\nodata}  \\
	8	& L1228N   & 3.9 & 0.1 &   \multicolumn{2}{c}{\nodata}  &  \multicolumn{2}{c}{\nodata} &  \multicolumn{2}{c}{\nodata}  \\
	9	& L1228N   & 15 & 1 & 1.1 & 1.4 & 2.0 & 0.3 & 0.89 & 0.01 &  1, 4 \\
	17	& L1174	   & 4.2 & 0.1 &   \multicolumn{2}{c}{\nodata}  &  \multicolumn{2}{c}{\nodata} &  \multicolumn{2}{c}{\nodata}  \\
\enddata
	\tablerefs{1 - \citet{2006young}, 2 - \citet{2006lee}, 3 - \citet{2002visser}, 4 - \citet{2008kauffmann} }

	\tablenotetext{a}{The 850 and 450~\micron\ aperture radius was 20\arcsec\ for \citet{2006young} and 25\arcsec\ for \cite{2002visser}. The \citet{2006young} flux for L1251B was portioned between the sources by \citet{2006lee} based on the relative brightness of the sources at 24~\micron. The \citet{2008kauffmann} 1200~\micron\ fluxes were measured in a 20\arcsec radius aperture. The 160~\micron\ fluxes were measured in a 40\arcsec\ radius aperture. }
	\tablenotetext{b}{The errors for these sources are taken as the same as the relative errors for the combined source, L1251E-1, from \citet{2006young}.}
	
\end{deluxetable}
\clearpage	
\begin{deluxetable}{rccc}
\tablecaption{\label{tab:sed1} Class 0/I derived submillimeter parameters}
\tablecolumns{4}
\tablewidth{0pt}
\tablehead{
	\colhead{Id.} & \colhead{$T_{\rm dust}$} & \colhead{$L_{\rm submm}$} & \colhead{$L_{\rm submm}/L_{\rm bol}$} \\
	& \colhead{[K]} & \colhead{[$L_{\odot}$]} &  \colhead{[\%]}\\
}
\startdata
 1 &      17. &   0.080 &     5.0 \\
 49 &      15. &   0.054 &     9.4 \\
 63 &      18. &   0.13 &     4.3 \\
 64 &      17. &   0.13 &     9.1 \\
 90 &      16. &   0.50 &     3.1 \\
 134 &      18. &   0.25 &     2.8 \\
 135 &      19. &   0.17 &   0.37 \\
 143 &      19. &   0.23 &   0.69 \\
\enddata
\end{deluxetable}

\clearpage
\begin{deluxetable}{l r@{.}l r r@{.}l r@{.}l r@{.}l r@{.}l }
\tablecaption{\label{tab:class2model} SED modelling parameters for Class II YSOs}
\tablecolumns{12}
\tabletypesize{\scriptsize}
\tablewidth{0pt}
\tablehead{ 
	\colhead{Index} & \multicolumn{2}{c}{$A_{\rm V}$} &  \colhead{$\lambda_{\rm norm}$} & \multicolumn{2}{c}{$L_{\rm star}$} & \multicolumn{2}{c}{$\log(L_{\rm disc}/L_{\rm star})$} & \multicolumn{2}{c}{$\lambda_{\rm turnoff}$} & \multicolumn{2}{c}{$\alpha_{\rm excess}$} \\
	 & \multicolumn{2}{c}{ } &  \colhead{[\micron]} & \multicolumn{2}{c}{[$L_{\odot}$]} & \multicolumn{2}{c}{ } & \multicolumn{2}{c}{[\micron]} & \multicolumn{2}{c}{ }  \\ }
\startdata
 2 &     2&3 & $J$ &   0&31 &   0&20 &     1&7 &  -0&28 \\
 3 &     9&5 & $J$ &     3&5 &  -0&29 &     3&6 &  -0&55 \\
 6 &   0&50 & $J$ &   0&030 &  -0&76 &     5&7 &    -1&0 \\
 10 &     3&8 & $K_s$ &   0&0040 &  -0&045 &     3&6 &  -0&70 \\
 11 &     9&0 & $J$ &   0&29 &  -0&38 &     2&2 &  -0&87 \\
 12 &      13&  & $J$ &   0&099 &     1&1 &     2&2 &  -0&74 \\
 13 &     3&0 & $J$ &   0&25 &  -0&93 &     5&7 &  -0&86 \\
 14 &     7&7 & $J$ &   0&21 &  -0&67 &     3&6 &    -1&3 \\
\enddata
\end{deluxetable}

\clearpage
\begin{deluxetable}{l r@{.}l r r@{.}l r@{.}l r@{.}l r@{.}l }
\tablecaption{\label{tab:class3model} SED modelling parameters for Class III YSOs}
\tablecolumns{12}
\tabletypesize{\scriptsize}
\tablewidth{0pt}
\tablehead{ 
	\colhead{Index} & \multicolumn{2}{c}{$A_{\rm V}$} &  \colhead{$\lambda_{\rm norm}$} & \multicolumn{2}{c}{$L_{\rm star}$} & \multicolumn{2}{c}{$\log(L_{\rm disc}/L_{\rm star})$} & \multicolumn{2}{c}{$\lambda_{\rm turnoff}$} & \multicolumn{2}{c}{$\alpha_{\rm excess}$} \\
	 & \multicolumn{2}{c}{ } &  \colhead{[\micron]} & \multicolumn{2}{c}{[$L_{\odot}$]} & \multicolumn{2}{c}{ } & \multicolumn{2}{c}{[\micron]} & \multicolumn{2}{c}{ }  \\ }
\startdata
 21 &     1&3 & $J$ &   0&46 &  -0&57 &      24&  & \multicolumn{2}{c}{\nodata } \\
 58 &     2&9 & $J$ &   0&35 &  -0&99 &     7&9 & \multicolumn{2}{c}{\nodata } \\
 62 &     4&5 & $J$ &      150&  &    -2&1 &      25&  & \multicolumn{2}{c}{\nodata } \\
 95 &     1&4 & $J$ &   0&16 &    -1&1 &     7&9 & \multicolumn{2}{c}{\nodata } \\
 107 &     1&8 & $J$ &   0&36 &  -0&41 &     7&9 & \multicolumn{2}{c}{\nodata } \\
 108 &     4&5 & $J$ &   0&19 &    -1&0 &     5&7 & \multicolumn{2}{c}{\nodata } \\
 112 &   0&00 & $J$ &   0&35 &  -0&75 &     2&2 & \multicolumn{2}{c}{\nodata } \\
 116 &   0&00 & $J$ &   0&15 &  -0&62 &     7&9 & \multicolumn{2}{c}{\nodata } \\
 117 &     3&3 & $J$ &      25&  &    -2&2 &     2&2 & \multicolumn{2}{c}{\nodata } \\
 119 &   0&39 & $J$ &   0&82 &    -3&2 &     2&2 & \multicolumn{2}{c}{\nodata } \\
 129 &   0&00 & $J$ &     2&7 &    -1&5 &      24&  &   0&80 \\
 136 &     4&8 & $J$ & \multicolumn{2}{c}{\nodata } & \multicolumn{2}{c}{\nodata } &     1&7 & \multicolumn{2}{c}{\nodata } \\
 139 &     1&4 & $J$ &   0&37 &    -2&1 &     7&9 & \multicolumn{2}{c}{\nodata } \\
\enddata
\end{deluxetable}

\clearpage

\begin{deluxetable}{ l rrrrrr rr r rr rrr  }
	\tablecaption{\label{tab:cores} Statistics of star forming dark clouds in Cepheus}
	\tablecolumns{15}
	\tablewidth{0pt}
	\tabletypesize{\scriptsize}
	\tablehead{
		\colhead{Dark Cloud} & \multicolumn{6}{c}{\spitzer\ identified YSOs\tablenotemark{a}} & \multicolumn{2}{c}{non-S. YSOs\tablenotemark{b}} & \colhead{$N_{\rm YSO}$} & \multicolumn{2}{c}{$A_V$} & \colhead{Area\tablenotemark{c}} & \colhead{$M_{\rm cloud}$} & \colhead{SFE} \\
		
		& \colhead{I} & \colhead{F} & \colhead{II} & \colhead{III} & \colhead{Tot.} & \colhead{I/Tot.} & \colhead{Inc.} & \colhead{Tot.} & \colhead{$A_V>5$} & \colhead{Max} & \colhead{Mean} & & & \\
		
		& & & & & & \colhead{[\%]} & & & & \colhead{[mag]} & \colhead{[mag]} & \colhead{[pc$^{-2}$]} & \colhead{[$M_{\odot}$]} & \colhead{[\%]} \\
		
		}
\startdata
																												
\sidehead{Prestellar Cores}																														
\tableline																														
L1155		&	0	&	0	&	0	&	0	&	0	&	0	&	0	&	0	&	0	&	\nodata	&	\nodata	&	\nodata	&	\nodata	&	\nodata	\\
L1241		&	0	&	0	&	0	&	1	&	1	&	0	&	0	&	0	&	0	&	11.2	&	6.2	&	0.058	&	150	&	0	\\
L1247		&	0	&	0	&	0	&	0	&	0	&	0	&	0	&	0	&	0	&	8.9	&	6.2	&	0.015	&	38	&	0	\\
\tableline																														
Sub Total		&	0	&	0	&	0	&	1	&	1	&		&	0	&	0	&	0	&		&		&	0.073	&	188	&		\\
\sidehead{Protostellar Cores (peak $\rho_{*} <$ disruption density)}																														
\tableline																														
L1148+L1147		&	1	&	1	&	2	&	0	&	4	&	25	&	0	&	0	&	2	&	8.8	&	6.0	&	0.051	&	150	&	1	\\
L1152		&	1	&	0	&	2	&	0	&	3	&	33	&	0	&	0	&	0	&	\nodata	&	\nodata	&	\nodata	&	\nodata	&	\nodata	\\
L1157		&	0	&	0	&	0	&	0	&	0	&	0	&	1	&	1	&	0	&	\nodata	&	\nodata	&	\nodata	&	\nodata	&	\nodata	\\
L1172		&	1	&	1	&	3	&	1	&	6	&	17	&	0	&	0	&	4	&	13.1	&	7.0	&	0.040	&	107	&	2	\\
\enddata

\tablenotetext{a}{The number of \spitzer\ identified YSO candidates by Class (as determined by the $\alpha_{\rm IR}$ spectral index) associated with the named core (within the projected $A_V=1$ contour) or cluster (within the relevant tight or loose contour).}
\tablenotetext{b}{The number of non-\spitzer\ identified YSOs as listed in Table \ref{tab:simbadYSOs}. The column labeled `Inc.' lists the number of non-\spitzer\ identified YSOs that were coincident with entries in the Cepheus catalog. The column labeled `Tot.' lists the total number of YSOs missed by the \spitzer\ YSO schemes in that region.}
\tablenotetext{c}{The area of the region used to calculate the mean $A_V$, the cloud mass $M_{\rm cloud}$, and the star formation efficiency (SFE). For a core, this is the area that is enclosed by the $A_V=5$ contour. For a loose and tight cluster, this is the area that is enclosed by the $\rho_{*} = 1.0 $ and $ 25 \textrm{ M}_{\odot}\textrm{ pc}^{-3}$ contours respectively.}
\end{deluxetable}
\clearpage
\begin{deluxetable}{l r@{:}c@{:}l r@{:}c@{:}l c l c l }
	\tablecaption{\label{tab:simbadYSOs} Non-identified SIMBAD YSOs}
	\tablecolumns{11}
	\tablewidth{0pt}
	\tabletypesize{\scriptsize}
	\tablehead{
		\colhead{Source Name\tablenotemark{a}} & \multicolumn{3}{c}{Right Ascension} & \multicolumn{3}{c}{Declination} & \colhead{Origin\tablenotemark{b}} & \colhead{Region} & \colhead{\spitzer} & \colhead{\spitzer} \\
		 & \multicolumn{3}{c}{(2000)} & \multicolumn{3}{c}{(2000)} &  &  & \colhead{Catalog.\tablenotemark{c}} & \colhead{YSO Id\tablenotemark{d}} \\ }
		
	\startdata
	L1157 			& 20 & 39 & 06.2 & +68 & 02 & 15 & S  &  L1157		& Y & 134 \\
	PV Cep			& 20 & 45 & 53.9 & +67 & 57 & 39 & S  &  Off-cloud	& Y & 135 \\
	IRAS F20597+6800	& 21 & 00 & 18.4 & +68 & 12 & 02 & S  &  NGC 7023	& N & \nodata \\
	L1228 S1(b)		& 21 & 01 & 25.1 & +77 & 06 & 54 & K  &  L1228		& Y & 136 \\
	NGC 7023 RS 2		& 21 & 01 & 27.1 & +68 & 10 & 38 & K  &  NGC 7023	& Y & 137 \\
	IRAS F21022+7651	& 21 & 01 & 29.4 & +77 & 03 & 45 & S  &  L1228		& N & \nodata \\
	HD 200775		& 21 & 01 & 36.9 & +68 & 09 & 48 & S  &  NGC 7023	& Y & 138 \\
	NGC 7023 RS 3		& 21 & 01 & 43.6 & +68 & 09 & 36 & K  &  NGC 7023	& Y & 139 \\
\enddata
	
	\tablenotetext{a}{Name of the source in either the SIMBAD database or \citet{2008kun}.}
	\tablenotetext{b}{The origin of the YSO candidate. S - a YSO or pre-main sequence listing in the SIMBAD database, K - a T Tauri star from \citet{2008kun}}
	\tablenotetext{c}{Does the source have an entry in the Cepheus \spitzer\ catalog? In this instance, the only condition for inclusion is to have a line entry in the \spitzer\ catalog irrespective of whether that source was formally detected in any particular band. }
	\tablenotetext{d}{The YSO identification number from Table \ref{tab:ysofluxes} for sources that are found in the \spitzer\ Cepheus catalog. }
\end{deluxetable}
\clearpage
\begin{deluxetable}{lccccccccc}
	\tablecaption{\label{tab:regionstats} Star formation statistics for the Cepheus Flare}
	\tablecolumns{10}
	\tablewidth{0pt}
	\tabletypesize{\scriptsize}
	\tablehead{
		\colhead{Boundary} & \colhead{$N_{\rm YSO}$} & \colhead{$\Omega$\tablenotemark{a}} & \colhead{$A$\tablenotemark{b}} & \colhead{$N_{\rm YSO}/\Omega$} & \colhead{$N_{\rm YSO}/A$} & \colhead{SFR\tablenotemark{c}} & \colhead{SFR$/A$} & \colhead{M$_{\rm cloud}$} & \colhead{M$_{\rm cloud}/A$} \\ 
		\colhead{ } & \colhead{ } & \colhead{[deg$^{2}$]} & \colhead{[pc$^{2}$]} & 
		\colhead{[deg$^{-2}$]} & \colhead{[pc$^{-2}$]} & \colhead{[M$_\odot$ Myr$^{-1}$]} & \colhead{[M$_\odot$ Myr$^{-1}$ pc$^{-2}$]} & \colhead{[$M_\odot$]} & \colhead{[M$_\odot$ pc$^{-2}$]} \\ 
	}
	\startdata
		$A_V < 1$ &  19 &  6.2 & 170 & 3.1 & 0.11 & \nodata & \nodata & \nodata & \nodata \\
		$A_V > 1$ & 114 &  2.1 & 59 &  53 & 1.9 & 29 & 0.49 &\nodata & \nodata \\
		$A_V > 5$ &  92	&  0.40 & 11 & 230 & 8.4 & 58 & 2.1 & 1003 & 91 \\
	\enddata
	\tablenotetext{a}{The solid angle enclosed by the contour}
	\tablenotetext{b}{The cross sectional area assuming an average distance of 300~pc (1 sqr deg = 27.4~pc$^2$)}
	\tablenotetext{c}{The star formation rate (SFR) calculated assuming a star formation timescale of 2~Myr and an average stellar mass of 0.5~M$_\odot$}
\end{deluxetable}

\clearpage
\begin{deluxetable}{lllll}
	\tablecaption{\label{tab:outflows} Known outflows in Cepheus }
	\tablecolumns{5}
	\tablewidth{0pt}
	\tabletypesize{\scriptsize}
	\tablehead{ \colhead{Region} & \colhead{Lit. Name} & \colhead{Assoc. YSO} & \colhead{HH No.} & \colhead{Example References} \\ }
	
	\startdata
	Off-core	& PV Cep	& 135	& 215, 315, 415 & \citealt{2004goodman} \\
	L1152		& GM 3-12		& 3	& 376 & \citealt{1997reipurth,2004movsessian} \\
	L1152		& L1152 SMM	& 1	&\nodata & \citealt{1996bontemps} \\
	L1157		& L1157 IRS	& 134	& 375 	& \citealt{1992umemoto, 2007looney} \\
	L1172D		& L1172 SMM 1	& 49	& \nodata & \citealt{1988myers} \\
	L1174		& HD 200775	& 136	& \nodata & \citealt{1998fuente} \\
	L1221		& L1221 IRS 1	& 105	& \nodata & \citealt{1991umemoto, 2005lee} \\
	L1228N		& HH 200 IRS	& 7	& 200	& \citealt{1995bally} \\
	L1228N		& HH 199 1 & 9	& 199		& \citealt{1995bally} \\
	L1251B		& \nodata & Undetermined & 189	& \citealt{1994eiroa} \\
	L1251A		& \nodata	& 143   & 149  & \citealt{1989sato, 1992balazs} \\ 
	\enddata
\end{deluxetable}

\end{document}